\documentclass[10pt, a4paper]{article} 
\usepackage[square,numbers,sort&compress]{natbib}
\usepackage{booktabs}
\usepackage[shortlabels]{enumitem}
\usepackage{setspace}
\usepackage{subfloat}
\usepackage{mathtools}
\usepackage{amsmath,amsfonts,amssymb,amsthm,mathrsfs, bm}
\usepackage[utf8]{inputenc}
\usepackage[swedish, english]{babel}
\usepackage{csquotes}
\usepackage{subfig}
\usepackage{graphicx}
\usepackage{supertabular}                                             
\usepackage[table]{xcolor} 
\usepackage{textgreek} 
\usepackage{graphicx}
\usepackage{epstopdf}
\usepackage{microtype}
\usepackage{multirow}
\usepackage{xcolor}
\usepackage{authblk}
\usepackage{datetime2}	
\usepackage[margin=1in]{geometry} 
\usepackage{setspace}	
\theoremstyle{plain}
\newtheorem{theorem}{Theorem}[section]
\newtheorem{lemma}{Lemma}[section]

\newtheorem{assumption}{Assumption}[section]

\usepackage{amsmath,amsfonts,amssymb,amsthm,mathrsfs, bm}
\usepackage{stmaryrd}

\usepackage{graphicx}
\usepackage{tikz}
\usepackage[utf8]{inputenc}
\usepackage{pgfplots} 
\usepackage{pgfgantt}
\usepackage{pdflscape}
\pgfplotsset{compat=newest} 
\pgfplotsset{plot coordinates/math parser=false}

\usepackage{endnotes}
\usepackage{commath}
\usepackage{gensymb}
\usepackage{mathtools}
\usepackage{tikz} 
\usetikzlibrary{calc,quotes,angles}
\usepackage{tkz-euclide}
\usetikzlibrary{automata,positioning} 

\usepackage{xcolor}
\usepackage{tikz,tkz-euclide,siunitx}
\usepackage{etoolbox} 

\usetikzlibrary{quotes,arrows.meta,angles,calc,shadings,positioning,babel}

\newtheorem{definition}{Definition}[section]

\newtheorem{remark}{Remark}[section]

\usepackage{endnotes}
\usepackage{commath}
\usepackage{gensymb}
\usepackage{tcolorbox}

\makeatletter
\patchcmd{\tkz@DrawLine}{\begingroup}{\begingroup\makeatletter}{}{}
\makeatother

\allowdisplaybreaks
\DeclareMathOperator{\D}{D}

\DeclareMathOperator{\diag}{\mathrm{diag}}
\DeclareMathOperator{\sgn}{sgn}

\DeclareMathOperator{\esup}{ess\, sup}

\makeatletter
\newcommand\makebig[2]{%
  \@xp\newcommand\@xp*\csname#1\endcsname{\bBigg@{#2}}%
  \@xp\newcommand\@xp*\csname#1l\endcsname{\@xp\mathopen\csname#1\endcsname}%
  \@xp\newcommand\@xp*\csname#1r\endcsname{\@xp\mathclose\csname#1\endcsname}%
}
\makeatother

\providecommand*{\ped}[1]{%
\ensuremath{_\textnormal{#1}}}
\providecommand*{\ap}[1]{%
\ensuremath{^\textnormal{#1}}}
\providecommand*{\eu}%
{\ensuremath{\mathrm{e}}}
\providecommand*{\im}%
{\ensuremath{\mathrm{i}}}
\providecommand*{\GammaF}%
{\ensuremath{\mathrm{\Gamma}}}
\providecommand*{\BetaF}%
{\ensuremath{\mathrm{\Beta}}}

\makebig{biggg} {3.0}
\makebig{Biggg} {3.5}
\makebig{bigggg}{4.0}
\makebig{Bigggg}{4.5}
\makebig{biggggg}{5.0}
\makebig{Biggggg}{5.5}
\usepackage{longtable}
\usepackage{xcolor}
\DeclareMathSymbol{\Gamma}{\mathalpha}{letters}{"00}
\DeclareMathSymbol{\Delta}{\mathalpha}{letters}{"01}
\DeclareMathSymbol{\Theta}{\mathalpha}{letters}{"02}
\DeclareMathSymbol{\Lambda}{\mathalpha}{letters}{"03}
\DeclareMathSymbol{\Xi}{\mathalpha}{letters}{"04}
\DeclareMathSymbol{\Pi}{\mathalpha}{letters}{"05}
\DeclareMathSymbol{\Sigma}{\mathalpha}{letters}{"06}
\DeclareMathSymbol{\Upsilon}{\mathalpha}{letters}{"07}
\DeclareMathSymbol{\Phi}{\mathalpha}{letters}{"08}
\DeclareMathSymbol{\Psi}{\mathalpha}{letters}{"09}
\DeclareMathSymbol{\Omega}{\mathalpha}{letters}{"0A}

\definecolor{matblue}{rgb}{0.0000,0.4470,0.7410}
\definecolor{matred}{rgb}{0.8500,0.3250,0.0980}
\definecolor{matyellow}{rgb}{0.9290,0.6940,0.1250}
\definecolor{matpurple}{rgb}{0.4940,0.1840,0.5560}
\definecolor{matgreen}{rgb}{0.4660,0.6740,0.1880}
\definecolor{matcyan}{rgb}{0.3010,0.7450,0.9330}
\definecolor{matmaroon}{rgb}{0.6350,0.0780,0.1840}

\usepackage{empheq}
\usepackage{hyperref}
\hypersetup{
    colorlinks = true,
    linkbordercolor = {white},
            linkcolor=blue,
            bookmarks=true,
            filecolor=blue,
            urlcolor=blue,
            citecolor=blue
}

\newtcolorbox[auto counter]{modelbox}[2][]{%
  colback=white, colframe=black,
  fonttitle=\bfseries,
  title=Models~\thetcbcounter: #2,
  label=#1
}

\begin{document}

\title{Hyperbolic dynamic friction models\\ for viscoelastic sliding and rolling contact\footnote{This document is the author version of the manuscript.}}
\date{}
\author[a,b]{Luigi Romano\thanks{Corresponding author. Email: luigi.romano@liu.se.}}
\affil[a]{\footnotesize{Division of Vehicular Systems, Department of Electrical Engineering, Linköping University, SE-581 83 Linköping, Sweden}}
\affil[b]{\footnotesize{Control Systems Technology Group, Department of Mechanical Engineering, Eindhoven University of Technology, Groene Loper 1, 5612 AZ Eindhoven, the Netherlands}}

\maketitle

\begin{abstract}
This paper considers the sliding and rolling contact between viscoelastic bodies. Combining linear viscoelastic rheologies for bristle-like elements with nonlinear dynamic friction models, and under the assumption of small deformations and displacement gradients, it derives a class of \emph{viscoelasto-kinematic equations}, formulated as a system of semilinear \emph{partial differential equations} (PDEs) governing the evolution of the frictional force, bristle deformations, and internal state variables at the interface between the contacting bodies. The resulting system is analysed mathematically, demonstrating that linear viscoelasticity preserves the hyperbolic character of the PDE systems typically encountered in rolling contact. The proposed theory is illustrated through representative examples of both sliding and rolling contact, highlighting that these two processes, whilst often treated as distinct, may in fact exhibit closely related underlying dynamics. Overall, the framework provides a general theoretical setting applicable to a broad class of viscoelastic frictional systems.
\end{abstract}
\section*{Keywords}
Sliding and rolling friction; friction modelling; contact mechanics; hyperbolic PDEs

\section{Introduction}\label{intro}
Friction is a key physical phenomenon governing the response of many mechanical and mechatronic components \cite{Persson1,Persson2}, with applications spanning servo systems, pneumatic and hydraulic actuators \cite{Flores1,Flores3,NonlinearDynId,2D}, robotic and tribological elements \cite{SphereNoSlip1,CarboneTrans,TransModel,Belt1,Frendo1,Frendo2,Frendo3,bearing1,bearing2,bearing3,bearing4,bearing5}, and interfaces involving rolling contact \cite{Knothe,KalkerBook,2000RCP,Guiggiani,Mavros,LibroMio,Gauterin,Gauterin2}.
Because of its ubiquity in engineering applications, friction has been the subject of extensive theoretical investigation, leading to a proliferation of analytical modelling frameworks aimed at capturing its complex behaviour. In particular, a major distinction is often drawn between sliding and rolling regimes. These are typically regarded as distinct phenomena and, in the context of dynamic friction modelling, are described using different mathematical formalisms, with sliding friction commonly represented through systems of \emph{ordinary differential equations} (ODEs), and rolling friction formulated in terms of \emph{partial differential equations} (PDEs).

Amongst the many dynamic modelling approaches proposed for sliding friction, the LuGre model \cite{Astrom1,Olsson,Astrom2} constitutes a particularly influential reference point. Inspired by rate- and state-dependent formulations originally proposed in geophysics by Rice and Ruina \cite{Rice}, it provides a relatively simple and computationally efficient framework that has become widely adopted, largely due to its ability to reproduce a broad spectrum of qualitative frictional phenomena.
Departing from the original LuGre structure, several extensions have been proposed in the literature \cite{Integrated,Leuven,Elasto1,Elasto2,GMS} to better reproduce additional physical effects, in particular those related to memory and hysteresis in the pre-sliding regime. The Leuven model \cite{Integrated,Leuven}, for instance, introduces explicit nonlocal hysteresis effects, albeit at the cost of a substantially more involved implementation. An alternative development is provided by the Generalised Maxwell-slip (GMS) class of models \cite{GMS}, which incorporates rate-state-type dynamics within the slip phases of classical Maxwell-slip elements. These formulations improve the representation of displacement-driven hysteresis and memory effects whilst preserving a comparatively compact structural form. Similar to LuGre, but grounded on a solid physical foundation, the recently introduced FrBD models can successfully capture viscoelastic sliding contact and have also been extended to address rolling friction \cite{FrBD,FrBDroll,FrBDvisc}. More recently, bristle-based frameworks have also found application to texture-dynamics-exploiting control of dry friction \cite{MengaSugg}. However complex, all these descriptions share a common mathematical structure: they are expressed as systems of nonlinear ODEs governing the evolution of internal bristle-like states, in combination with constitutive rheological relations that map bristle deformation to the resulting friction force. The finite-dimensional (i.e., ODE) nature of these sliding friction models is often rooted in the common misconception that associates sliding friction with the absence of material advection at the contact interface. This assumption is valid, for instance, when a soft block of material slides and spins over a rigid plane, but it clearly breaks down when the substrate is compliant or when both contacting bodies are deformable.

Concerning instead rolling friction, in which forces and deformations are clearly advected through the contact patch as a consequence of the rolling motion, the dominant theoretical framework is that developed by Kalker, who pioneered the analysis of similarly elastic bodies in rolling contact \cite{Kalker51,Kalker5,KalkerPhD,KalkerBook}, with application to cylinders and wheel-rail interactions. Kalker's models are typically formulated as PDEs of hyperbolic type, and more specifically as transport equations, which may include nonlinear (semilinear) terms.
Extensions to viscoelastic rolling contact, developed for instance by Hunter \cite{Hunter}, Goryacheva \cite{GoryachevaP,Goryacheva4,Goryacheva1,Goryacheva2,GoryachevaBook}, and Kalker himself \cite{Panek1,Panek2}, lead to more involved systems of equations, which nonetheless retain the same inherent transport-like structure \cite{Guler,Julia1,Julia2,
Nielsen,Alfredsson}. This peculiar feature is a consequence of the fact that, in standard formulations of rolling contact, the interacting bodies are supposed to advect internal variables with comparable velocities. This key modelling assumption enormously simplifies the analytical treatment of the problem, but can also introduce conceptual and technical difficulties.
Such issues are particularly evident in the so-called \emph{brush models} widely used in road vehicle dynamics, and essentially corresponding to Kalker's simplified theory, in which the full elastic Green's functions (Cerruti--Boussinesq operators) are replaced by local Winkler-type constitutive laws \cite{KalkerBook,Johnson,KalkerSimp}. In this context, it is not uncommon to find ambiguities between the translational velocity of the vehicle and the effective rolling speed of the tyre in the definition of slip and spin variables \cite{Pacejka2,Jazar}. Brush-type models are also widely used in the modelling of wheel-rail interactions \cite{Kalker4,Vollebregt3,Gross,Alonso1,Alonso2,Ciavarella1,Ciavarella2,Ciavarella3,Al-Bender,USB,LibroMio,Meccanica2,SphericalWheel,LuGreSpin,Tribology}, and even the railway engineering community does not seem to be entirely immune to confusion \cite{Knothe}. 

For similar elastic bodies, or in the case of strongly incommensurate elastic pairs (where one body is significantly more rigid than the other), these inconsistencies, whilst conceptually dangerous, are not fatal. Indeed, in such settings, the kinematics of the material particles (or, equivalently, of the bristle-like elements advected within the contact region) are largely independent of the assumed rheological law. However, when rolling contact occurs between elastically dissimilar bodies, this decoupling no longer holds, and the constitutive behaviour of the contact pair enters explicitly into the kinematic relationships, as recently demonstrated in \cite{KinematicsMio}, where the notion of \emph{elasto-kinematic} equations was introduced. In particular, the analysis conducted in \cite{KinematicsMio} was restricted to purely elastic bodies, and adopted a Winkler-type approximation to describe the tangential interaction. In many practical applications, however, rolling and sliding contact involve highly viscoelastic materials, such as the polymers commonly used in tyres \cite{bookRheol1,bookRheol2,bookRheol3}. In these scenarios, more advanced constitutive descriptions that explicitly account for viscoelastic effects become necessary \cite{JMPS1,JMPS2,JMPS3}, in order to capture internal stress relaxation over a broad frequency range \cite{Rheology2,Rheology2-3,Rheology3}. Motivated by this, the present paper extends the analysis of \cite{KinematicsMio} by considering both sliding and rolling friction for viscoelastic contact pairs.

\begin{figure}
\centering
\includegraphics[width=1\linewidth]{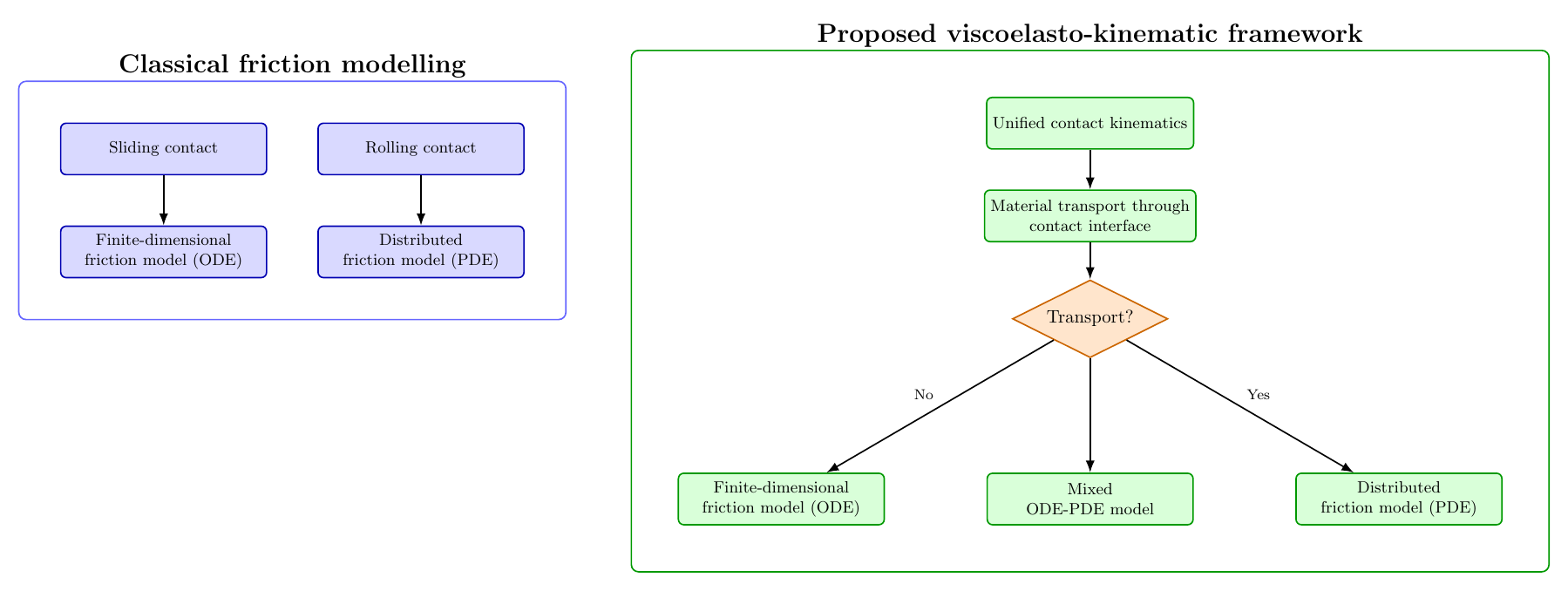} 
\caption{Distinction between the classical friction modelling approaches, based on ODE (sliding) and PDE (rolling) models, and the proposed viscoelasto-kinematic framework.}
\label{fig:mathFrame2}
\end{figure}

More specifically, restricting attention to the classical derivative-based treatment \cite{Fractional1,Fractional2,Fractional3} of linear viscoelasticity, the present work extends the theory initiated in \cite{KinematicsMio} by adopting rheological descriptions equivalent to Generalised Kelvin-Voigt (GKV) and Generalised Maxwell (GM) elements, coupled with nonlinear evolution equations -- borrowed from established friction models -- to describe the dynamics of bristle-like elements at the contact interface.
Within this framework, the governing \emph{viscoelasto-kinematics} PDEs of the frictional process can be derived in the form of a semilinear hyperbolic system, describing the evolution of tangential contact forces together with the internal state variables associated with the viscoelastic behaviour of the bristle-like elements. In this context, it is worth clarifying that, for the sake of simplicity, the analysis is confined to planar contact configurations and to situations involving localised contact, such as non-conformal, for which the half-space approximation is appropriate; however, the derivation of the viscoelasto-kinematic equations presented in this paper would extend analogously to the full three-dimensional setting (see, e.g., \cite{KinematicsMio}).
A key finding of this work is that distributed and mixed formulations, expressed in terms of PDEs or ODE--PDE systems, also arise naturally in the context of sliding friction. This perspective helps clarify the traditional dichotomy between sliding models formulated in terms of ODEs and rolling models described by PDEs, suggesting that this distinction is artificial.
In fact, beyond the mathematical formulation itself, the proposed framework provides a new conceptual viewpoint on friction modelling: rather than constructing different mathematical formulations for different contact configurations, the present approach shows that sliding, rolling, and mixed rolling-sliding processes are governed by the same viscoelasto-kinematic principles. Consequently, the governing equations, their dimensionality, and their characteristic structure become direct consequences of the underlying contact kinematics and constitutive behaviour, rather than modelling assumptions introduced \emph{a priori}. The distinction between the classical friction modelling approaches and the proposed viscoelasto-kinematic framework is provided pictorially in Fig. \ref{fig:mathFrame2}.

The results presented in this paper build upon and extend those contained in \cite{FrBDvisc,KinematicsMio}: compared to \cite{FrBDvisc}, a more general formulation for viscoelastic contact pairs is developed, enabling the modelling of relative deformation between the contacting bodies under both sliding and rolling motion; compared to \cite{KinematicsMio}, the framework is extended from purely elastic to viscoelastic contact, enabling a treatment of richer rheological and friction laws. 
More precisely, the contributions of this work can be summarised as follows:
\begin{enumerate}[(i)]
\item Systematic extension of dynamic viscoelastic friction models to distributed (PDE) and mixed (ODE--PDE) frameworks,
\item Derivation of viscoelasto-kinematic equations governing the coupled evolution of frictional forces, material deformation, and internal state variables,
\item Application of the theory to classical sliding and rolling contact problems, highlighting the close analogies between sliding and rolling friction.
\end{enumerate}
Within the framework of Winkler-type constitutive laws, the theoretical developments of this paper offer a unified perspective on sliding and rolling friction processes, which, although characterised by distinct physical features, share an identical mathematical structure in terms of linear or semilinear hyperbolic PDE systems. In this context, it is worth emphasising that the objective of this work is not to develop a complete predictive contact solver for rolling and sliding frictional contact problems. Thus, several complex physical mechanisms -- including coupling between normal and tangential interactions, thermo-viscoelastic effects, and nonlinear rheological behaviours -- are deliberately neglected. Such simplifications are standard in the derivation of tractable analytical theories \cite{KalkerBook,2000RCP}, and their incorporation would not only prevent the derivation of concise mathematical results but also obfuscate the main conclusions and physical insights of the analysis. 

The remainder of the manuscript is organised as follows. Section~\ref{sect:basic} discusses basic aspects of dynamic friction modelling and reviews existing formulations that are encompassed by the proposed framework. Section~\ref{sect:linearVisc0} provides a concise overview of the linear viscoelastic models adopted in this work, and proceeds to the derivation of the viscoelasto-kinematic equations for sliding and rolling contact. The resulting system of hyperbolic PDEs, governing the evolution of friction forces and internal state variables, is presented in Sect.~\ref{sect:DynamicDer}, where its hyperbolic structure is analysed and appropriate \emph{boundary conditions} (BCs) and \emph{initial conditions} (ICs) are formulated following a suitable transformation into Riemann variables. Applications of the proposed theory to simple sliding and rolling line contact problems for isotropic bodies in the absence of spin are illustrated in Sect.~\ref{sect:appl}, using an extension of the FrBD$_{n+1}$-GKV model to viscoelastic contact pairs. Finally, the main findings of the paper are summarised in Sect.~\ref{sect:conc}, together with an outlook on future research directions. Some technical results are deferred to Appendix~\ref{app:techn}, whilst Appendix~\ref{app:FrBD} contains additional details on the FrBD$_{n+1}$-GKV model and establishes its passivity. Finally, Appendix~\ref{app:numerical} provides a comprehensive description of the numerical implementations, together with a convergence analysis for both sliding and rolling contact problems.

\subsection*{Notation}
In this paper, $\mathbb{R}$ denotes the set of real numbers; $\mathbb{R}_{>0}$ and $\mathbb{R}_{\geq 0}$ indicate the set of positive real numbers excluding and including zero, respectively. The set of positive integer numbers is indicated with $\mathbb{N}$, whereas $\mathbb{N}_{0}$ denotes the extended set of positive integers including zero, i.e., $\mathbb{N}_{0} = \mathbb{N} \cup \{0\}$.
The set of $n\times m$ matrices with values in $\mathbb{F}$ ($\mathbb{F} = \mathbb{R}$, $\mathbb{R}_{>0}$, $\mathbb{R}_{\geq0}$) is denoted by $\mathbf{M}_{n\times m}(\mathbb{F})$ (abbreviated as $\mathbf{M}_{n}(\mathbb{F})$ whenever $m=n$). $\mathbf{Sym}_n(\mathbb{R})$ and $\mathbf{GL}_n(\mathbb{R})$ represent the groups of symmetric and invertible matrices with values in $\mathbb{R}$; the identity matrix on $\mathbb{R}^n$ is indicated with $\mathbf{I}_n$. A positive-definite matrix is noted as $\mathbf{M}_n(\mathbb{R}) \ni \mathbf{Q} \succ \mathbf{0}$; a positive semidefinite one as $\mathbf{M}_n(\mathbb{R}) \ni \mathbf{Q} \succeq \mathbf{0}$. 
The standard Euclidean norm on $\mathbb{R}^n$ is indicated with $\norm{\cdot}_2$; operator norms are simply denoted by $\norm{\cdot}$.
Given a domain $\Omega$ with closure $\overline{\Omega}$, $L^p(\Omega;\mathcal{Z})$ and $C^k(\overline{\Omega};\mathcal{Z})$ ($p, k \in \{1, 2, \dots, \infty\}$) denote respectively the spaces of $L^p$-integrable functions and $k$-times continuously differentiable functions on $\overline{\Omega}$ with values in $\mathcal{Z}$ (for $T = \infty$, the interval $[0,T]$ is identified with $\mathbb{R}_{\geq 0}$). In particular, $L^2(\Omega;\mathbb{R}^n)$ denotes the Hilbert space of square-integrable functions on $\Omega$ with values in $\mathbb{R}^n$, endowed with inner product $\langle \bm{\zeta}_1, \bm{\zeta}_2 \rangle_{L^2(\Omega;\mathbb{R}^n)} = \int_\Omega \bm{\zeta}_1^{\mathrm{T}}(\bm{x})\bm{\zeta}_2(\bm{x}) \dif \bm{x}$ and induced norm $\norm{\bm{\zeta}(\cdot)}_{L^2(\Omega;\mathbb{R}^n)}$. The Hilbert space $H^1(\Omega;\mathbb{R}^n)$ consists of functions $\bm{\zeta}\in L^2(\Omega;\mathbb{R}^n)$ whose weak derivative also belongs to $L^2(\Omega;\mathbb{R}^n)$. For a function $f :\Omega \to \mathbb{R}$, the sup norm is defined as $\norm{f(\cdot)}_\infty \triangleq \esup_{\Omega} \abs{f(\cdot)}$; $f : \Omega \to \mathbb{R}$ belongs to the space $L^\infty(\Omega;\mathbb{R})$ if $\norm{f(\cdot)}_\infty < \infty$.



\section{Basic considerations on dynamic friction modelling}\label{sect:basic}
This section covers some general aspects of the dynamic modelling of friction. Specifically, Sect.~\ref{sect:generalStruct} introduces a general class of dynamic equations for describing frictional processes, whilst Sect.~\ref{sect:Review} provides a brief overview of existing dynamic friction models.

\subsection{General structure of dynamic friction models}\label{sect:generalStruct}
\begin{figure}
\centering
\includegraphics[width=1\linewidth]{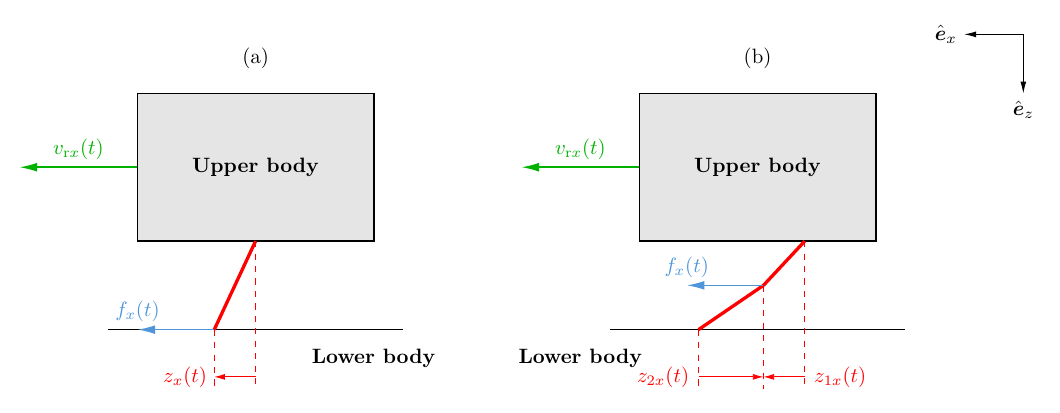} 
\caption{A schematic representation of the friction model: (a) configuration with a rigid substrate; (b) configuration with a deformable substrate. The problem is studied in a right-handed reference frame $(O;x,y,z)$ with unit vectors $(\hat{\bm{e}}_x, \hat{\bm{e}}_y, \hat{\bm{e}}_z)$.}
\label{fig:LumpModel}
\end{figure}

To discuss the general structure of dynamic friction models, the contact configurations illustrated in Fig~\ref{fig:LumpModel} may be considered. The first configuration, depicted in Fig.~\ref{fig:LumpModel} (a), corresponds to a viscoelastic body (upper body) sliding over a rigid flat substrate (lower body), whereas the second one, represented in Fig.~\ref{fig:LumpModel} (b), involves two viscoelastic bodies that are both deformable in the tangential direction. 

Accordingly, the rigid relative velocity (in the tangential, or equivalently planar direction) between the two objects is indicated by $\mathbb{R}^2 \ni \bm{v}\ped{r} = [v_{\textnormal{r}x}\; v_{\textnormal{r}y}]^{\mathrm{T}} \triangleq \bm{v}_1-\bm{v}_2$, where $\mathbb{R}^2 \ni \bm{v}_1 = [v_{1x}\; v_{1y}]^{\mathrm{T}}$ and $\mathbb{R}^2 \ni \bm{v}_2 = [v_{2x}\; v_{2y}]^{\mathrm{T}}$ denote the absolute velocities of the upper and lower bodies, respectively. To both bodies, massless bristles are attached, whose deflections are given by $\mathbb{R}^2 \ni \bm{z}_1 = [z_{1x}\; z_{1y}]^{\mathrm{T}}$ and $\mathbb{R}^2 \ni \bm{z}_2 = [z_{2x}\; z_{2y}]^{\mathrm{T}}$. These bristles are traditionally interpreted as material particles or micro-asperities at the interface of the two bodies \cite{Antali}. The relative deformation between two bristle elements in contact is denoted by $\mathbb{R}^2 \ni \bm{z} = [z_x\; z_y]^{\mathrm{T}} = \bm{z}_1-\bm{z}_2$. By deflecting, the bristles generate nondimensional forces (that is, normalised with respect to a reference normal load $p \in \mathbb{R}_{\geq 0}$) which oppose the local sliding motion, and are indicated respectively as $\mathbb{R}^2 \ni \bm{f}_1 = [f_{1x}\; f_{1y}]^{\mathrm{T}}$ and $\mathbb{R}^2 \ni \bm{f}_2 = [f_{2x}\; f_{2y}]^{\mathrm{T}}$.

Renaming $\mathbb{R}^2 \ni \bm{f} = [f_x\; f_y]^{\mathrm{T}}\triangleq \bm{f}_1 = -\bm{f}_2$, a standard dynamic model for friction may then be formulated as \cite{Antali}
\begin{subequations}\label{eq:frictionGeneral}
\begin{align}
\bm{f} & = \bm{F}(\bm{z}, \bm{\zeta}, \bm{v}\ped{r}), \\
\dot{\bm{z}} & = \bm{h}_0(\bm{z}, \bm{f},\bm{\zeta}, \bm{v}\ped{r}), \label{eq:h_0}\\
\dot{\bm{\zeta}}_1 & = \bm{h}_1(\bm{z}, \bm{f},\bm{\zeta}, \bm{v}\ped{r}), \label{eq:h_1}\\
\dot{\bm{\zeta}}_2 & = \bm{h}_2(\bm{z}, \bm{f},\bm{\zeta}, \bm{v}\ped{r}), \quad t \in (0,T), \label{eq:h_2}
\end{align}
\end{subequations}
where $\bm{\zeta}_2 \in \mathbb{R}^{2n_2}$ are vectors of internal state variables which dictate the evolution of the bristle deformations and generated force, and $\mathbb{R}^{2n}\ni \bm{\zeta} \triangleq [\bm{\zeta}_1^{\mathrm{T}}\; \bm{\zeta}_2^{\mathrm{T}}]^{\mathrm{T}}$, with $n = n_1+n_2$. The functions $\bm{F} : \mathbb{R}^{2(n+2)} \to \mathbb{R}^2$, $\bm{h}_0 : \mathbb{R}^{2(n+3)} \to \mathbb{R}^2$, $\bm{h}_1 :  \mathbb{R}^{2(n+3)} \to \mathbb{R}^{2n_1}$, and $\bm{h}_2 :  \mathbb{R}^{2(n+3)} \to \mathbb{R}^{2n_2}$ in Eq.~\eqref{eq:frictionGeneral}, which govern the dynamics of the system, are typically postulated independently of each other. In particular, $\bm{F}(\bm{z}, \bm{\zeta}, \bm{v}\ped{r})$, $\bm{h}_1(\bm{z}, \bm{f},\bm{\zeta}, \bm{v}\ped{r})$, and $\bm{h}_2(\bm{z}, \bm{f},\bm{\zeta}, \bm{v}\ped{r})$ are commonly deduced starting from rheological representations of the bristle element, whereas $\bm{h}_0(\bm{z},\bm{f},\bm{\zeta},\bm{v}\ped{r})$ is usually specified on empirical grounds, even though recent studies have sought to derive it rigorously from first principles. It is worth mentioning that the functions appearing in Eq.~\eqref{eq:frictionGeneral} may additionally depend on other (possibly time-varying) parameters, including the reference pressure $p$, but their specification is inessential at this stage and therefore omitted for notational convenience. 

Equations~\eqref{eq:frictionGeneral} can describe both lumped and distributed frictional contact, provided that the time derivatives are interpreted appropriately. In the absence of advection within the contact region, these derivatives coincide with ordinary time derivatives. Conversely, when advection phenomena are present -- as is typically the case in rolling contact -- the time derivatives may be replaced by material (Eulerian) derivatives, yielding a system of hyperbolic PDEs. In fact, this extension is straightforward when all state variables are advected with the same transport velocity. In viscoelastic contact pairs, however, forces and deformations may be transported at different velocities, rendering the interpretation of the time derivatives in Eq.~\eqref{eq:frictionGeneral} a nontrivial issue. Addressing this issue requires an appropriate specification of the functions $\bm{F} : \mathbb{R}^{2(n+2)} \to \mathbb{R}^2$, $\bm{h}_1 : \mathbb{R}^{2(n+3)} \to \mathbb{R}^{2n_1}$, and $\bm{h}_2 : \mathbb{R}^{2(n+3)} \to \mathbb{R}^{2n_2}$, which, in this paper, are derived in Sect.~\ref{Viscoelastic contact pairs} within the classic framework of linear viscoelasticity. Conversely, the specific structure of $\bm{h}_0 : \mathbb{R}^{2(n+3)} \to \mathbb{R}^2$ plays no role in determining the lumped or distributed character of the friction dynamics. As a general rule, the present work disregards non-smooth models and assumes $\bm{h}_0 \in C^0(\mathbb{R}^{2(n+3)} ;\mathbb{R}^2)$. Some choices of $\bm{h}_0 : \mathbb{R}^{2(n+3)} \to \mathbb{R}^2$ that satisfy this regularity requirement and are based on well-established friction formulations are reviewed next in Section~\ref{sect:Review}.

\subsection{Review of dynamic friction models}\label{sect:Review}
The literature on dynamic friction models is inhabited by a plethora of formulations, postulating different expressions for $\bm{h}_0(\bm{z}, \bm{f},\bm{\zeta}, \bm{v}\ped{r})$. This paper restricts itself to a class of sufficiently smooth functions, like those adopted by the Dahl, LuGre, FrBD, and Bouc-Wen models. 

\subsubsection{Dahl and LuGre models}\label{sect:LuGreandDahl}
The two-dimensional versions of the Dahl and LuGre friction models \cite{Tsiotras1,Tsiotras3} postulate the function $\bm{h}_0 : \mathbb{R}^{2(n+3)} \to \mathbb{R}^2$ as
\begin{align}\label{eq:h0LuGre}
\bm{h}_0(\bm{z},\bm{f},\bm{\zeta},\bm{v}\ped{r}) = \bm{h}_0(\bm{z},\bm{v}\ped{r})\triangleq -\mathbf{M}^{-2}(\bm{v}\ped{r})\norm{\mathbf{M}(\bm{v}\ped{r})\bm{v}\ped{r}}_{2,\varepsilon}\mathbf{\Sigma}_0\bm{z}-\bm{v}\ped{r},
\end{align}
where $\mathbf{M}(\bm{v}\ped{r})$ is a positive definite matrix of friction coefficients, i.e., $\mathbf{Sym}_2(\mathbb{R}) \ni \mathbf{M}(\bm{y}) \succ \mathbf{0}$ for all $\bm{y}\in \mathbb{R}^2$, $\mathbf{Sym}_2(\mathbb{R}) \ni \mathbf{\Sigma}_0 \succ \mathbf{0}$ is a matrix of micro-stiffness coefficients, and $\norm{\cdot}_{2,\varepsilon} \in C^0(\mathbb{R}^2;\mathbb{R}_{\geq 0})$ is a regularisation of the Euclidean norm $\norm{\cdot}_2$ for $\varepsilon\in \mathbb{R}_{>0}$, often converging uniformly to $\norm{\cdot}_2$ in $C^0(\mathbb{R}^2;\mathbb{R}_{\geq 0})$ for $\varepsilon \to 0$ (e.g., $\norm{\bm{y}}_{2,\varepsilon }= \sqrt{\norm{\bm{y}}_2^2 +\varepsilon}$), and with $\norm{\cdot}_{2,\varepsilon} \in C^1(\mathbb{R}^2;\mathbb{R}_{\geq 0})$ for $\varepsilon \in \mathbb{R}_{>0}$.
It should be clarified that Eq.~\eqref{eq:h0LuGre} adopts the opposite sign convention for $\bm{v}\ped{r}$ compared to the formulation presented in \cite{Tsiotras1,Tsiotras3}, which is consistent with the observations reported in \cite{Antali}.

\subsubsection{FrBD models}
Structurally similar to Dahl and LuGre, but physically justified, the recently developed FrBD models \cite{FrBD,FrBDroll,FrBDvisc} assume $\bm{h}_0 : \mathbb{R}^{2(n+3)} \to \mathbb{R}^2$ of the form
\begin{align}\label{eq:h0FrBD}
\bm{h}_0(\bm{z},\bm{f},\bm{\zeta},\bm{v}\ped{r}) =\bm{h}_0(\bm{f},\bm{v}\ped{r}) \triangleq -\mathbf{M}^{-2}(\bm{v}\ped{r})\norm{\mathbf{M}(\bm{v}\ped{r})\bm{v}\ped{r}}_{2,\varepsilon}\bm{f}-\bm{v}\ped{r},
\end{align}
where the matrix $\mathbf{M}(\bm{v}\ped{r})$ has the same meaning as in Sect.~\ref{sect:LuGreandDahl}. Compared to Eq.~\eqref{eq:h0LuGre}, it may be noted that Eq.~\eqref{eq:h0FrBD} replaces the term $\mathbf{\Sigma}_0\bm{z}$, which essentially corresponds to the elastic component of the generated bristle force, with $\bm{f}$. This has important repercussions for the model's passivity, in both its lumped and distributed formulations. In this context, it should be mentioned that the term $\mathbf{M}^{-2}(\bm{v}\ped{r})\norm{\mathbf{M}(\bm{v}\ped{r})\bm{v}\ped{r}}_{2,\varepsilon}$ in Eq.~\eqref{eq:h0FrBD} may be replaced by any well-behaved matrix $\mathbf{Sym}_2(\mathbb{R}) \ni \mathbf{G}(\bm{v}\ped{r}) \succeq \mathbf{0}$ if the steady-state friction force equals $\bm{f} = -\mathbf{G}^{-1}(\bm{v}\ped{r})\bm{v}\ped{r}$.

\subsubsection{Bouc-Wen model}
The two-dimensional generalisation of the Bouc-Wen model \cite{Park} prescribes the function $\bm{h}_0 : \mathbb{R}^{2(n+3)} \to \mathbb{R}^2$ as
\begin{align}
\bm{h}_0(\bm{z},\bm{f},\bm{\zeta},\bm{v}\ped{r}) =\bm{h}_0(\bm{f},\bm{v}\ped{r}) \triangleq -\mathbf{G}(\bm{z},\bm{v}\ped{r})\bm{z}-A\bm{v}\ped{r} ,
\end{align}
where $A \in \mathbb{R}_{>0}$ and the matrix $\mathbf{G}(\bm{z},\bm{v}\ped{r}) \in \mathbf{Sym}_2(\mathbb{R})$ reads
\begin{align}
\mathbf{G}(\bm{z},\bm{v}\ped{r}) \triangleq \begin{bmatrix} g_x(\bm{z},\bm{v}\ped{r}) & 0 \\ 0 & g_y(\bm{z},\bm{v}\ped{r}) \end{bmatrix},
\end{align}
where
\begin{subequations}\label{eq:funcGxGy}
\begin{align}
g_x(\bm{z},\bm{v}\ped{r}) &\triangleq \beta_1\abs{v_{\textnormal{r}x}z_x} +\beta_2\abs{v_{\textnormal{r}y}z_y}- \gamma (v_{\textnormal{r}x}z_x+v_{\textnormal{r}y}z_y), \\
g_y(\bm{z},\bm{v}\ped{r}) &\triangleq \beta_2\abs{v_{\textnormal{r}x}z_x} +\beta_1\abs{v_{\textnormal{r}y}z_y} - \gamma (v_{\textnormal{r}x}z_x+v_{\textnormal{r}y}z_y),
\end{align}
\end{subequations}
with $\beta_1,\beta_2 \in \mathbb{R}_{>0}$ and $\gamma \in \mathbb{R}$. The absolute values appearing in Eq.~\eqref{eq:funcGxGy} may be regularised adopting a similar rationale as in Sect.~\ref{sect:LuGreandDahl}.

\subsubsection{Frictionless contact}
In the case of frictionless contact, the function $\bm{h}_0 : \mathbb{R}^{2(n+3)} \to \mathbb{R}^2$ may be specified as
\begin{align}\label{eq:h0frictionless}
\bm{h}_0(\bm{z},\bm{f},\bm{\zeta},\bm{v}\ped{r}) = \bm{h}_0(\bm{v}\ped{r})\triangleq -\bm{v}\ped{r}.
\end{align}
Equation~\eqref{eq:h0frictionless} may be obtained, for instance, after linearising Eqs.~\eqref{eq:h0LuGre} and~\eqref{eq:h0FrBD} respectively around $(\bm{z},\bm{v}\ped{r}) = (\bm{0},\bm{0})$ and $(\bm{f},\bm{v}\ped{r}) = (\bm{0},\bm{0})$ (for $\varepsilon \to 0$), and agrees with the expression found when considering linear regimes corresponding to other, possibly non-dynamic, models of friction.

\section{Viscoelastic contact pairs and viscoelasto-kinematics}\label{sect:2Dext}

Suitable expressions for $\bm{F}(\bm{z},\bm{\zeta},\bm{v}\ped{r})$, $\bm{h}_1(\bm{z},\bm{f},\bm{\zeta},\bm{v}\ped{r})$, and $\bm{h}_2(\bm{z},\bm{f},\bm{\zeta},\bm{v}\ped{r})$ in Eq.~\eqref{eq:frictionGeneral} may be specified starting from an assumed rheological model for the bristle elements. Postulating linear viscoelastic constitutive equations also permits deducing the viscoelasto-kinematic equations governing the dynamics of the relative bristle deflection and frictional force.

The linear rheological models considered in this paper are detailed in Sect.~\ref{sect:linearVisc}, whilst Sect.~\ref{sect:viscokin} is dedicated to the derivation of the viscoelasto-kinematics.

\subsection{Linear viscoelasticity}\label{sect:linearVisc0}
Limited to individual bristle elements, Sect.~\ref{sect:linearVisc} introduces a class of linear viscoelastic rheologies that can be easily cast in state-space representation. The extension to the case of viscoelastic contact pairs is then provided in Sect.~\ref{Viscoelastic contact pairs}.

\subsubsection{Linear viscoelastic rheological models}\label{sect:linearVisc}

\begin{figure}
\centering
\includegraphics[width=0.7\linewidth]{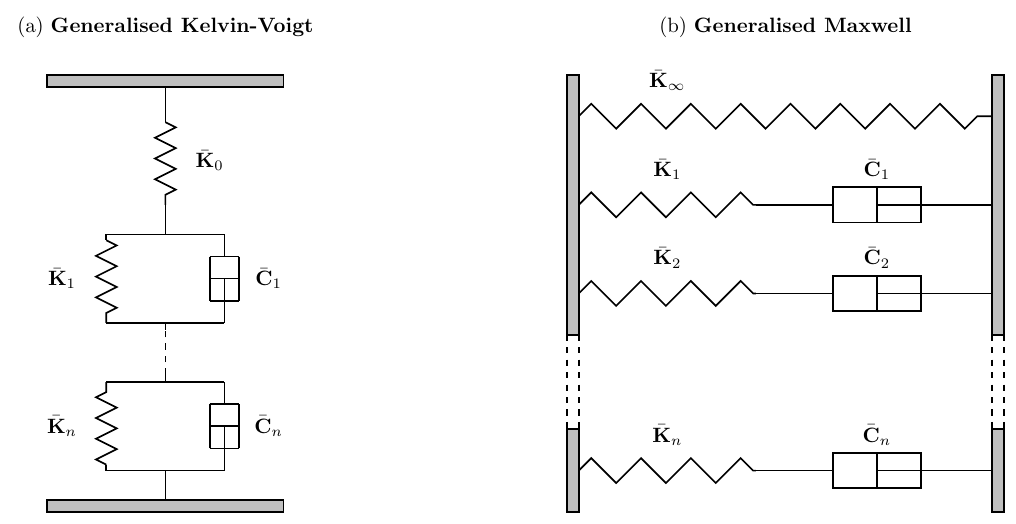} 
\caption{A schematic representation of the Generalised Kelvin-Voigt (GKV) and Generalised Maxwell (GM) rheological models. The matrix $\bar{\mathbf{K}}_0$ and $\bar{\mathbf{K}}_\infty$ collect the normalised micro-stiffnesses of the elastic branches, modelled as linear springs. The matrices $\bar{\mathbf{K}}_i$ and $\bar{\mathbf{C}}_i$ denote the normalised micro-stiffness and micro-damping matrices of the dissipative element $i$, $i \in \{1,\dots, n\}$.}
\label{fig:Dashpot}
\end{figure}

Within the framework of linear viscoelasticity, complex rheological models are traditionally constructed by combining elastic and viscous elements in series and parallel configurations, as illustrated in Fig.~\ref{fig:Dashpot}. A prototypical example is the GKV model with $n+1$ branches (Fig.~\ref{fig:Dashpot}(a)), which can be written as
\begin{subequations}\label{eq:force0GKV_Stress}
\begin{align}
\bm{f} & = \bar{\mathbf{K}}_0\bm{z} - \bar{\mathbf{K}}_0\sum_{i=1}^n \bm{z}_i, \label{eq:force00GKV} \\
\dot{\bm{z}}_i & = -\bm{\tau}_i^{-1}\bm{z}_i + \bar{\mathbf{C}}_i^{-1}\bm{f}, \quad i \in \{1,\dots,n\}, \; t \in (0,T),
\end{align}
\end{subequations}
where $\bm{z}\in \mathbb{R}^2$ denotes the total deformation of the element, $\bm{z}_i \in \mathbb{R}^2$, $i \in \{1,\dots,n\}$, the deformations of the dissipative branches, $\mathbf{Sym}_2(\mathbb{R}) \ni \bar{\mathbf{K}}_0 \succ \mathbf{0}$ is the stiffness matrix of the elastic branch, $\mathbf{Sym}_2(\mathbb{R}) \ni \bar{\mathbf{C}}_i \succ \mathbf{0}$, $i \in \{1,\dots,n\}$, are matrices of normalised damping coefficients, and $\bm{\tau}_i \in \mathbf{M}_2(\mathbb{R})$, $i \in \{1,\dots,n\}$, are matrices of relaxation times, given by
\begin{align}\label{eq:relaxM}
\bm{\tau}_i \triangleq \bar{\mathbf{K}}_i^{-1}\bar{\mathbf{C}}_i, \quad i \in \{1,\dots,n\},
\end{align}
being $\mathbf{Sym}_2(\mathbb{R}) \ni \bar{\mathbf{K}}_i \succ \mathbf{0}$, $i \in \{1,\dots,n\}$, matrices of normalised stiffnesses.

Analogously, the Generalised Maxwell (GM) model illustrated in Fig.~\ref{fig:Dashpot}(b) describes a viscoelastic rheology obtained by connecting an elastic element in parallel with dissipative branches. Denoting the deformations of the dampers by $\bm{z}_i \in \mathbb{R}^2$, $i \in \{1,\dots,n\}$, it may be formulated as
\begin{subequations}\label{eq:force0GM_Stress}
\begin{align}
\bm{f} & = \bar{\mathbf{K}}_0\bm{z} - \sum_{i=1}^n \bar{\mathbf{K}}_i \bm{z}_i, \label{eq:force00GM} \\
\dot{\bm{z}}_i & = -\bm{\tau}_i^{-1}\bm{z}_i + \bm{\tau}_i^{-1}\bar{\mathbf{K}}_0^{-1}\bigggl(\bm{f}+\sum_{i=1}^n \bar{\mathbf{K}}_i\bm{z}_i\bigggr), \quad i \in \{1,\dots,n\}, \; t \in (0,T),
\end{align}
\end{subequations}
with $\mathbf{Sym}_2(\mathbb{R}) \ni \bar{\mathbf{K}}_0 \succ \mathbf{0}$ defined as $\bar{\mathbf{K}}_0 \triangleq \bar{\mathbf{K}}_\infty + \sum_{i=1}^n \bar{\mathbf{K}}_i$, $\mathbf{Sym}_2(\mathbb{R}) \ni \bar{\mathbf{K}}_\infty \succ \mathbf{0}$ representing the stiffness matrix of the elastic branch, the matrices $\mathbf{Sym}_2(\mathbb{R}) \ni \bar{\mathbf{K}}_i \succ \mathbf{0}$, $i \in \{0,\dots,n\}$, and $\mathbf{Sym}_2(\mathbb{R}) \ni \bar{\mathbf{C}}_i \succ \mathbf{0}$, $i \in \{1,\dots,n\}$, having the same meaning as before, and $\bm{\tau}_i \in \mathbf{M}_2(\mathbb{R})$, $i \in \{1,\dots,n\}$, reading as in Eq.~\eqref{eq:relaxM}.
Collecting the internal state variables into the vector $\mathbb{R}^{2n} \ni \bm{\zeta} \triangleq [\bm{z}_1^{\mathrm{T}}\; \bm{z}_2^{\mathrm{T}}\; \dots \; \bm{z}_n^{\mathrm{T}}]^{\mathrm{T}}$, the GKV and GM models may be cast in state-space form with input $\bm{f}$ and output $\bm{z}$:
\begin{subequations}\label{eq:ss_0}
\begin{align}
\bm{f}
&= \bar{\mathbf{K}}_{0}\bm{z} + \mathbf{C}\bm{\zeta},\label{eq:f1_0} \\
\dot{\bm{\zeta}}
&= \mathbf{A}\bm{\zeta} + \mathbf{B}\bm{f}, \quad t \in (0,T),
\end{align}
\end{subequations}
with the matrices $\mathbf{A} \in \mathbf{M}_{2n}(\mathbb{R})$, $\mathbf{B}\in \mathbf{M}_{2n\times2}(\mathbb{R})$, and $\mathbf{C}\in \mathbf{M}_{2\times 2n}$ opportunely specified. The state-space representation~\eqref{eq:ss_0} of a GKV or GM element is stable and dissipative, with a rational transfer function. In fact, both models admit equivalent realisations in terms of the other one \cite{Rheology2,Rheology2-3}.

Owing to these premises, the present work focuses on rheological models of the form~\eqref{eq:ss_0}, which can be employed to describe the rheology of bristle elements. The extension to the more general case of viscoelastic contact pairs is then developed in Sect.~\ref{Viscoelastic contact pairs}.

\subsubsection{Viscoelastic contact pairs}\label{Viscoelastic contact pairs}
Based on the discussion initiated in Sects.~\ref{sect:generalStruct} and~\ref{sect:linearVisc}, the rheology of the bristle elements attached to the first (upper) body is modelled as a stable linear system in the fashion of Eq.~\eqref{eq:ss_0}, that is,
\begin{subequations}\label{eq:ss1}
\begin{align}
\bm{f}_1
&= \bar{\mathbf{K}}_{0,1}\bm{z}_1 + \mathbf{C}_1\bm{\zeta}_1,\label{eq:f1} \\
\frac{\D_1\bm{\zeta}_1}{\D t}
&= \mathbf{A}_1\bm{\zeta}_1 + \mathbf{B}_1\bm{f}_1, \quad t\in(0,T). \label{eq:dyn1}
\end{align}
\end{subequations}
Similarly, a stable linear behaviour is assumed for the second (lower) body:
\begin{subequations}\label{eq:ss2}
\begin{align}
\bm{f}_2 & = \bar{\mathbf{K}}_{0,2}\bm{z}_2 + \mathbf{C}_2\bm{\zeta}_2, \label{eq:f2} \\
\frac{\D_2\bm{\zeta}_2}{\D t} & = \mathbf{A}_2\bm{\zeta}_2 + \mathbf{B}_2\bm{f}_2, \quad t\in(0,T). \label{eq:dyn2}
\end{align}
\end{subequations}
In Eqs.~\eqref{eq:ss1} and~\eqref{eq:ss2}, $\bm{\zeta}_1 \in \mathbb{R}^{2n_1}$ and $\bm{\zeta}_2 \in \mathbb{R}^{2n_2}$ are internal state variables already introduced in Sect.~\ref{sect:generalStruct}, the matrices $\mathbf{Sym}_2(\mathbb{R}) \ni \bar{\mathbf{K}}_{0,1}, \bar{\mathbf{K}}_{0,2} \succ \mathbf{0}$ contain normalised stiffnesses, and the structure of $\mathbf{A}_i \in \mathbf{M}_{2n_i}(\mathbb{R})$, $\mathbf{B}_{i} \in \mathbf{M}_{2n_i\times2}(\mathbb{R})$, and $\mathbf{C}_i \in \mathbf{M}_{2\times 2n_i}(\mathbb{R})$, $i \in \{1,2\}$, may be deduced starting from adopted realisation of the bristle rheology, which may differ between the first and second body. In the main part of the manuscript, the following Assumption~\ref{ass:comm}, inherited from the elastic framework introduced in \cite{KinematicsMio}, is postulated.
\begin{assumption}[Coaxiality \cite{KinematicsMio}]\label{ass:comm}
The matrices $\bar{\mathbf{K}}_{0,1}\succ \mathbf{0}$ and $\bar{\mathbf{K}}_{0,2}\succ \mathbf{0}$ commute, that is, $\bar{\mathbf{K}}_{0,1}\bar{\mathbf{K}}_{0,2} = \bar{\mathbf{K}}_{0,2}\bar{\mathbf{K}}_{0,1}$. 
\end{assumption}
The above coaxiality/commutativity condition is automatically satisfied for isotropic materials and for anisotropic materials sharing common principal elastic directions. In particular, the matrices $\bar{\mathbf{K}}_{0,1}$ and $\bar{\mathbf{K}}_{0,2}$ are frequently assumed to be diagonal in brush- and LuGre-like models of rolling contact \cite{Tsiotras1,Tsiotras3}, whereas the substrate is almost always supposed to be infinitely rigid or elastically isotropic. Conversely, in similarly elastic contact pairs, the two bodies often share common principal elastic directions, as it happens in wheel-rail interactions. Overall, Assumption~\ref{ass:comm} is physically realistic for many practical applications, and slightly simplifies the mathematical treatment of the problem; however, it will be removed in Appendix~\ref{app:noncoaxial} without altering the main conclusions of the paper.

The state-space representations in Eqs.~\eqref{eq:ss1} and~\eqref{eq:ss2} are propaedeutic to the derivation of the viscoelasto-kinematic equations, carried out in Sect.~\ref{sect:viscokin}. In this context, it should be emphasised that, compared to Eq.~\eqref{eq:ss_0}, the time derivatives in Eqs.~\eqref{eq:ss1} and~\eqref{eq:ss2} have been replaced by the material (Eulerian) ones $\frac{\D_1\bm{\zeta}_1}{\D t}$ and $\frac{\D_2\bm{\zeta}_2}{\D t}$, as it will be better clarified in Sect.~\ref{sect:Eulerian}. Another crucial observation is that the normalised forces generated by the bristle elements of the two bodies must be opposite, i.e., $\bm{f}_1 = -\bm{f}_2$. Therefore, adopting the same notation as in Sect.~\ref{sect:generalStruct}, Eqs.~\eqref{eq:f1} and~\eqref{eq:f2} yield
\begin{subequations}\label{eq:sicjd}
\begin{align}
\bm{z}_1 & = \bar{\mathbf{K}}_{0,1}^{-1}(\bm{f}-\mathbf{C}_1\bm{\zeta}_1), \\
\bm{z}_2 & =- \bar{\mathbf{K}}_{0,2}^{-1}(\bm{f}+\mathbf{C}_2\bm{\zeta}_2),
\end{align}
\end{subequations}
together with
\begin{align}\label{eq:fz_zeta}
\bm{f} = \bar{\mathbf{K}}_0\bm{z} + \bar{\mathbf{C}}\bm{\zeta},
\end{align}
where $\bar{\mathbf{K}}_0 \in \mathbf{Sym}_2(\mathbb{R})$ and $ \bar{\mathbf{C}} \in \mathbf{M}_{2\times 2n}(\mathbb{R})$ read
\begin{subequations}
\begin{align}
\bar{\mathbf{K}}_0 & \triangleq \Bigl(\bar{\mathbf{K}}_{0,1}^{-1} + \bar{\mathbf{K}}_{0,2}^{-1}\Bigr)^{-1}, \label{eq:K0eq} \\
\bar{\mathbf{C}} & \triangleq \bar{\mathbf{K}}_0\begin{bmatrix} \bar{\mathbf{K}}_{0,1}^{-1}\mathbf{C}_1 & -\bar{\mathbf{K}}_{0,2}^{-1}\mathbf{C}_2\end{bmatrix}.
\end{align}
\end{subequations}
Equations~\eqref{eq:dyn1},~\eqref{eq:dyn2}, and~\eqref{eq:fz_zeta} may be cast in the form~\eqref{eq:frictionGeneral} with
\begin{subequations}\label{eq:fishic}
\begin{align}
\bm{F}(\bm{z},\bm{\zeta},\bm{v}\ped{r}) & = \bm{F}(\bm{z},\bm{\zeta}) \triangleq  \bar{\mathbf{K}}_0\bm{z} + \bar{\mathbf{C}}\bm{\zeta}, \\
\bm{h}_1(\bm{z},\bm{f},\bm{\zeta},\bm{v}\ped{r}) & = \bm{h}_1(\bm{f},\bm{\zeta}) \triangleq \mathbf{A}_1\bm{\zeta}_1 + \mathbf{B}_1\bm{f}, \label{eq:h_1spec}\\
\bm{h}_2(\bm{z},\bm{f},\bm{\zeta},\bm{v}\ped{r}) & = \bm{h}_2(\bm{f},\bm{\zeta}) \triangleq \mathbf{A}_2\bm{\zeta}_2 - \mathbf{B}_2\bm{f}. \label{eq:h_2spec}
\end{align}
\end{subequations}
Combined with a suitable expression for the function $\bm{h}_0 : \mathbb{R}^{2(n+3)} \to \mathbb{R}^2$, Eq.~\eqref{eq:fishic} completely characterises the dynamics of the frictional interaction, provided that the derivatives $\dot{\bm{z}}$, $\dot{\bm{\zeta}}_1$, and $\dot{\bm{\zeta}}_2$ are interpreted correctly in Eq.~\eqref{eq:frictionGeneral}.

Furthermore, manipulating Eqs.~\eqref{eq:f1} and~\eqref{eq:f2}, the following identities may be inferred:
\begin{subequations}\label{eq:z1z2_z}
\begin{align}
\bm{z}_1 & = \mathbf{S}_0\bm{\zeta}+\mathbf{S}_1\bm{z}, \\
\bm{z}_2 & = \mathbf{S}_0\bm{\zeta}-\mathbf{S}_2\bm{z}, 
\end{align}
\end{subequations}
with $\mathbf{S}_0\in \mathbf{M}_{2\times 2n}(\mathbb{R})$ and $\mathbf{Sym}_2(\mathbb{R}) \ni \mathbf{S}_1,\mathbf{S}_2 \succ \mathbf{0}$ defined as
\begin{subequations}\label{eq:matricesS}
\begin{align}
\mathbf{S}_0 & \triangleq -\bigl(\bar{\mathbf{K}}_{0,1} + \bar{\mathbf{K}}_{0,2}\bigr)^{-1}\begin{bmatrix}\mathbf{C}_1 & \mathbf{C}_2 \end{bmatrix}, \\
\mathbf{S}_1 & \triangleq \bigl(\bar{\mathbf{K}}_{0,1} + \bar{\mathbf{K}}_{0,2}\bigr)^{-1}\bar{\mathbf{K}}_{0,2} = \bar{\mathbf{K}}_{0,2}\bigl(\bar{\mathbf{K}}_{0,1} + \bar{\mathbf{K}}_{0,2}\bigr)^{-1}, \\
\mathbf{S}_2 & \triangleq \bigl(\bar{\mathbf{K}}_{0,1} + \bar{\mathbf{K}}_{0,2}\bigr)^{-1}\bar{\mathbf{K}}_{0,1} = \bar{\mathbf{K}}_{0,1}\bigl(\bar{\mathbf{K}}_{0,1} + \bar{\mathbf{K}}_{0,2}\bigr)^{-1},
\end{align}
\end{subequations}
and satisfying identically $\mathbf{S}_1\mathbf{S}_2 = \mathbf{S}_2\mathbf{S}_1$ and $\mathbf{S}_1 + \mathbf{S}_2 = \mathbf{I}_2$. Equation~\eqref{eq:z1z2_z} allows expressing the deformation of the bristles attached to the contacting bodies as a function of their relative difference, plus the internal state variables governing the dynamics of the force generation. 


\subsection{Viscoelasto-kinematics}\label{sect:viscokin}
To completely describe the sliding and rolling contact process, it is necessary to correctly identify the derivatives appearing in Eq.~\eqref{eq:frictionGeneral}. Specifically, the variables $\bm{\zeta}_1$ and $\bm{\zeta}_2$ are advected with transport velocities that may be determined independently from the kinematics of the two bodies, as explained in Sect.~\ref{sect:Eulerian}. Conversely, the quantity $\bm{z}$ represents a relative deformation, and its dynamics must be inferred by combining kinematic relationships and constitutive equations. This is detailed in Sect.~\ref{subsect:deriv}, where the governing viscoelasto-kinematics of the system are derived.

\subsubsection{Apparent contact area and Eulerian derivatives}\label{sect:Eulerian}
\begin{figure}
\centering
\includegraphics[width=1\linewidth]{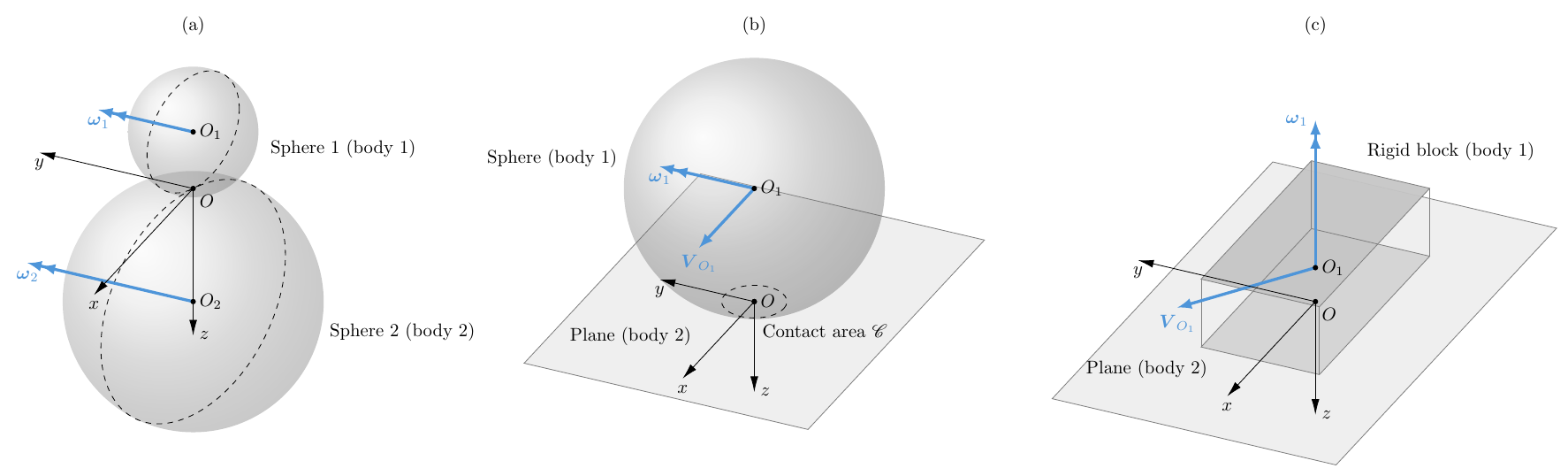} 
\caption{Sliding and rolling contact problem between: (a) two spheres with angular velocities $\bm{\omega}_1, \bm{\omega}_2 \in \mathbb{R}^3$; (b) a sphere translating and rolling over a stationary plane, where $\bm{V}_{O_1}\in \mathbb{R}^3$ denotes the translational velocity of its centre, and $\bm{\omega}_1\in \mathbb{R}^3$ its rolling velocity; (c) a rigid block translating and rotating over a plane, where $\bm{V}_{O_1} \in \mathbb{R}^3$ denotes the velocity of the centre of the block, and $\bm{\omega}_1 \in \mathbb{R}^3$ its angular velocity (with a single component around the vertical axis).}
\label{fig:RollingBodies}
\end{figure}

The following discussion is adapted from \cite{FrBDroll}. As illustrated in Fig.~\ref{fig:RollingBodies}, the sliding and rolling contact problem between two bodies is typically studied in a contact-fixed reference frame $(O;x,y,z)$, with the $x$-axis (longitudinal) oriented along the main sliding or rolling direction, the $z$-axis (vertical) pointing into one of the two bodies (often the lower one), and the $y$-axis (lateral) defined to complete a right-handed coordinate system. The origin $O$ coincides with the centroid of the (apparent), possibly time-varying, contact area $\mathscr{C}(t)$, which is often supposed to be independent of the tangential (longitudinal and lateral) interactions between the two bodies, and solely determined by the normal contact configuration \cite{KalkerBook}. The apparent contact area, namely, the domain $\mathscr{C}(t)$, can be defined starting from the contact areas of the two bodies, $\mathscr{C}_1(t), \mathscr{C}_2(t) \subset \mathbb{R}^2$, with boundaries $\partial \mathscr{C}_1(t), \partial \mathscr{C}_2(t)$ and interiors $\mathring{\mathscr{C}}_1(t), \mathring{\mathscr{C}}_2(t)$. In this paper, the contact areas are assumed to be compact subsets of $\mathbb{R}^2$, consistent with the half-space approximation \cite{KalkerBook}, which accommodates planar as well as localised interactions, which are typically encountered in non-conformal contact problems. Additionally, it appears natural to require $\mathscr{C}_1(t)$ and $\mathscr{C}_2(t)$ to be regular closed, i.e., $\mathscr{C}_1(t) = \overline{\mathring{\mathscr{C}}_1}(t)$ and $\mathscr{C}_2(t) = \overline{\mathring{\mathscr{C}}_2}(t)$. Identical assumptions are postulated on the apparent contact area $\mathring{\mathscr{C}}(t)$ with boundary $\partial \mathscr{C}(t) = \mathscr{C}(t)\setminus\mathring{\mathscr{C}}(t)$, and such that $\mathscr{C}(t) = \overline{\mathring{\mathscr{C}}}(t)$. In this context, it should be mentioned that, albeit the apparent contact area may, in principle, be deduced from the knowledge of $\mathscr{C}_1(t)$ and $\mathscr{C}_2(t)$, the situation in practice is often reversed: in typical applications, $\mathscr{C}(t)$ is prescribed directly by assuming a given geometric shape, which is generally chosen to satisfy the assumptions stated above.

Both the situations of reciprocal rolling, as depicted in Fig.~\ref{fig:RollingBodies}(a), and translational and rolling contact, as shown in Figs.~\ref{fig:RollingBodies}(b) and (c), may be considered: in the first case, the shape of the contact area may vary over time, but the origin $O$ is typically fixed; in the second and third scenarios, the reference frame moves together with one of the two bodies (usually, the upper one).

Inside the contact areas of the two bodies, a generic quantity is advected with a certain transport velocity, becoming thus a function of both time and space, that is, $\mathbb{R}^2 \ni \bm{\xi}_1 = \bm{\xi}_1(\bm{x}_1,t)$ for $\bm{x}_1 \in \mathscr{C}_1(t)$, and $\mathbb{R}^2\ni\bm{\xi}_2 = \bm{\xi}_2(\bm{x}_2,t)$ for $\bm{x}_2\in \mathscr{C}_2(t)$. Hence, according to the Eulerian approach, the material time derivatives of $\bm{\xi}_1(\bm{x}_1,t)$ and $\bm{\xi}_2(\bm{x}_2,t)$ read, respectively,
\begin{subequations}\label{eq:eulerian}
\begin{align}
\dot{\bm{\xi}}_1 =\frac{\D_1\bm{\xi}_1(\bm{x}_1,t)}{\D t}& = \dpd{\bm{\xi}_1(\bm{x}_1,t)}{t} + \bigl(\bm{V}_1(\bm{x}_1,t)\cdot\nabla_{\bm{x}_1}\bigr)\bm{\xi}_1(\bm{x}_1,t), \label{eq:grad1} \\
\dot{\bm{\xi}}_2 =\frac{\D_2\bm{\xi}_2(\bm{x}_2,t)}{\D t}& = \dpd{\bm{\xi}_2(\bm{x}_2,t)}{t} + \bigl(\bm{V}_2(\bm{x}_2,t)\cdot\nabla_{\bm{x}_2}\bigr)\bm{\xi}_2(\bm{x}_2,t), \label{eq:grad2}
\end{align}
\end{subequations}
where $\mathbb{R}^2 \ni \bm{V}_1(\bm{x}_1,t) = [V_{1x}(\bm{x}_1,t) \; V_{1y}(\bm{x}_1,t)]^{\mathrm{T}}$ and $\mathbb{R}^2 \ni \bm{V}_2(\bm{x}_2,t) = [V_{2x}(\bm{x}_2,t) \; V_{2y}(\bm{x}_2,t)]^{\mathrm{T}}$ represent the transport velocities, typically with $\bm{V}_1(\bm{x}_1,t) \not = \bm{V}_2(\bm{x}_2,t)$, and $\mathbb{R}^2 \ni \nabla_{\bm{x}_1} \triangleq [ \pd{}{x_1}\; \pd{}{y_1}]^{\mathrm{T}}$, $\mathbb{R}^2 \ni \nabla_{\bm{x}_2} \triangleq [ \pd{}{x_2}\; \pd{}{y_2}]^{\mathrm{T}}$ denote the tangential gradients.

\subsubsection{Viscoelasto-kinematic equations}\label{subsect:deriv}
Equation~\eqref{eq:eulerian} applies to physical quantities associated with the two different bodies in sliding and rolling contact, including the internal variables $\bm{\zeta}_1 = \bm{\zeta}_1(\bm{x}_1,t)$ and $\bm{\zeta}_2= \bm{\zeta}_2(\bm{x}_2,t)$. Conversely, the deflection $\bm{z} = \bm{z}(\bm{x},t)= \bm{z}_1(\bm{x}_1,t)-\bm{z}_2(\bm{x}_2,t)$ is a hybrid quantity, whose dynamics must be appropriately deduced by combining kinematic relationships with rheological equations. To this end, it is beneficial to express all the relevant quantities as functions of a single set of space variables. The following Assumption~\ref{ass:s,allGrad}, adapted from \cite{KalkerBook}, serves exactly this purpose.
\begin{assumption}[Small displacement and displacement gradients]\label{ass:s,allGrad}
The unstressed reference state may be chosen so that the displacements $\bm{z}_i(\bm{x}_i,t)$, $i \in \{1,2\}$, are small compared to the typical diameter of the bodies. In the same reference state, the displacement gradients are smaller than the unity, that is, $\norm{\nabla_{\bm{x}_i}\bm{z}_i(\bm{x}_i,t)^{\mathrm{T}}} < \epsilon \ll 1$, $i \in \{1,2\}$.
\end{assumption}
The next crucial observation is that, after the deformation has occurred, the following relationship needs to hold automatically for two particles in contact:
\begin{align}
\bm{x}_1 + \bm{z}_1(\bm{x}_1,t) = \bm{x}_2 + \bm{z}_2(\bm{x}_2,t), \quad \bm{x}_1 \in \mathscr{C}_1(t), \;    \bm{x}_2 \in \mathscr{C}_2(t), \; t \in (0,T).
\end{align}
At this point, following the same rationale as in \cite{KinematicsMio}, it is possible to define a new common coordinate as
\begin{align}\label{eq:x_1-z_2}
\bm{x} \triangleq \bm{x}_1-\bm{z}_2(\bm{x}_2,t) = \bm{x}_2-\bm{z}_1(\bm{x}_1,t).
\end{align}
According to Eq.~\eqref{eq:x_1-z_2}, the coordinate $\bm{x}$ may be interpreted as the common material point of the contact interface after deformation, expressed in a symmetric \emph{pull-back} way from either body, that is, by removing the opposite body's deformation.
In turn, Eq.~\eqref{eq:x_1-z_2} provides
\begin{align}\label{eq:gradientNEs}
\begin{split}
\nabla_{\bm{x}_1}\bm{x}^{\mathrm{T}} & = \Bigl[ \mathbf{I}_2 + \nabla_{\bm{x}_2}\bm{z}_2(\bm{x}_2,t)^{\mathrm{T}}\Bigr]^{-1}\Bigl(\mathbf{I}_2-\nabla_{\bm{x}_2}\bm{z}_2(\bm{x}_2,t)^{\mathrm{T}}\nabla_{\bm{x}_1}\bm{z}_1(\bm{x}_1,t)^{\mathrm{T}}\Bigr) = \mathbf{I}_2 + \mathcal{O}(\epsilon), \\
\nabla_{\bm{x}_2}\bm{x}^{\mathrm{T}} & = \Bigl[ \mathbf{I}_2 + \nabla_{\bm{x}_1}\bm{z}_1(\bm{x}_1,t)^{\mathrm{T}}\Bigr]^{-1}\Bigl(\mathbf{I}_2-\nabla_{\bm{x}_1}\bm{z}_1(\bm{x}_1,t)^{\mathrm{T}}\nabla_{\bm{x}_2}\bm{z}_2(\bm{x}_2,t)^{\mathrm{T}}\Bigr) = \mathbf{I}_2 + \mathcal{O}(\epsilon).
\end{split}
\end{align}
Thus, considering two generic quantities $\bm{\xi}_1(\bm{x}_1,t), \bm{\xi}_2(\bm{x}_2,t) \in \mathbb{R}^2$ and invoking~\eqref{eq:gradientNEs}, it may be concluded that
\begin{subequations}\label{eq:gradients}
\begin{align}
\nabla_{\bm{x}_1}\bm{\xi}_1(\bm{x}_1,t)^{\mathrm{T}} &= \nabla_{\bm{x}}\bm{\xi}_1(\bm{x},t)^{\mathrm{T}}\nabla_{\bm{x}_1}\bm{x}^{\mathrm{T}} = \nabla_{\bm{x}}\bm{\xi}_1(\bm{x},t)^{\mathrm{T}}\bigl[\mathbf{I}_2 +\mathcal{O}(\epsilon)\bigr] =  \nabla_{\bm{x}}\bm{\xi}_1(\bm{x},t)^{\mathrm{T}}+ \mathcal{O}(\epsilon), \\
\nabla_{\bm{x}_2}\bm{\xi}_2(\bm{x}_2,t)^{\mathrm{T}} & = \nabla_{\bm{x}}\bm{\xi}_2(\bm{x},t)^{\mathrm{T}}\nabla_{\bm{x}_2}\bm{x}^{\mathrm{T}} = \nabla_{\bm{x}}\bm{\xi}_2(\bm{x},t)^{\mathrm{T}}\bigl[\mathbf{I}_2 + \mathcal{O}(\epsilon)\bigr] =  \nabla_{\bm{x}}\bm{\xi}_2(\bm{x},t)^{\mathrm{T}} + \mathcal{O}(\epsilon),
\end{align}
\end{subequations}
where $\mathbb{R}^2 \ni \nabla_{\bm{x}} \triangleq [ \pd{}{x}\; \pd{}{y}]^{\mathrm{T}}$.
Hence, neglecting higher-order terms in Eq.~\eqref{eq:eulerian} gives
\begin{subequations}\label{eq:eulerian2}
\begin{align}
\dot{\bm{\xi}}_1 =\frac{\D_1\bm{\xi}_1(\bm{x}_1,t)}{\D t}& \approx \frac{\D_1\bm{\xi}_1(\bm{x},t)}{\D t} = \dpd{\bm{\xi}_1(\bm{x},t)}{t} + \bigl(\bm{V}_1(\bm{x},t)\cdot\nabla_{\bm{x}}\bigr)\bm{\xi}_1(\bm{x},t), \\
\dot{\bm{\xi}}_2 =\frac{\D_2\bm{\xi}_2(\bm{x}_2,t)}{\D t}& \approx \frac{\D_2\bm{\xi}_2(\bm{x},t)}{\D t} = \dpd{\bm{\xi}_2(\bm{x},t)}{t} + \bigl(\bm{V}_2(\bm{x},t)\cdot\nabla_{\bm{x}}\bigr)\bm{\xi}_2(\bm{x},t). 
\end{align}
\end{subequations}
The above relationships~\eqref{eq:gradients} and~\eqref{eq:eulerian2} define a first-order small-gradient model. More precisely, the displacement gradients are $\mathcal{O}(\epsilon)$ under Assumption~\ref{ass:s,allGrad}, whilst the spatial gradients of the force and internal state fields have been supposed to be bounded independently of $\epsilon$ on the considered time interval, which allows neglecting products between the $\mathcal{O}(\epsilon)$ Jacobian corrections and these gradients. Such an approximation is not intended to resolve discontinuities or layers whose gradients grow as $\mathcal{O}(\epsilon^{-1})$, essentially restricting the results advocated in the sequel within the framework of a first-order approximation. 
Consequently, recalling~\eqref{eq:sicjd}, Eq.~\eqref{eq:eulerian2} with $\bm{\xi}_1(\bm{x}_1,t) = \bm{z}_1(\bm{x}_1,t)$ and $\bm{\xi}_2(\bm{x}_2,t) = \bm{z}_2(\bm{x}_2,t)$ permits to deduce the following expression for $\dot{\bm{z}}$:
\begin{align}\label{eq:zdot}
\begin{split}
\dot{\bm{z}} &  \approx \dfrac{\D_1 \bm{z}_1(\bm{x},t)}{\D t}-\dfrac{\D_2 \bm{z}_2(\bm{x},t)}{\D t} \\
& = \bar{\mathbf{K}}_0^{-1}\dpd{\bm{f}(\bm{x},t)}{t} +\Bigl[ \bigl(\bm{V}_1(\bm{x},t)\cdot \nabla_{\bm{x}}\bigr)\bar{\mathbf{K}}_{0,1}^{-1} + \bigl(\bm{V}_2(\bm{x},t)\cdot \nabla_{\bm{x}}\bigr)\bar{\mathbf{K}}_{0,2}^{-1}\Bigr]\bm{f}(\bm{x},t) \\
& \quad -\bar{\mathbf{K}}_{0,1}^{-1}\mathbf{C}_1\dfrac{\D_1\bm{\zeta}_1(\bm{x},t)}{\D t} +\bar{\mathbf{K}}_{0,2}^{-1}\mathbf{C}_2\dfrac{\D_2\bm{\zeta}_2(\bm{x},t)}{\D t}\\
& = \bar{\mathbf{K}}_0^{-1}\dpd{\bm{f}(\bm{x},t)}{t} + \Bigr[\bigl(\bm{V}_1(\bm{x},t)\cdot \nabla_{\bm{x}}\bigr)\bar{\mathbf{K}}_{0,1}^{-1} + \bigl(\bm{V}_2(\bm{x},t)\cdot \nabla_{\bm{x}}\bigr)\bar{\mathbf{K}}_{0,2}^{-1}\Bigl]\bm{f}(\bm{x},t) \\
& \quad -\bar{\mathbf{K}}_{0,1}^{-1}\mathbf{C}_1\bm{h}_1\bigl(\bm{f}(\bm{x},t),\bm{\zeta}(\bm{x},t)\bigr) + \bar{\mathbf{K}}_{0,2}^{-1}\mathbf{C}_2\bm{h}_2\bigl(\bm{f}(\bm{x},t),\bm{\zeta}(\bm{x},t)\bigr).
\end{split}
\end{align}
Finally, combining Eq.~\eqref{eq:z1z2_z} with~\eqref{eq:x_1-z_2} gives
\begin{subequations}\label{eq:x1x2}
\begin{align}
\bm{x}_1 & = \bm{x}-\mathbf{S}_2\bm{z}(\bm{x},t) + \mathbf{S}_0\bm{\zeta}(\bm{x},t), \\
\bm{x}_2 & = \bm{x}+\mathbf{S}_1\bm{z}(\bm{x},t)+ \mathbf{S}_0\bm{\zeta}(\bm{x},t),
\end{align}
\end{subequations}
and
\begin{align}\label{eq:xS_1x_1S_2x_2}
\bm{x} = \mathbf{S}_1\bm{x}_1+\mathbf{S}_2\bm{x}_2-\mathbf{S}_0\bm{\zeta}(\bm{x},t),
\end{align}
which allow~\eqref{eq:x1x2} expressing the rigid relative velocity input $\mathbb{R}^2 \ni \bm{v}\ped{r}(\bm{z},\bm{x},t) = [v_{\textnormal{r}x}(\bm{z},\bm{x},t) \; v_{\textnormal{r}y}(\bm{z},\bm{x},t)]^{\mathrm{T}}$ as
\begin{align}\label{vDY}
\bm{v}\ped{r}(\bm{z},\bm{\zeta},\bm{x},t) &\triangleq \bm{v}_1(\bm{x}_1,t)-\bm{v}_2(\bm{x}_2,t) = \bm{v}_1(\bm{x}-\mathbf{S}_2\bm{z} + \mathbf{S}_0\bm{\zeta},t)-\bm{v}_2(\bm{x}+\mathbf{S}_1\bm{z} + \mathbf{S}_0\bm{\zeta},t).
\end{align}
Equations~\eqref{eq:eulerian2},~\eqref{eq:zdot}, and~\eqref{vDY} constitute the viscoelasto-kinematic equations for sliding and rolling friction contact pairs. Two important comments are in order.

First, Eq.~\eqref{eq:zdot} structurally resembles the corresponding equation derived in \cite{KinematicsMio}, which was formulated directly in terms of the bristle displacement $\bm{z}(\bm{x},t)$ under the assumption of linear elasticity. In this context, it is worth noting that reformulating Eq.~\eqref{eq:zdot} in terms of $\bm{z}(\bm{x},t)$ would introduce additional higher-order coupling terms with the internal variables, producing a contribution $[(\bm{V}_1(\bm{x},t)-\bm{V}_2(\bm{x},t))\cdot \nabla_{\bm{x}}]\mathbf{S}_0\bm{\zeta}(\bm{x},t)$ originating from the viscoelastic components of the bristle rheology. This would render the analysis of the resulting PDE system significantly more involved than in the formulation expressed in terms of the bristle force $\bm{f}(\bm{x},t)$. Moreover, as clarified in Sects.~\ref{ects:hyper} and~\ref{sect:BCsandICs}, the bristle force constitutes the natural and most appropriate state variable in the viscoelastic setting.

Second, Eq.~\eqref{eq:xS_1x_1S_2x_2} expresses the coordinate $\bm{x}$ as a weighted average of $\bm{x}_1$ and $\bm{x}_2$ via the matrices $\mathbf{S}_1$ and $\mathbf{S}_2$ -- which are completely determined by the elastic part of the rheology -- plus a term that arises from the viscoelastic component. In the elastic case, the quantity $\mathbf{S}_0\bm{\zeta}(\bm{x},t)$ in Eq.~\eqref{eq:xS_1x_1S_2x_2} disappears, and the definition of the variable $\bm{x}$ reduces exactly to that introduced in \cite{KinematicsMio}. Overall, Eq.~\eqref{eq:xS_1x_1S_2x_2} appears to be novel in the literature.

\section{Dynamic friction models for viscoelastic contact pairs}\label{sect:DynamicDer}
When combined with one of the models reviewed in Sect.~\ref{sect:Review}, the mathematical machinery developed in Sect.~\ref{sect:viscokin} permits to derive the PDEs governing the frictional sliding and rolling dynamics of the viscoelastic contact pair. The full set of equations is detailed next in Sect.~\ref{Sect.ModelEw}. The mathematical character of the resulting PDEs is then analysed in Sect.~\ref{ects:hyper}, whereas appropriate BCs and ICs are formulated in Sect.~\ref{sect:BCsandICs}. 

\subsection{Model equations}\label{Sect.ModelEw}
In the sequel, the model equations are divided into governing PDEs (Sect.~\ref{sect:PDEs}) and equilibrium equations (Sect.~\ref{sect:globalEq}). The former set describes the evolution of the bristle force and internal variables as a function of the rigid relative velocity input; the latter allows for computing the global forces and moments produced by the frictional interaction.

\subsubsection{Governing PDEs}\label{sect:PDEs}
Inserting Eqs.~\eqref{eq:zdot} and~\eqref{vDY} into~\eqref{eq:h_0} provides
\begin{align}\label{eq:PDEf}
\begin{split}
 \dpd{\bm{f}(\bm{x},t)}{t} + \Bigr[\bigl(\bm{V}_1(\bm{x},t)\cdot \nabla_{\bm{x}}\bigr)\mathbf{S}_1 + \bigl(\bm{V}_2(\bm{x},t)\cdot \nabla_{\bm{x}}\bigr)\mathbf{S}_2\Bigl]\bm{f}(\bm{x},t)  & = \tilde{\bm{h}}_0\bigl(\bm{f}(\bm{x},t),\bm{\zeta}(\bm{x},t),\bm{x},t\bigr), \\
& \qquad \bm{x}\in \mathring{\mathscr{C}}(t), \; t \in (0,T),
\end{split}
\end{align}
with $\tilde{\bm{h}}_0 : \mathbb{R}^{2(n+2)}\times \mathbb{R}_{\geq 0} \to \mathbb{R}^2$ defined as
\begin{align}
\begin{split}
\tilde{\bm{h}}_0(\bm{f},\bm{\zeta},\bm{x},t) &\triangleq \bar{\mathbf{K}}_0\bm{h}_0\Bigl(\bar{\mathbf{K}}_0^{-1}(\bm{f}-\bar{\mathbf{C}}\bm{\zeta}), \bm{f}, \bm{\zeta}, \bm{v}\ped{r}\bigl(\bar{\mathbf{K}}_0^{-1}(\bm{f}-\bar{\mathbf{C}}\bm{\zeta}),\bm{\zeta},\bm{x},t\bigr)\Bigr) \\
& \quad + \mathbf{S}_1\mathbf{C}_1\bm{h}_1(\bm{f},\bm{\zeta})-\mathbf{S}_2\mathbf{C}_2\bm{h}_2(\bm{f},\bm{\zeta}).
\end{split}
\end{align}
Equation~\eqref{eq:PDEf} describes the evolution of the tangential forces, depending on the postulated friction model~\eqref{eq:h_0}. It needs to be supplemented with the rheological equations governing the dynamics of the internal state variables. In particular, Eqs.~\eqref{eq:eulerian2} and~\eqref{eq:h_1},~\eqref{eq:h_2} give
\begin{subequations}\label{eq:PDEzeta1zeta2}
\begin{align}
\dpd{\bm{\zeta}_1(\bm{x},t)}{t} + \bigl(\bm{V}_1(\bm{x},t)\cdot\nabla_{\bm{x}}\bigr)\bm{\zeta}_1(\bm{x},t)
&= \bm{h}_1\bigl(\bm{f}(\bm{x},t),\bm{\zeta}(\bm{x},t)\bigr), \\
\dpd{\bm{\zeta}_2(\bm{x},t)}{t} + \bigl(\bm{V}_2(\bm{x},t)\cdot\nabla_{\bm{x}}\bigr)\bm{\zeta}_2(\bm{x},t)
&=  \bm{h}_2\bigl(\bm{f}(\bm{x},t),\bm{\zeta}(\bm{x},t)\bigr), \quad \bm{x}\in \mathring{\mathscr{C}}(t), \; t \in (0,T).
\end{align}
\end{subequations}
Introducing the vector $\mathbb{R}^{2(n+1)} \ni \bm{u}(\bm{x},t) \triangleq [\bm{f}^{\mathrm{T}}(\bm{x},t)\; \bm{\zeta}^{\mathrm{T}}(\bm{x},t)]^{\mathrm{T}}$, the PDEs~\eqref{eq:PDEf} and~\eqref{eq:PDEzeta1zeta2} may be conveniently restated as
\begin{align}\label{eq:PDEsystem}
\dpd{\bm{u}(\bm{x},t)}{t} + \mathbf{\Lambda}_x(\bm{x},t)\dpd{\bm{u}(\bm{x},t)}{x} + \mathbf{\Lambda}_y(\bm{x},t)\dpd{\bm{u}(\bm{x},t)}{y} = \bm{h}\bigl(\bm{u}(\bm{x},t),\bm{x},t\bigr),\quad \bm{x}\in \mathring{\mathscr{C}}(t), \; t \in (0,T),
\end{align}
with the matrices $\mathbf{\Lambda}_x(\bm{x},t), \mathbf{\Lambda}_y(\bm{x},t) \in \mathbf{Sym}_{2(n+1)}(\mathbb{R})$ given by
\begin{subequations}\label{eq:Lambndas}
\begin{align}
\mathbf{\Lambda}_x(\bm{x},t) \triangleq \begin{bmatrix}\mathbf{S}_1V_{1x}(\bm{x},t)+\mathbf{S}_2V_{2x}(\bm{x},t) & \mathbf{0} & \mathbf{0} \\ \mathbf{0} & \mathbf{I}_{2n_1}V_{1x}(\bm{x},t) & \mathbf{0} \\ \mathbf{0} & \mathbf{0} & \mathbf{I}_{2n_2}V_{2x}(\bm{x},t) \end{bmatrix}, \\
\mathbf{\Lambda}_y(\bm{x},t) \triangleq \begin{bmatrix}\mathbf{S}_1V_{1y}(\bm{x},t)+\mathbf{S}_2V_{2y}(\bm{x},t) & \mathbf{0} & \mathbf{0} \\ \mathbf{0} & \mathbf{I}_{2n_1}V_{1y}(\bm{x},t) & \mathbf{0} \\ \mathbf{0} & \mathbf{0} & \mathbf{I}_{2n_2}V_{2y}(\bm{x},t) \end{bmatrix},
\end{align}
\end{subequations}
and $\bm{h} : \mathbb{R}^{2(n+2)}\times \mathbb{R}_{\geq 0} \to \mathbb{R}^{2(n+1)}$ defined as
\begin{align}\label{eq:funcH2}
\bm{h}(\bm{u},\bm{x},t) \triangleq \begin{bmatrix} \tilde{\bm{h}}_0(\bm{u},\bm{x},t)\\ \bm{h}_1(\bm{u}) \\ \bm{h}_2(\bm{u}) \end{bmatrix}.
\end{align}
Equations~\eqref{eq:PDEf}-\eqref{eq:PDEzeta1zeta2} and~\eqref{eq:PDEsystem}-\eqref{eq:funcH2} appear to be novel: in contrast to the classic models for rolling contact systems with friction, where the contact pairs are assumed either elastically similar or very dissimilar, with one body being much more rigid than the other, they are not simple transport equations. However, coherently with the findings of \cite{KinematicsMio}, they are hyperbolic, as demonstrated later on in Sect.~\ref{ects:hyper0}. Another important feature of Eqs.~\eqref{eq:PDEsystem}-\eqref{eq:funcH2} is that the distributed or lumped character of the resulting friction dynamics is not postulated \emph{a priori}, but emerges naturally from the contact kinematics. In particular, the advection terms appearing in the governing equations are entirely determined by the transport of material particles through the contact region. Consequently, situations in which one or more bodies do not experience material transport inside the contact patch may lead to a partial or complete loss of advection, thereby reducing the distributed nature of the friction dynamics. This observation provides a unified interpretation of both classical finite-dimensional friction models and distributed contact formulations within a common viscoelasto-kinematic framework.

Before moving to the mathematical analysis of the PDE system~\eqref{eq:PDEsystem}-\eqref{eq:funcH2}, Sect.~\ref{sect:globalEq} is devoted to the calculation of the global forces and moments generated by the frictional process. 

\subsubsection{Forces and moments (global equilibrium)}\label{sect:globalEq}
Starting with the solution of Eqs.~\eqref{eq:PDEsystem}-\eqref{eq:funcH2}, it is possible to determine the total forces and moments acting on the two bodies in sliding and rolling contact. Concerning the tangential forces, these are denoted as $\mathbb{R}^2 \ni \bm{F}_{\bm{x},1}(t) = [F_{x,1}(t)\; F_{y,1}(t)]^{\mathrm{T}}$ and $\mathbb{R}^2 \ni \bm{F}_{\bm{x},2}(t) = [F_{x,2}(t)\; F_{y,2}(t)]^{\mathrm{T}}$ for the lower and upper body, respectively. Thus, setting $\mathbb{R}^2 \ni \bm{F}_{\bm{x}}(t) = [F_x(t)\; F_y(t)]^{\mathrm{T}} \triangleq \bm{F}_{\bm{x},1}(t) = -\bm{F}_{\bm{x},2}(t)$, and considering a variable reference normal pressure $p\in C^0(\mathscr{C}\times[0,T];\mathbb{R}_{\geq 0})$, the following formula may be deduced:
\begin{align}\label{eq:Force}
\begin{split}
\bm{F}_{\bm{x}}(t) & = \iint_{\mathscr{C}_1(t)}p(\bm{x},t)\bm{f}(\bm{x},t) \dif \bm{x}_1 = \iint_{\mathscr{C}(t)}p(\bm{x},t)\bm{f}(\bm{x},t) \abs{\det\nabla_{\bm{x}}\bm{x}_1^{\mathrm{T}}}\dif \bm{x} \\
& = \iint_{\mathscr{C}(t)}p(\bm{x},t)\bm{f}(\bm{x},t)\dif \bm{x} + \mathcal{O}(\epsilon), \quad t\in [0,T],
\end{split}
\end{align}
where Eq.~\eqref{eq:grad1} has been used. An expression equivalent to Eq.~\eqref{eq:Force} may be derived by integrating over $\mathscr{C}_2(t)$.

Concerning the vertical moment $\mathbb{R} \ni M_z(t) \triangleq M_{z,1}(t) = -M_{z,2}(t)$, where $M_{z,1}(t) \in \mathbb{R}$ and $M_{z,2}(t) \in \mathbb{R}$ indicate the moments acting on the upper and lower body, respectively, it is instead convenient to introduce the quantity $\mathbb{R}^2 \ni \bm{\phi}(\bm{x},t) = [\phi_{x}(\bm{x},t)\; \phi_{y}(\bm{x},t)]^{\mathrm{T}}$ as
\begin{align}\label{eq:phiDID}
\bm{\phi}(\bm{x},t) \triangleq (\mathbf{S}_1-\mathbf{S}_2)\bm{z}(\bm{x},t) +2 \mathbf{S}_0\bm{\zeta}(\bm{x},t),
\end{align}
so that $\bm{x}_1 + \bm{z}_1(\bm{x},t) = \bm{x}_2 + \bm{z}_2(\bm{x},t) = \bm{x} + \bm{\phi}(\bm{x},t)$. Accordingly, the total vertical moment acting on the upper body may be computed as
\begin{align}\label{eq:Mz}
\begin{split}
M_z(t) & = \iint_{\mathscr{C}_1(t)} p(\bm{x},t)\Bigl[\bigl(x+\phi_{x}(\bm{x},t)\bigr)f_y(\bm{x},t)-\bigl(y+\phi_{y}(\bm{x},t)\bigr)f_x(\bm{x},t)\Bigr] \dif \bm{x}_1 \\
& = \iint_{\mathscr{C}(t)}p(\bm{x},t)\Bigl[\bigl(x+\phi_{x}(\bm{x},t)\bigr)f_y(\bm{x},t)-\bigl(y+\phi_{y}(\bm{x},t)\bigr)f_x(\bm{x},t)\Bigr]\abs{\det\nabla_{\bm{x}}\bm{x}_1^{\mathrm{T}}} \dif \bm{x} \\
& = \iint_{\mathscr{C}(t)}p(\bm{x},t)\Bigl[\bigl(x+\phi_{x}(\bm{x},t)\bigr)f_y(\bm{x},t)-\bigl(y+\phi_{y}(\bm{x},t)\bigr)f_x(\bm{x},t)\Bigr] \dif \bm{x} + \mathcal{O}(\epsilon), \quad t \in [0,T].
\end{split}
\end{align}
owing again to Eq.~\eqref{eq:grad1}.

Alternatively, neglecting the terms related to $\bm{\phi}(\bm{x},t)$, Eq.~\eqref{eq:Mz} may be approximated as
\begin{align}\label{eq:Mzapprox}
M_z(t) & \approx \iint_{\mathscr{C}(t)}p(\bm{x},t)\bigl[xf_y(\bm{x},t)-yf_x(\bm{x},t)\bigr] \dif \bm{x}, \quad t \in [0,T].
\end{align}
Exactly as for Eq.~\eqref{eq:Force}, formulae equivalent to Eqs.~\eqref{eq:Mz} and~\eqref{eq:Mzapprox} may also be inferred by integrating over $\mathscr{C}_2(t)$. It is worth clarifying that Eq.~\eqref{eq:Mzapprox} is standard in the literature, whereas~\eqref{eq:Mz} appears to be novel.

Before moving to the mathematical characterisation of the PDE system~\eqref{eq:PDEsystem}, some comments are in order. First, within the framework of pure elasticity, the term $\mathbf{S}_0\bm{\zeta}(\bm{x},t)$ in Eq.~\eqref{eq:phiDID} vanishes, and the quantity $\bm{\phi}(\bm{x},t)$ is completely determined by the relative bristle deformation $\bm{z}(\bm{x},t)$, as discovered in \cite{KinematicsMio}. If, moreover, $\mathbf{S}_1 = \mathbf{S}_2$, as it happens in the case of similar elastic bodies, $\bm{\phi}(\bm{x},t) = \bm{0}$ and the coordinate $\bm{x}$ becomes the average between $\bm{x}_1$ and $\bm{x}_2$, which is consistent with the definition adopted in the classic works on rolling contact. Additionally, the limit situations $\mathbf{S}_1 = \bm{0}$ and $\mathbf{S}_2 = \bm{0}$ imply respectively $\bm{\phi}(\bm{x},t) = \bm{z}_2(\bm{x},t)$ and $\bm{\phi}(\bm{x},t) = \bm{z}_1(\bm{x},t)$, as known already from \cite{KinematicsMio}. 

Second, one of the most important consequences of the proposed viscoelasto-kinematic formulation is that the distributed or lumped nature of the friction dynamics is no longer imposed as a modelling assumption, but instead emerges naturally from the actual transport of material particles through the contact interface. This observation provides a unified interpretation of classical lumped friction models and distributed rolling contact models, which appear as different operating regimes of the same underlying continuum description rather than fundamentally different classes of models. In particular, whenever material transport through the contact patch disappears, the convective terms vanish automatically, and the governing equations reduce to finite-dimensional dynamics. Conversely, whenever material continuously enters and exits the contact region, the same formulation naturally generates distributed-parameter dynamics. Therefore, no \emph{ad hoc} distinction between sliding, rolling, or mixed rolling-sliding contact is required, since the mathematical structure follows directly from the contact kinematics.

\subsection{Hyperbolicity and Riemann variables}\label{ects:hyper}
As anticipated in Sect.~\ref{sect:PDEs}, although the PDE system~\eqref{eq:PDEsystem}-\eqref{eq:funcH2} is not a simple vector-valued transport equation, it nevertheless retains the classical hyperbolic character of rolling contact systems. This observation is formalised in Sect.~\ref{ects:hyper0}. Subsequently, Sect.~\ref{sect:Riemann} introduces a convenient change of variables in Riemann coordinates, which restores the standard transport-like form of the equations typically encountered in the literature.

\subsubsection{Hyperbolicity}\label{ects:hyper0}

In order to prescribe appropriate BCs for Eqs.~\eqref{eq:PDEsystem}-\eqref{eq:funcH2}, it is necessary to first examine their mathematical structure. Not surprisingly, the PDE system~\eqref{eq:PDEsystem}-\eqref{eq:funcH2} is hyperbolic, as stated in Lemma~\ref{lemma:hyper} below. The result is established under the assumption of a time-independent contact domain, i.e., $\mathscr{C}(t)=\mathscr{C}$. The extension to the case of a time-dependent contact region is discussed in Appendix~\ref{sect:appVarying}.

\begin{lemma}[Hyperbolicity]\label{lemma:hyper}
Under Assumptions~\ref{ass:comm} and~\ref{ass:s,allGrad}, the PDE system~\eqref{eq:PDEsystem}-\eqref{eq:funcH2} is \emph{symmetric hyperbolic}.
\begin{proof}
To prove the claim, it is necessary to show that, for all $\mathbb{R}^2 \ni \bm{\alpha} = (\alpha_x, \alpha_y)$, the matrix
\begin{align}\label{eq:MatrLambda}
\mathbf{\Lambda}(\bm{x},t) \triangleq \alpha_x\mathbf{\Lambda}_x(\bm{x},t) +  \alpha_y\mathbf{\Lambda}_y(\bm{x},t), \quad (\bm{x},t) \in\mathscr{C}\times[0,T],
\end{align}
is symmetric. This follows immediately from the fact that $\mathbf{\Lambda}_x(\bm{x},t), \mathbf{\Lambda}_y(\bm{x},t)$ are symmetric matrices, given by linear combinations of symmetric matrices $\mathbf{S}_1,\mathbf{S}_2$. Hence, for all $\bm{\alpha} \in \mathbb{R}^2$, their linear combination is also symmetric.
\end{proof}
\end{lemma}
From Lemma~\ref{lemma:hyper}, it may be concluded that viscoelasticity does not alter the hyperbolic nature of the frictional processes, as intuitively expected from \cite{KinematicsMio}. In fact, it is even possible to retrieve a transport-like character by introducing an appropriate change of variables. This is the subject of the next Sect.~\ref{sect:Riemann}.

\subsubsection{Riemann variables}\label{sect:Riemann}
At the level of higher-order derivatives, the only essential coupling in Eqs.~\eqref{eq:PDEsystem}-\eqref{eq:funcH2} arises from the elastic component of the bristle rheologies, encoded in the matrices $\mathbf{S}_1$ and $\mathbf{S}_2$. In particular, as is apparent from Eq.~\eqref{eq:Lambndas}, the matrices $\mathbf{\Lambda}_x(\bm{x},t)$ and $\mathbf{\Lambda}_y(\bm{x},t)$ share the same structural form and are simultaneously diagonalisable, as a consequence of the structure of $\mathbf{S}_1$ and $\mathbf{S}_2$.

In this context, it is profitable to consider the following rotation matrix: 
\begin{align}\label{eq:MR}
\mathbf{R} = \begin{bmatrix} \cos\theta & -\sin\theta \\ \sin\theta & \cos\theta \end{bmatrix}, \quad \textnormal{with} \quad \tan(2\theta) \triangleq \dfrac{2s_{1xy}}{s_{1xx}-s_{1yy}} = \dfrac{2s_{2xy}}{s_{2xx}-s_{2yy}},
\end{align} 
which is completely determined by the reciprocal elastic behaviour between the contacting bodies.
Clearly, in Eq.~\eqref{eq:MR}, $\mathbf{R} \in \mathbf{SO}_2(\mathbb{R})$ with $\mathbf{R}^{-1} \equiv \mathbf{R}^{\mathrm{T}}$ (it should be observed that, when $\mathbf{S}_1 = s_1\mathbf{I}_2$ for some $s_1 \in \mathbb{R}_{>0}$, every orthogonal basis is admissible and it is possible to choose $\mathbf{R} = \mathbf{I}_2$ without loss of generality). 
Accordingly, the change of variables $\mathbb{R}^2\ni \tilde{\bm{f}}(\bm{x},t) = [\tilde{f}_1(\bm{x},t) \; \tilde{f}_2(\bm{x},t)]^{\mathrm{T}}\triangleq \mathbf{R}^{\mathrm{T}}\bm{f}(\bm{x},t)$ transforms the PDE~\eqref{eq:PDEf} into 
\begin{align}\label{eq:PDEphi}
\begin{split}
 \dpd{\tilde{\bm{f}}(\bm{x},t)}{t} + \begin{bmatrix} \tilde{\bm{V}}_1(\bm{x},t)\cdot\nabla_{\bm{x}}  & 0 \\ 0 & \tilde{\bm{V}}_2(\bm{x},t)\cdot\nabla_{\bm{x}} \end{bmatrix}\tilde{\bm{f}}(\bm{x},t)  & = \mathbf{R}^{\mathrm{T}}\tilde{\bm{h}}_0\bigl(\mathbf{R}\tilde{\bm{f}}(\bm{x},t),\bm{\zeta}(\bm{x},t),\bm{x},t\bigr), \quad \bm{x}\in \mathring{\mathscr{C}}(t), \; t \in (0,T),
\end{split}
\end{align}
where the principal velocities $\mathbb{R}^2\ni\tilde{\bm{V}}_1(\bm{x},t) = [\tilde{V}_{1x}(\bm{x},t)\; \tilde{V}_{1y}(\bm{x},t)]^{\mathrm{T}}$ and $\mathbb{R}^2 \ni \tilde{\bm{V}}_2(\bm{x},t) = [\tilde{V}_{2x}(\bm{x},t)\; \tilde{V}_{2y}(\bm{x},t)]^{\mathrm{T}}$ are given by
\begin{subequations}\label{eq:princVel}
\begin{align}
\tilde{\bm{V}}_1(\bm{x},t) & = \lambda_{1}(\mathbf{S}_1)\bm{V}_1(\bm{x},t)+\lambda_{1}(\mathbf{S}_2)\bm{V}_{2}(\bm{x},t), \\
\tilde{\bm{V}}_2(\bm{x},t) & = \lambda_{2}(\mathbf{S}_1)\bm{V}_1(\bm{x},t)+\lambda_{2}(\mathbf{S}_2)\bm{V}_{2}(\bm{x},t),
\end{align}
\end{subequations}
with the eigenvalues $\lambda_{i}(\mathbf{S}_1), \lambda_{i}(\mathbf{S}_2) \in (0,1)$, $i \in \{1,2\}$, determined according to
\begin{subequations}\label{eq:eigens}
\begin{align}
\lambda_{1}(\mathbf{S}_1) & = 1-\lambda_{1}(\mathbf{S}_2) = \dfrac{s_{1xx}+s_{1yy}}{2} + \dfrac{1}{2}\sqrt{(s_{1xx}-s_{1yy})^2+4s_{1xy}^2}, \\
\lambda_{2}(\mathbf{S}_1) & = 1-\lambda_{2}(\mathbf{S}_2) = \dfrac{s_{1xx}+s_{1yy}}{2} - \dfrac{1}{2}\sqrt{(s_{1xx}-s_{1yy})^2+4s_{1xy}^2}.
\end{align}
\end{subequations}
Therefore, introducing the matrix $\mathbf{Q} \in \mathbf{GL}_{2(n+1)}(\mathbb{R})$ as
\begin{align}
\mathbf{Q} \triangleq \begin{bmatrix} \mathbf{R} & \mathbf{0} & \mathbf{0} \\ \mathbf{0} & \mathbf{I}_{2n_1} & \mathbf{0} \\ \mathbf{0} & \mathbf{0} & \mathbf{I}_{2n_2}\end{bmatrix},
\end{align}
and applying the transformation $\mathbb{R}^{2(n+1)} \ni \tilde{\bm{u}}(\bm{x},t) \triangleq \mathbf{Q}^{\mathrm{T}}\bm{u}(\bm{x},t) = [\tilde{\bm{f}}^{\mathrm{T}}(\bm{x},t)\; \bm{\zeta}^{\mathrm{T}}(\bm{x},t)]^{\mathrm{T}}$  yields the system in Riemann variables
	\begin{align}\label{eq:PDEsystemR}
	\dpd{\tilde{\bm{u}}(\bm{x},t)}{t} + \tilde{\mathbf{\Lambda}}_x(\bm{x},t)\dpd{\tilde{\bm{u}}(\bm{x},t)}{x} + \tilde{\mathbf{\Lambda}}_y(\bm{x},t)\dpd{\tilde{\bm{u}}(\bm{x},t)}{y} = \tilde{\bm{h}}\bigl(\tilde{\bm{u}}(\bm{x},t),\bm{x},t\bigr),\quad \bm{x}\in \mathring{\mathscr{C}}(t), \; t \in (0,T),
	\end{align}
with the matrices $\tilde{\mathbf{\Lambda}}_x(\bm{x},t),\tilde{\mathbf{\Lambda}}_y(\bm{x},t) \in \mathbf{Sym}_{2(n+1)}(\mathbb{R})$ given by
\begin{subequations}\label{eq:Lambdatilde}
\begin{align}
\begin{split}
\tilde{\mathbf{\Lambda}}_x(\bm{x},t)& \triangleq \mathbf{Q}^{\mathrm{T}}\mathbf{\Lambda}_x(\bm{x},t)\mathbf{Q}   \triangleq \begin{bmatrix}\tilde{V}_{1x}(\bm{x},t) & 0 & \mathbf{0} & \mathbf{0} \\ 0 & \tilde{V}_{2x}(\bm{x},t) & \mathbf{0} & \mathbf{0} \\ \mathbf{0} & \mathbf{I}_{2n_1}V_{1x}(\bm{x},t) & \mathbf{0} & \mathbf{0} \\\mathbf{0} &  \mathbf{0} & \mathbf{0} & \mathbf{I}_{2n_2}V_{2x}(\bm{x},t) \end{bmatrix}, 
\end{split} \\
\tilde{\mathbf{\Lambda}}_y(\bm{x},t)& \triangleq \mathbf{Q}^{\mathrm{T}}\mathbf{\Lambda}_y(\bm{x},t)\mathbf{Q}   \triangleq \begin{bmatrix}\tilde{V}_{1y}(\bm{x},t) & 0 & \mathbf{0} & \mathbf{0} \\ 0 & \tilde{V}_{2y}(\bm{x},t) & \mathbf{0} & \mathbf{0} \\ \mathbf{0} & \mathbf{I}_{2n_1}V_{1y}(\bm{x},t) & \mathbf{0} & \mathbf{0} \\\mathbf{0} &  \mathbf{0} & \mathbf{0} & \mathbf{I}_{2n_2}V_{2y}(\bm{x},t) \end{bmatrix},
\end{align}
\end{subequations}
and the function $\tilde{\bm{h}} : \mathbb{R}^{2(n+2)}\times \mathbb{R}_{\geq 0} \to \mathbb{R}^{2(n+1)}$ reading $\tilde{\bm{h}}(\tilde{\bm{u}},\bm{x},t) \triangleq \mathbf{Q}^{\mathrm{T}}\bm{h}(\mathbf{Q}\tilde{\bm{u}},\bm{x},t)$. The transformation to Riemann coordinates provides a particularly transparent interpretation of the frictional dynamics: in these variables, each component of the contact force propagates along characteristic directions associated with the principal elastic interaction modes of the contact pair. Therefore, the matrices $\mathbf{S}_1$ and $\mathbf{S}_2$ not only determine the partition of the elastic deformation between the two bodies, but also govern the characteristic transport structure of the resulting friction model. In particular, in Eq.~\eqref{eq:PDEsystemR}, the vector of nondimensional bristle forces, $\bm{f}(\bm{x},t)$, is replaced by $\tilde{\bm{f}}(\bm{x},t)$, collecting the force components expressed in the principal directions determined by the matrices $\mathbf{S}_1$ and $\mathbf{S}_2$. In this context, it is worth clarifying that the PDE system~\eqref{eq:PDEsystemR}-\eqref{eq:Lambdatilde} may be interpreted as an interconnection of four vector-valued transport equations, with different transport velocities.

By enforcing appropriate BCs and ICs (see Sect.~\ref{sect:BCsandICs}), and limited to the case of time-invariant contact areas and velocities $\bm{V}_1(\bm{x},t) = \bm{V}_1(\bm{x})$ and $\bm{V}_2(\bm{x},t) = \bm{V}_2(\bm{x})$, the well-posedness of the system in Riemann variables, as described by Eqs.~\eqref{eq:PDEsystemR} and~\eqref{eq:Lambdatilde}, may be proved as done in Appendix~\ref{sect:wellP}.

\subsection{Boundary and initial conditions}\label{sect:BCsandICs}

In order to specify the BCs for the PDE system in Riemann variables~\eqref{eq:PDEsystemR}, and consequently also for the original one described by Eqs.~\eqref{eq:PDEsystem}-\eqref{eq:funcH2}, it is beneficial to define the notions of \emph{leading edge} $\mathscr{L}_i(t)$, \emph{neutral edge} $\mathscr{N}_i(t)$, and \emph{trailing edge} $\mathscr{T}_i(t)$, $i \in \{1,2\}$, for the upper and lower bodies as follows \cite{FrBDroll}:
\begin{subequations}\label{eq:Inflow0Outflow}
\begin{align}
\mathscr{L}_i(t) &\triangleq \Bigl\{\bm{x}\in \partial \mathscr{C}(t) \mathrel{\Big|}\bigl[\bm{V}_i(\bm{x},t)-\bm{V}_{\partial \mathscr{C}}(\bm{x},t)\bigr] \cdot \hat{\bm{n}}_{\partial \mathscr{C}}(\bm{x},t) < 0\Bigr\}, \\
\mathscr{N}_i(t) &\triangleq \Bigl\{\bm{x}\in \partial \mathscr{C}(t) \mathrel{\Big|} \bigl[\bm{V}_i(\bm{x},t)-\bm{V}_{\partial \mathscr{C}}(\bm{x},t)\bigr] \cdot \hat{\bm{n}}_{\partial \mathscr{C}}(\bm{x},t) = 0\Bigr\}, \\
\mathscr{T}_i(t) &\triangleq \Bigl\{\bm{x}\in \partial \mathscr{C}(t) \mathrel{\Big|} \bigl[\bm{V}_i(\bm{x},t)-\bm{V}_{\partial \mathscr{C}}(\bm{x},t)\bigr] \cdot \hat{\bm{n}}_{\partial \mathscr{C}}(\bm{x},t) > 0\Bigr\}, \quad i \in \{1,2\}, 
\end{align}
\end{subequations}
where $\hat{\bm{n}}_{\partial \mathscr{C}}(\bm{x},t)\in \mathbb{R}^2$ denotes the outward unit normal to $\partial \mathscr{C}(t)$, and $\bm{V}_{\partial \mathscr{C}}(\bm{x},t) \in \mathbb{R}^2$ it velocity. 
The sets defined according to Eq.~\eqref{eq:Inflow0Outflow} are purely geometrical, and completely determined by the real kinematics of the contacting bodies.

Following a similar rationale, it is also possible to define
\begin{subequations}\label{eq:Inflow0Outflow2}
\begin{align}
\tilde{\mathscr{L}}_i(t) &\triangleq \Bigl\{\bm{x}\in \partial \mathscr{C}(t) \mathrel{\Big|}\bigl[\tilde{\bm{V}}_i(\bm{x},t)-\bm{V}_{\partial \mathscr{C}}(\bm{x},t)\bigr] \cdot \hat{\bm{n}}_{\partial \mathscr{C}}(\bm{x},t) < 0\Bigr\}, \\
\tilde{\mathscr{N}}_i(t) &\triangleq \Bigl\{\bm{x}\in \partial \mathscr{C}(t) \mathrel{\Big|} \bigl[\tilde{\bm{V}}_i(\bm{x},t)-\bm{V}_{\partial \mathscr{C}}(\bm{x},t)\bigr] \cdot \hat{\bm{n}}_{\partial \mathscr{C}}(\bm{x},t) = 0\Bigr\}, \\
\tilde{\mathscr{T}}_i(t) &\triangleq \Bigl\{\bm{x}\in \partial \mathscr{C}(t) \mathrel{\Big|} \bigl[\tilde{\bm{V}}_i(\bm{x},t)-\bm{V}_{\partial \mathscr{C}}(\bm{x},t)\bigr] \cdot \hat{\bm{n}}_{\partial \mathscr{C}}(\bm{x},t) > 0\Bigr\}, \quad i \in \{1,2\},
\end{align}
\end{subequations}
which may be interpreted as the viscoelasto-kinematic leading, neutral, and trailing edges.

Starting with Eqs.~\eqref{eq:Inflow0Outflow} and~\eqref{eq:Inflow0Outflow2}, appropriate BCs for the PDEs in Riemann variables may be prescribed as
\begin{subequations}\label{eq:BCs}
\begin{align}
\tilde{f}_1(\bm{x},t) & = 0, \quad \bm{x} \in \tilde{\mathscr{L}}_1(t), \; t \in (0,T), \label{eq:BC1_tilfed} \\
\tilde{f}_2(\bm{x},t) & = 0, \quad \bm{x} \in \tilde{\mathscr{L}}_2(t), \; t \in (0,T), \label{eq:BC2_tilfed} \\
\bm{\zeta}_1(\bm{x},t) & = \bm{0}, \quad \bm{x} \in \mathscr{L}_1(t), \; t \in (0,T), \label{eq:BC1_zeta}\\
\bm{\zeta}_2(\bm{x},t) & = \bm{0}, \quad \bm{x} \in \mathscr{L}_2(t), \; t \in (0,T).\label{eq:BC2_zeta}
\end{align}
\end{subequations}
When expressed in the original variable $\bm{f}(\bm{x},t)$ the first two BCs~\eqref{eq:BC1_tilfed} and~\eqref{eq:BC2_tilfed} become
\begin{subequations}
\begin{align}
f_x(\bm{x},t)\cos\theta+f_y(\bm{x},t)\sin\theta & = 0, \quad \bm{x} \in \tilde{\mathscr{L}}_1(t), \; t \in (0,T), \label{eq:BC1_tilfed0} \\
f_y(\bm{x},t)\cos\theta-f_x(\bm{x},t)\sin\theta & = 0, \quad \bm{x} \in \tilde{\mathscr{L}}_2(t), \; t \in (0,T). \label{eq:BC2_tilfed0}
\end{align}
\end{subequations} 
The BCs~\eqref{eq:BCs} (or alternatively~\eqref{eq:BC1_tilfed0} and~\eqref{eq:BC2_tilfed0} with~\eqref{eq:BC1_zeta} and~\eqref{eq:BC2_zeta}) ensure dissipativity, and permit establishing the well-posedness of the frictional problem (see, e.g., Appendix~\ref{sect:wellP}). Moreover, they generalise those known from the classical Winkler-type treatments of rolling contact \cite{FrBDvisc,KalkerBook,2000RCP}, and appear to be novel in the related literature.

For the system in Riemann coordinates, the ICs may be instead simply postulated as
\begin{align}
\tilde{\bm{u}}(\bm{x},0) = \tilde{\bm{u}}_0(\bm{x}), \quad \bm{x}\in \tilde{\mathscr{C}}.
\end{align}
The corresponding ICs for the system in the original variables may be then inferred as $\bm{u}_0(\bm{x}) = \mathbf{Q}\tilde{\bm{u}}_0(\bm{x})$ .

Before presenting applications of the proposed theory to classical sliding and rolling contact problems, some important considerations are gathered in the following Remark~\ref{remark:BCs}.

\begin{remark}[Boundary conditions]\label{remark:BCs}
The boundary conditions~\eqref{eq:BC1_tilfed} and~\eqref{eq:BC2_tilfed} enforce the continuity of the principal stresses $\tilde{\bm{f}}(\bm{x},t)$ at the transition from the free surfaces of the contacting bodies into the contact region. This is consistent with the observations reported by Kalker \cite{KalkerBook}. Clearly, if $\tilde{\mathscr{L}}_1(t)\equiv\tilde{\mathscr{L}}_2(t)$, as is the case in most practical applications, the Cartesian components of the stresses also vanish automatically. By contrast, the bristle deformations associated with the two bodies do not necessarily vanish, unless $\tilde{\mathscr{L}}_1(t)\equiv\tilde{\mathscr{L}}_2(t)\equiv\mathscr{L}_1(t)\equiv\mathscr{L}_2(t)$. In this regard, it is worth noting that some friction models based on lower-order constitutive relationships (e.g., the LuGre model) impose zero deformation at the leading edge. Such a condition is generally incompatible with the boundary conditions~\eqref{eq:BC1_tilfed} and~\eqref{eq:BC2_tilfed}.
\end{remark}

\section{Application to sliding and rolling contact}\label{sect:appl}
The hyperbolic friction models derived in Sect.~\ref{sect:DynamicDer} possess broad applicability. This section illustrates their use in a set of simple line contact problems based on the one-dimensional version of the FrBD formulation presented in Appendix~\ref{app:FrBD}. 
Both steady-state and transient sliding and rolling contact problems are examined to highlight the principal similarities and differences between these two classes of processes.

To simplify the analysis, the matrices $\bar{\mathbf{K}}_{0,1}$ and  $\bar{\mathbf{K}}_{0,2}$ are assumed to be isotropic, namely $\bar{\mathbf{K}}_{0,1} = \bar{k}_{0,1}\mathbf{I}_2$ and $\bar{\mathbf{K}}_{0,2} = \bar{k}_{0,2}\mathbf{I}_2$. Consequently, $\bar{\mathbf{K}}_0 = \bar{k}_0\mathbf{I}_2$, with $\mathbb{R}_{>0}\ni \bar{k}_0 = \frac{\bar{k}_{0,1}\bar{k}_{0,2}}{\bar{k}_{0,1} + \bar{k}_{0,2}}$ and, similarly, $\mathbf{S}_1 = s_1\mathbf{I}_2$, and $\mathbf{S}_2 = s_2\mathbf{I}_2$. It is convenient to introduce the parameter $(0,1) \ni s \triangleq s_2$, with $s_1 = (1-s)$.

Furthermore, assuming that all the remaining matrices appearing in Eqs.~\eqref{eq:PDEsinternalstates} and~\eqref{eq:PDEf_FrBD} are diagonal, the longitudinal and lateral dynamics become completely decoupled and may therefore be analysed independently. In particular, the matrix $\mathbf{M}(\bm{v}\ped{r})$ in Eq.~\eqref{eq:h0FrBD} is modelled as $\mathbf{M}(\bm{v}\ped{r}) = \mu(\bm{v}\ped{r})\mathbf{I}_2$, with the friction coefficient $\mu \in C^0(\mathbb{R}^2; \mathbb{R}_{>0})$ postulated as
\begin{align}\label{eq:muExample}
\mu(\bm{v}\ped{r}) = \mu\ped{d} + (\mu\ped{s}-\mu\ped{d})\exp\Biggl(-\biggl(\dfrac{\norm{\bm{v}\ped{r}}_2}{v\ped{S}}\biggr)^{\delta\ped{S}}\Biggr)+ \mu\ped{v}(\bm{v}\ped{r}),
\end{align}
where $\mu\ped{s},\mu\ped{d} \in \mathbb{R}_{>0}$ denote the static and dynamic friction coefficients, $v\ped{S} \in \mathbb{R}_{>0}$ the Stribeck velocity, $\delta\ped{S} \in \mathbb{R}_{\geq 0}$ the Stribeck exponent, and $\mu\ped{v} : \mathbb{R}^2 \to \mathbb{R}_{\geq 0}$ captures the viscous friction. 

In the remainder of this section, attention is also restricted to longitudinal sliding and rolling contact in the absence of spin, with the intent of elucidating the fundamental features of the two phenomena whilst keeping the mathematical treatment tractable. In fact, the study of more complex situations involving combined slip and spin conditions possesses its own dignity and will be the subject of future works. The next Sect.~\ref{sect:sliding} investigates both stationary and sliding friction problems, whereas Sect.~\ref{sect:roll} is dedicated to rolling contact.

\subsection{Sliding friction}\label{sect:sliding}
Within the scope of the present work, sliding friction deserves particular attention because of its distinctive mathematical structure, which combines hyperbolic PDEs with local ODEs that may be interpreted as degenerate distributed equations. To illustrate this feature, the prototypical configuration shown in Fig.~\ref{fig:RollingBodies}(c) may be considered, where a (possibly rigid), rectangular block of length $L \in \mathbb{R}_{>0}$ slides over a viscoelastic substrate with velocity $\mathbb{R}^2 \ni \bm{V}_{O_1}(t) = [V_x(t)\; 0 ]^{\mathrm{T}}$.

During the sliding motion, new substrate material particles continuously enter the contact region $\mathbb{R} \supset \mathscr{C}(t)  = \mathscr{C}= [-\frac{L}{2}, \frac{L}{2}]$ with velocity $\mathbb{R}^2 \ni \bm{V}_2(\bm{x},t) = -\bm{V}_{O_1}(t) = -[V_x(t)\; 0]^{\mathrm{T}}$; by contrast, the bristles attached to the block remain the same throughout the motion and are not advected, so that $\mathbb{R}^2 \ni \bm{V}_{1}(\bm{x},t) = \bm{0}$. Consequently, the rigid relative velocity reduces to $\mathbb{R}^2 \ni \bm{v}_{\mathrm{r}x}(\bm{x},t) = [V_x(t)\; 0]^{\mathrm{T}}$. Thus, assuming $V_x(t) \in [V\ped{min},V\ped{max}]$, with $0 < V\ped{min} \leq V\ped{max}$, introducing the coordinate transformation $[0,L]\ni \xi \triangleq \frac{L}{2}-x$, and defining the longitudinal state vector as $\mathbb{R}^{n+1} \ni \bm{u}_x(\xi,t) \triangleq [f_x(\xi,t)\; z_{1,1x}(\xi,t)\; \dots \; z_{n_1,1x}(\xi,t)\; z_{1,2x}(\xi,t)\; \dots \; z_{n_2,2x}(\xi,t)]^{\mathrm{T}}$, the longitudinal components of Eqs.~\eqref{eq:PDEsinternalstates} and~\eqref{eq:PDEf_FrBD} (with $\varepsilon = 0$) simplify to 
\begin{subequations}\label{eq:PDEsinternalstatesS} 
\begin{align}
\begin{split}\label{eq:fxSlid}
 \dpd{f_x(\xi,t)}{t}  +\tilde{V}_x(t)\dpd{f_x(\xi,t)}{\xi}  & =  -\bar{k}_0\dfrac{V_x(t)}{\mu\bigl(V_x(t)\bigr)}f_x(\xi,t)-\bar{k}_0V_x(t) \\
& \quad  - \bar{k}_0\sum_{i=1}^{n_1}\dfrac{\D_1 z_{i,1x}(\xi,t)}{\D t}+ \bar{k}_0\sum_{i=1}^{n_2}\dfrac{\D_2 z_{i,2x}(\xi,t)}{\D t}, 
\end{split} \\
\begin{split}
\dfrac{\D_1 z_{i,1x}(\xi,t)}{\D t} & = \dpd{z_{i,1x}(\xi,t)}{t}  = -\dfrac{z_{i,1x}(\xi,t)}{\tau_{i,1}} + \dfrac{f_x(\xi,t)}{\bar{c}_{i,1}}, \quad i \in \{1,\dots,n_1\}, \label{€q:z1Slidx}
\end{split}\\
\begin{split}
\dfrac{\D_2 z_{i,2x}(\xi,t)}{\D t} & = \dpd{z_{i,2x}(\xi,t)}{t} +V_x(t)\dpd{z_{i,2x}(\xi,t)}{\xi} \\
& = -\dfrac{z_{i,2x}(\xi,t)}{\tau_{i,2}} - \dfrac{f_x(\xi,t)}{\bar{c}_{i,2}}, \quad i \in \{1,\dots,n_2\}, \; (\xi,t) \in (0,L) \times (0,T), \label{€q:z2Slidx}
\end{split} \\
f_x(0,t) & = 0, \quad t \in (0,T),  \label{eq:BCfxSlid}\\
z_{i,2x}(0,t) & = 0, \quad i \in \{1,\dots,n_2\}, \; t \in (0,T), \label{eq:BCz_2Slid}\\
\bm{u}_x(\xi,0) & = \bm{u}_{x,0}(\xi), \quad \xi \in (0,L),
\end{align}
\end{subequations}
with $\mathbb{R}_{>0}\ni \tilde{V}_x(t) = sV_x(t)$, and the model parameters satisfying $\bar{k}_0 \in \mathbb{R}_{>0}$, $\tau_{i,1},\bar{c}_{i,1} \in \mathbb{R}_{>0}$, $i \in \{1,\dots,n_1\}$, and $\tau_{i,2},\bar{c}_{i,2} \in \mathbb{R}_{>0}$, $i \in \{1,\dots,n_2\}$. 

The system described by Eq.~\eqref{eq:PDEsinternalstatesS} constitutes an interconnection of hyperbolic PDEs, namely~\eqref{eq:fxSlid} and~\eqref{€q:z2Slidx}, $i \in \{1,\dots,n_2\}$, together with local ODEs~\eqref{€q:z1Slidx}, $i \in \{1,\dots,n_1\}$. In particular, the ODE structure arises from the absence of advection of material particles within the block. Nevertheless, the equations~\eqref{€q:z1Slidx} remain spatially local, in the sense that their evolution depends on the spatial coordinate, in contrast with classical finite-dimensional friction models, which assume spatially uniform dynamics for forces and deformations. Considering a constant $V_x(t) = V_x$, well-posedness for the PDE system~\eqref{eq:PDEsinternalstatesS}  follows directly from Theorem~\ref{thm:wellP} in Appendix~\ref{sect:wellP}; alternatively, assuming $V_x \in C^1(\mathbb{R}_{\geq 0};[V\ped{min},V\ped{max}])$, with $0 < V\ped{min} \leq V\ped{max}$, a strategy similar to that in \cite{MScMath} may be adopted.

The main implications of the hyperbolic character of Eq.~\eqref{eq:PDEsinternalstatesS} are discussed in the sequel concerning both steady-state and transient sliding, which are addressed respectively in Sects.~\ref{sect:SteadySlid} and~\ref{sect:TransSlid}.

\subsubsection{Steady sliding}\label{sect:SteadySlid}
To gain some preliminary intuition about the steady-state sliding process, it is beneficial to first consider the purely elastic case, with $z_{i,1x}(\xi,t) = 0$, $i \in \{1,\dots,n_1\}$, and $z_{i,2x}(\xi,t) = 0$, $i \in \{1,\dots,n_2\}$ in Eq.~\eqref{eq:fxSlid}. This allows for deriving closed-form solutions for the friction force and global longitudinal force acting on the block. In particular, in stationary conditions ($\pd{f_x(\xi,t)}{t} = 0$), Eq.~\eqref{eq:fxSlid} equipped with the BC~\eqref{eq:BCfxSlid} admits the following solution: 
\begin{align}\label{eq:fxSlidSolEl}
f_x(\xi) & = -\mu(V_x)\bigggl[1-\exp\Biggl(-\dfrac{(1-s)\bar{k}_{0,1}\xi}{s\mu(V_x)}\Biggr)\bigggr], \quad \xi \in [0,L],
\end{align} 
where the identity $\mathbb{R}_{>0} \ni \bar{k}_{0,1} = \bar{k}_0/(1-s)$ has been used. The deformations of the bristles attached to the upper and lower bodies read, accordingly,
\begin{subequations}\label{eq:zSlidElSol}
\begin{align}
z_{1x}(\xi) & = \dfrac{f_x(\xi)}{\bar{k}_{0,1}} = -\dfrac{\mu(V_x)}{\bar{k}_{0,1}}\bigggl[1-\exp\Biggl(-\dfrac{(1-s)\bar{k}_{0,1}\xi}{s\mu(V_x)}\Biggr)\bigggr],\\
z_{2x}(\xi) & = -\dfrac{f_x(\xi)}{\bar{k}_{0,2}} = \dfrac{s\mu(V_x)}{(1-s)\bar{k}_{0,1}}\bigggl[1-\exp\Biggl(-\dfrac{(1-s)\bar{k}_{0,1}\xi}{s\mu(V_x)}\Biggr)\bigggr], \quad \xi \in [0,L].
\end{align}
\end{subequations}
The expressions reported in Eqs.~\eqref{eq:fxSlidSolEl} and~\eqref{eq:zSlidElSol} are in theoretical agreement with the distributed versions of the Dahl, LuGre, and FrBD models. It is important to observe that, for $s \in (0,1)$, both the friction force and the bristle deformations identically vanish at the leading edge ($\xi = 0$). In stationary conditions, this is not a mere consequence of the assumed elastic behaviour. Indeed, whilst for the friction force and lower bristle deflection this constraint is naturally enforced by the BCs~\eqref{eq:BCfxSlid} and~\eqref{eq:BCz_2Slid} for any rheological order of the constitutive equations, $\pd{z_{i,1x}(\xi,t)}{t} = 0$, $i \in \{1,\dots,n_1\}$, in Eq.~\eqref{€q:z1Slidx}, in conjunction with $f_x(0) = 0$, immediately implies $z_{1x}(\xi) = 0$.

Conversely, in the limit $s \to 0$, the PDEs~\eqref{eq:fxSlid} and~\eqref{€q:z2Slidx} degenerate into local ODEs, and the forces as well as the bristle deflections become spatially uniform, independently of the elastic behaviour of the block, whose material particles remain fixed with respect to the contact patch. In this regime, the classical ODE-type behaviour of finite-dimensional sliding friction models is thus recovered.

Similar considerations may be drawn by calculating the total longitudinal force acting on the sliding block. Assuming a constant reference pressure distribution $\mathbb{R}_{>0} \ni p = \frac{F_z}{L}$, where $F_z \in \mathbb{R}_{>0}$ denotes the applied vertical load, and integrating Eq.~\eqref{eq:fxSlidSolEl} according to~\eqref{eq:Force} yields
\begin{align}\label{eq:FxSlid}
\bar{F}_x(V_x,s) \triangleq \dfrac{F_x(V_x,s)}{\mu(V_x)F_z} & = -1 +\dfrac{s\mu(V_x)}{(1-s)\bar{k}_{0,1}L}\bigggl[1-\exp\Biggl(-\dfrac{(1-s)\bar{k}_{0,1}L}{s\mu(V_x)}\Biggr)\bigggr], \quad (V_x,s) \in [V\ped{min},V\ped{max}]\times(0,1).
\end{align}
The second term on the right-hand side of~\eqref{eq:FxSlid} is always positive. Consequently, the magnitude of the longitudinal force is always strictly smaller than $\mu(V_x)F_z$, which corresponds to the limiting value attained as $s\to 0$. In a similar context, it is also crucial to observe that $\lim_{V_x \to 0} F_x(V_x,s), \bar{F}_x(V_x,s) > 0$ due to the assumptions made on $\mu(V_x)$. This is an important distinction with rolling friction, as already known from the literature and further corroborated by the analysis conducted in Sect.~\ref{ect:SteadtRoll}.

\begin{figure}
\centering
\includegraphics[width=1\linewidth]{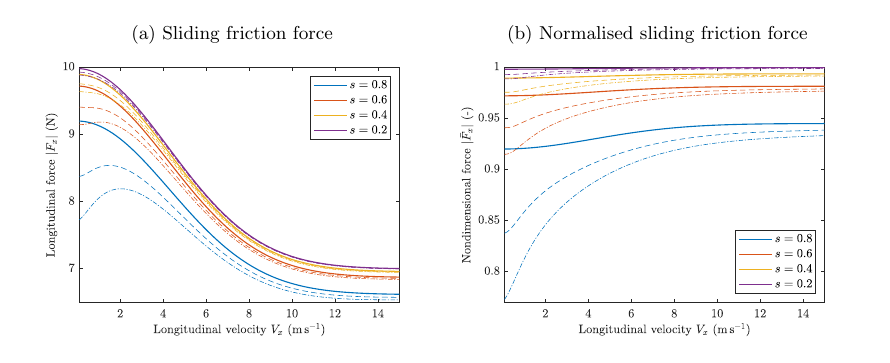} 
\caption{Steady-state longitudinal and normalised force $\abs{F_x(V_x,s)}$ and $\abs{\bar{F}_x(V_x,s)}$, respectively, for different values of the parameter $s=0.2,0.4,0.6$, and 0.8. Line styles: 0 dissipative branches (solid thick lines), 1 dissipative branch (dashed lines), 2 dissipative branches (dash-dotted lines). Model parameters as in Table~\ref{tab:parameters1}.}
\label{fig:Num1}
\end{figure}

The analytical arguments presented above remain valid when viscoelastic components are included. In particular, inspection of Eqs.~\eqref{eq:fxSlid} and~\eqref{€q:z1Slidx} show that the rheology of the block does not influence the steady-state response of the sliding system. Fig.~\ref{fig:Num1} compares the curves for $F_x(V_x,s)$ and $\bar{F}_x(V_x,s)$ for different values of the parameter $s=0.2,0.4,0.6$, and 0.8, whilst simultaneously illustrating the effect of increasing the number of dissipative branches in the substrate rheology. Two main conclusions can be drawn from Fig.~\ref{fig:Num1}. The first is that, as the parameter $s$ decreases, larger forces are generated, in agreement with the prediction already inferred from Eq.~\eqref{eq:FxSlid}. This behavior can be understood by observing that, for $s=0$, full sliding occurs within the contact area since the substrate becomes effectively rigid, and the bristles of the block cannot sustain spatially varying deformations due to the absence of advection. As a consequence, the maximum friction force is already attained in the vicinity of the leading edge, and remains constant through the contact length.
The second relates to the adoption of higher-order rheological models for the substrate. As the number of dissipative branches increases, damping effects associated with the distributed nature of the sliding process lead to spatial relaxation behaviours, namely to an attenuation of the bristle force not only in time but also in space, resulting in a reduced friction force at the interface. This is consistent with the findings of \cite{FrBDvisc}, where only one of the contacting bodies was assumed to be deformable. 

\begin{figure}
\centering
\includegraphics[width=1\linewidth]{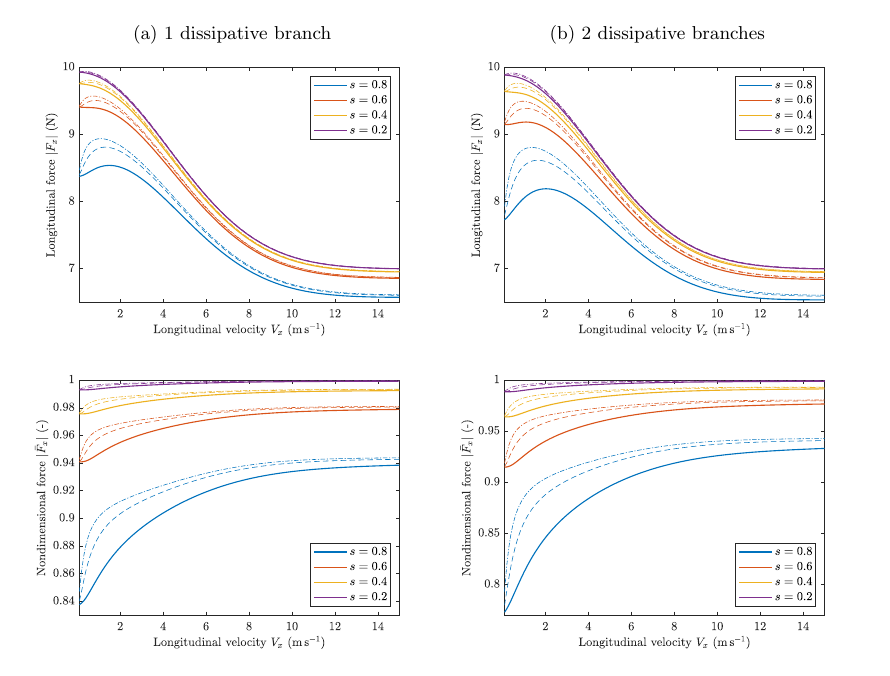} 
\caption{Steady-state longitudinal and normalised force $\abs{F_x(V_x,s)}$ and $\abs{\bar{F}_x(V_x,s)}$, respectively, for different values of the parameter $s=0.2,0.4,0.6$, and 0.8: (a) 1 dissipative branch; (b) 2 dissipative branches. Line styles: $\tau_{1,2} = \tau_{2,2} = 0.1$ s (solid thick lines), $\tau_{1,2} = \tau_{2,2} = 0.3$ s (dashed lines), $\tau_{1,2} = \tau_{2,2} = 0.6$ s (dash-dotted lines). Other model parameters as in Table~\ref{tab:parameters1}.}
\label{fig:Num2}
\end{figure}

The effect of different relaxation time constants on the steady-state response of the sliding frictional system was also investigated. Figures~\ref{fig:Num2}(a) and (b) report the trends of $F_x(V_x,s)$ and $\bar{F}_x(V_x,s)$ for different rheological orders of the substrate ($n_2 = 1$ and 2, respectively) and relaxation time constants $\tau_{i,2} = 0.1, 0.3$, and 0.6, for $i \in \{1,\dots,n_2\}$. It is evident from Fig.~\ref{fig:Num2} that, as the relaxation time constant increases, the force also increases, in agreement with the observations reported in \cite{FrBDvisc}.

Figures~\ref{fig:Num1} and~\ref{fig:Num2} were produced numerically in MATLAB\textsuperscript{\textregistered} environment, using a finite-difference scheme with 200 grid elements, and the parameter values listed in Table~\ref{tab:parameters1}, which are typical of rubber-like materials and inspired by \cite{FrBDvisc}. In particular, the damping coefficients for the dissipative branches of the bristle elements of the substrate were modelled as
\begin{align}\label{eq:cBSod}
c_{i,2} = \tau_{i,2}\bar{k}_{0,1}\dfrac{1-s}{s}, \quad i \in \{1,2\}.
\end{align}
Additional details on the numerical implementation are given in Appendix~\ref{app:numerical}. 

\begin{table}[h!]\centering 
\caption{Model parameters for sliding contact}
{\begin{tabular}{|c|l|c|c|}
\hline
Parameter & Description & Unit & Value \\
\hline 
$L$ & Contact length & m & 0.2\\
$\bar{k}_{0,1}$ &Micro-stiffness of body 1 (0th branch) & $\textnormal{m}^{-1}$ & 240 \\
$\bar{c}_{1,1}$ &Damping of body 1 (1st branch) & $\textnormal{m}^{-1}\,\textnormal{s}$ & 100 \\
$\bar{c}_{2,1}$ & Damping of body 1 (2nd branch) & $\textnormal{m}^{-1}\,\textnormal{s}$ & 50 \\
$\bar{c}_{1,2}$ &Damping of body 2 (1st branch) & $\textnormal{m}^{-1}\,\textnormal{s}$ & Eq. \eqref{eq:cBSod} \\
$\bar{c}_{2,2}$ & Damping of body 2 (2nd branch) & $\textnormal{m}^{-1}\,\textnormal{s}$ & Eq. \eqref{eq:cBSod} \\
$\tau_{1,1}$ & Relaxation time of body 1 (1st branch) & s &0.1 \\
$\tau_{2,1}$ & Relaxation time of body 1 (2nd branch) & s & 0.1 \\
$\tau_{1,2}$ & Relaxation time of body 2 (1st branch) & s &0.1 \\
$\tau_{2,2}$ & Relaxation time of body 2 (2nd branch) & s & 0.1 \\
$\mu\ped{s}$ & Static friction coefficient & -& 1\\
$\mu\ped{d}$ & Dynamic friction coefficient & -& 0.7\\
$\mu\ped{v}$ & Viscous friction coefficient & -& 0\\
$v\ped{S}$ & Stribeck velocity & $\textnormal{m}\,\textnormal{s}^{-1}$ & 6 \\
$\delta\ped{S}$ & Stribeck exponent & - & 2 \\
$F_{z}$ & Vertical force & N & 10\\
$\varepsilon$ & Regularisation parameter & $\textnormal{m}^2\,\textnormal{s}^{-2}$ & 0 \\ 
\hline
\end{tabular} }
\label{tab:parameters1}
\end{table}

\subsubsection{Transient sliding}\label{sect:TransSlid}

Phenomena similar to those identified in the steady contact regime can also be observed during transient sliding, albeit with minor differences introduced by the rheology of the block, which now actively contributes to the system response. In this context, it is worth noting that, according to Eq.~\eqref{€q:z1Slidx}, the deformation of the bristle elements attached to the block need not vanish at the leading edge, even though the corresponding tangential force and substrate deformation are identically zero. Indeed, enforcing the BC~\eqref{eq:BCfxSlid} and thus setting $f_x(0,t) = 0$ in Eq.~\eqref{€q:z1Slidx} provides
\begin{align}\label{eq:slidz1x0ppp}
\dpd{z_{i,1x}(0,t)}{t}  = -\dfrac{z_{i,1x}(0,t)}{\tau_{i,1}}, \quad i \in \{1,\dots,n_1\}, \; t \in (0,T), 
\end{align}
which implies that the bristle deformation at the leading edge of the block converges exponentially to zero, unless the internal states all start from an equilibrium configuration, in which case the solution to Eq.~\eqref{eq:slidz1x0ppp} is trivial. However, this is not necessarily the case, as can be readily understood by considering velocity-reversal scenarios.

\begin{figure}
\centering
\includegraphics[width=1\linewidth]{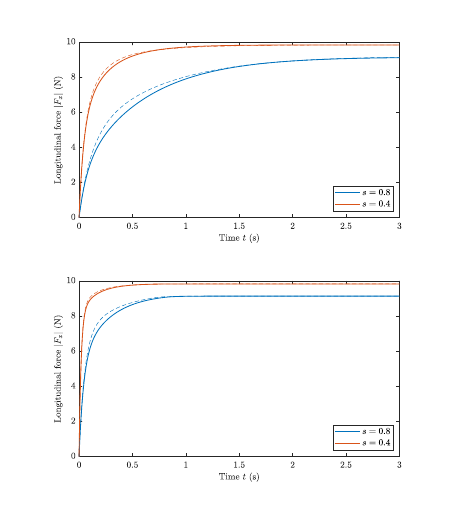} 
\caption{Transient system behaviour: (a) $V_x = 0.1$ $\textnormal{m}\,\textnormal{s}^{-1}$; (b) $V_x = 0.3$ $\textnormal{m}\,\textnormal{s}^{-1}$. Line styles: $\tau_{1,2} = \tau_{2,2} = 0.1$ s (solid thick lines), $\tau_{1,2} = \tau_{2,2} = 0.3$ s (dashed lines). Other model parameters as in Table~\ref{tab:parameters1}.}
\label{fig:Num3}
\end{figure}

Concerning instead the transient behaviour of the total friction force generated at the interface, Fig.~\ref{fig:Num3} depicts the evolution of $\abs{F_x(t)}$ for rheological orders $n_1=n_2 = 2$, two different values of the sliding velocity $V_x = 0.1$ and 0.3 $\textnormal{m}\,\textnormal{s}^{-1}$ (corresponding to Fig.~\ref{fig:Num3}(a) and (b), respectively), and $s = 0.4$ and 0.8. The simulation results reported in Fig.~\ref{fig:Num3} may be interpreted, for instance, as representative of velocity-controlled experiments involving rubber blocks slowly sliding over viscoelastic substrates.
The conclusions drawn from inspection of Fig.~\ref{fig:Num3} are essentially consistent with those obtained in the stationary regime: as $s$ decreases, larger forces are produced, and higher relaxation times lead to a slightly faster convergence toward the steady-state force values.
Figure~\ref{fig:Num3} was produced using a semi-discrete implementation combining the finite-difference spatial discretisation method introduced in Sect.~\ref{sect:SteadySlid} with \texttt{ode15s} for time resolution, as explained in Appendix~\ref{app:numerical}.

\subsection{Rolling friction}\label{sect:roll}

The rolling friction problem, although of clear interest from an application standpoint, is analysed in less detail than the sliding contact case, since the introduction of viscoelastic contact pairs does not alter the fundamental hyperbolic nature of the process, which is already well established in the literature. As in Sect.~\ref{sect:sliding}, attention is restricted to the line contact configuration, considering configurations similar to those illustrated in Figs.~\ref{fig:RollingBodies}(a) and (b). More specifically, a viscoelastic cylinder is considered, rolling over a substrate with rolling velocity $\mathbb{R}^2 \ni \bm{V}_1(\bm{x},t) = -[V\ped{r}(t)\; 0]^{\mathrm{T}}$, with $V\ped{r}(t) \in [V\ped{min},V\ped{max}]$, and translating longitudinally with forward velocity $\mathbb{R}^2 \ni \bm{V}_2(\bm{x},t) = [V_x(t)\; 0]^{\mathrm{T}}$, $V_x(t) \in [V\ped{min},V\ped{max}]$, with $0 < V\ped{min} \leq V\ped{max}$. Accordingly, the rigid relative velocity may be deduced as $\mathbb{R}^2 \ni \bm{v}\ped{r}(\bm{x},t) = [v_x(t)\; 0]^{\mathrm{T}} \triangleq [V_x(t)-V\ped{r}(t)\; 0]^{\mathrm{T}} $, and, after performing the same change of coordinate as in Sect.~\ref{sect:sliding}, the longitudinal components of Eqs.~\eqref{eq:PDEsinternalstates} and~\eqref{eq:PDEf_FrBD} become
\begin{subequations}\label{eq:PDEsinternalstatesR}
\begin{align}
\begin{split}\label{eq:fxRoll}
 \dpd{f_x(\xi,t)}{t}  + \tilde{V}_x(t)\dpd{f_x(\xi,t)}{\xi}  & =  -\bar{k}_0\dfrac{\abs{v_x(t)}_\varepsilon}{\mu\bigl(v_x(t)\bigr)}f_x(\xi,t)-\bar{k}_0v_x(t) \\
& \quad  - \bar{k}_0\sum_{i=1}^{n_1}\dfrac{\D_1 z_{i,1x}(\xi,t)}{\D t} +\bar{k}_0\sum_{i=1}^{n_2}\dfrac{\D_2 z_{i,2x}(\xi,t)}{\D t}, 
\end{split} \\
\begin{split}
\dfrac{\D_1 z_{i,1x}(\xi,t)}{\D t} & = \dpd{z_{i,1x}(\xi,t)}{t}+V\ped{r}(t)\dpd{z_{i,1x}(\xi,t)}{\xi}   = -\dfrac{z_{i,1x}(\xi,t)}{\tau_{i,1}} + \dfrac{f_x(\xi,t)}{\bar{c}_{i,1}}, \quad i \in \{1,\dots,n_1\}, \label{€q:z1Rollx}
\end{split}\\
\begin{split}
\dfrac{\D_2 z_{i,2x}(\xi,t)}{\D t} & = \dpd{z_{i,2x}(\xi,t)}{t} +V_x(t)\dpd{z_{i,2x}(\xi,t)}{\xi} \\
& = -\dfrac{z_{i,2x}(\xi,t)}{\tau_{i,2}} - \dfrac{f_x(\xi,t)}{\bar{c}_{i,2}}, \quad i \in \{1,\dots,n_2\}, \; (\xi,t) \in (0,L) \times (0,T), \label{€q:z2Rollx}
\end{split} \\
f_x(0,t) & = 0, \quad t \in (0,T),  \label{eq:BCfxRoll}\\
z_{i,1x}(0,t) & = 0, \quad i \in \{1,\dots,n_1\}, \;  t \in (0,T), \label{eq:BCz_1Roll}\\
z_{i,2x}(0,t) & = 0, \quad i \in \{1,\dots,n_2\}, \;  t \in (0,T), \label{eq:BCz_2Roll}\\
\bm{u}_x(\xi,0) & = \bm{u}_{x,0}(\xi), \quad \xi \in (0,L),
\end{align}
\end{subequations}
where $L \in \mathbb{R}_{>0}$ denotes the contact length, and $\mathbb{R}_{>0}\ni \tilde{V}_x(t) = V\ped{r}(t) + sv_x(t)$.

In contrast to Eq.~\eqref{eq:PDEsinternalstatesS}, there is no loss of advection in Eq.~\eqref{eq:PDEsinternalstatesR}, which retain their full PDE character for any $V\ped{r}(t), V_x(t) \in \mathbb{R}_{>0}$. This has important implications for both the steady-state and transient behaviour of the system, as discussed in the following Sects.~\ref{ect:SteadtRoll} and~\ref{sect:rollTrans}. Again, for constant velocities, existence and uniqueness properties for the solution of Eq.~\eqref{eq:PDEsinternalstatesR} may be established by invoking Theorem~\ref{thm:wellP} in Appendix~\ref{sect:wellP}. Requiring instead $V\ped{r}, V_x \in C^1(\mathbb{R}_{\geq 0};[V\ped{min},V\ped{max}])$, the results contained in \cite{MScMath} may be used.

\subsubsection{Steady rolling}\label{ect:SteadtRoll}

Exactly as done in Sect.~\ref{sect:SteadySlid}, focusing initially on the pure elastic case permits to deduce a closed-form expression for the normalised bristle force,
\begin{align}\label{eq:fxRollSolEl}
f_x(\xi) & = -\sgn_{\varepsilon}(v_x)\mu(v_x)\bigggl[1-\exp\Biggl(-\dfrac{(1-s)\bar{k}_{0,1}\abs{v_x}_\varepsilon}{(V\ped{r} + sv_x)\mu(v_x)}\xi\Biggr)\bigggr], \quad \xi \in [0,L],
\end{align} 
from which the bristle deflections of the cylindrical roller and substrate may be easily determined.

Moreover, considering a constant pressure distribution yields the following formula for the total longitudinal force generated at the interfaces
\begin{align}\label{eq:FxRoll}
\begin{split}
\bar{F}_x(v_x,V\ped{r},s) \triangleq \dfrac{F_x(v_x,V\ped{r},s)}{\mu(v_x)F_z} & = -\sgn_\varepsilon(v_x)\Bigggl(1 -\dfrac{(V\ped{r} + sv_x)\mu(v_x)}{(1-s)\bar{k}_{0,1}\abs{v_x}_\varepsilon L} \bigggl[1-\exp\Biggl(-\dfrac{(1-s)\bar{k}_{0,1}\abs{v_x}_\varepsilon L}{(V\ped{r} + sv_x)\mu(v_x)}\Biggr)\bigggr]\Bigggr), \\
& \qquad \qquad \qquad \qquad \qquad\qquad \qquad (v_x,V\ped{r},s) \in \mathbb{R}\times[V\ped{min},V\ped{max}]\times(0,1).
\end{split}
\end{align}
Although exhibiting evident structural similarities with Eq.~\eqref{eq:FxSlid}, the expression for $\bar{F}_x(v_x,V\ped{r},s)$ derived in Eq.~\eqref{eq:FxRoll} also highlights important fundamental differences between steady sliding and rolling. First, the force in Eq.~\eqref{eq:FxRoll} depends on three distinct arguments, since the relative velocity no longer coincides with the forward speed, but is instead determined by the kinematic difference between the rolling and translational velocities. This arises from the dual advection mechanisms occurring within the contact region, as material particles from both the cylinder and the substrate continuously enter and leave the contact patch.
As a consequence, the resulting longitudinal force vanishes for $v_x = 0$, as directly indicated by the presence of the term $\sgn_{\varepsilon}(v_x)$ in Eq.~\eqref{eq:FxRoll} (even in the absence of regularisation). On the other hand, the qualitative dependence on the parameter s remains similar to that observed in the steady sliding case.

\begin{figure}
\centering
\includegraphics[width=1\linewidth]{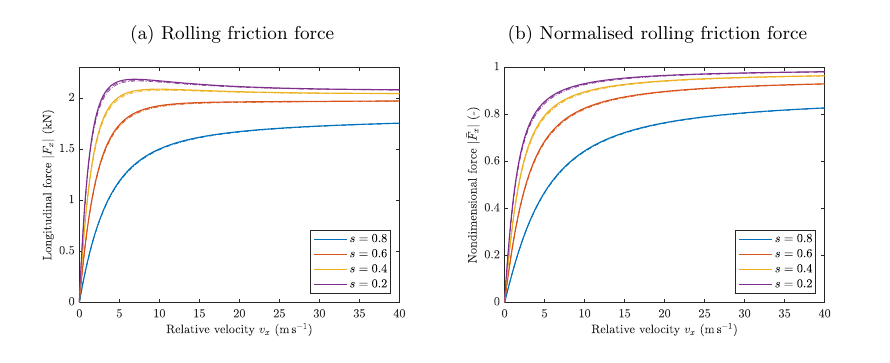} 
\caption{Steady-state longitudinal and normalised force $\abs{F_x(v_x,V\ped{r},s)}$ and $\abs{\bar{F}_x(v_x,V\ped{r},s)}$, respectively, for different values of the parameter $s=0.2,0.4,0.6$, and 0.8. Line styles: 0 dissipative branches (solid thick lines), 1 dissipative branch (dashed lines), 2 dissipative branches (dash-dotted lines). Model parameters as in Table~\ref{tab:parameters2}.}
\label{fig:Num4}
\end{figure}

Considering more complex, viscoelastic rheologies, these behaviours are visualised in Fig.~\ref{fig:Num4}, where the quantities $F_x(v_x,V\ped{r},s)$ and $\bar{F}_x(v_x,V\ped{r},s)$ are plotted for $s = 0.2, 0.4, 0.6$, and 0.8, and $n_1=n_2 = 0$, 1, and 2. Overall, the depicted trends are in theoretical agreement with the observations collected in Sect.~\ref{sect:SteadySlid}, with only a minor influence of higher-order rheologies on the generated forces. The parameter values used to produce Fig.~\ref{fig:Num4} are listed in Table~\ref{tab:parameters2}.

\begin{table}[h!]\centering 
\caption{Model parameters for rolling contact}
{\begin{tabular}{|c|l|c|c|}
\hline
Parameter & Description & Unit & Value \\
\hline 
$L$ & Contact length & m & 0.1\\
$\bar{k}_{0,1}$ &Micro-stiffness of body 1 (0th branch) & $\textnormal{m}^{-1}$ & 240 \\
$\bar{c}_{1,1}$ &Damping of body 1 (1st branch) & $\textnormal{m}^{-1}\,\textnormal{s}$ & 72 \\
$\bar{c}_{2,1}$ & Damping of body 1 (2nd branch) & $\textnormal{m}^{-1}\,\textnormal{s}$ & 78 \\
$\bar{c}_{1,2}$ &Damping of body 2 (1st branch) & $\textnormal{m}^{-1}\,\textnormal{s}$ & 18 \\
$\bar{c}_{2,2}$ & Damping of body 2 (2nd branch) & $\textnormal{m}^{-1}\,\textnormal{s}$ & 21.6 \\
$\tau_{1,1}$ & Relaxation time of body 1 (1st branch) & s &0.1 \\
$\tau_{2,1}$ & Relaxation time of body 1 (2nd branch) & s & 0.1 \\
$\tau_{1,2}$ & Relaxation time of body 2 (1st branch) & s &0.3 \\
$\tau_{2,2}$ & Relaxation time of body 2 (2nd branch) & s & 0.3 \\
$\mu\ped{s}$ & Static friction coefficient & -& 1.2\\
$\mu\ped{d}$ & Dynamic friction coefficient & -& 0.7\\
$\mu\ped{v}$ & Viscous friction coefficient & -& 0\\
$v\ped{S}$ & Stribeck velocity & $\textnormal{m}\,\textnormal{s}^{-1}$ & 3.49 \\
$\delta\ped{S}$ & Stribeck exponent & - & 0.6 \\
$F_{z}$ & Vertical force & N & 3000 \\
$\varepsilon$ & Regularisation parameter & $\textnormal{m}^2\,\textnormal{s}^{-2}$ & $10^{-12}$ \\ 
\hline
\end{tabular} }
\label{tab:parameters2}
\end{table}

\subsubsection{Transient rolling}\label{sect:rollTrans}

\begin{figure}
\centering
\includegraphics[width=1\linewidth]{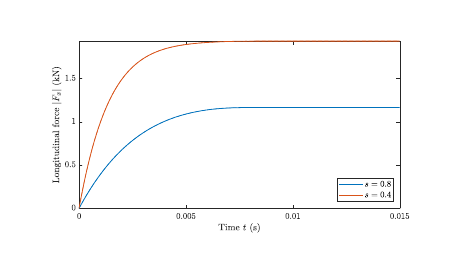} 
\caption{Transient system behaviour. Line styles: $\tau_{1,2} = \tau_{2,2} = 0.1$ s (solid thick lines), $\tau_{1,2} = \tau_{2,2} = 0.3$ s (dashed lines). Other model parameters as in Table~\ref{tab:parameters2}.}
\label{fig:Num5}
\end{figure}

For completeness, similar numerical tests to those reported in Sect.~\ref{sect:TransSlid} were conducted. Figure~\ref{fig:Num5} illustrates the relaxation behaviour predicted for the rolling system (with two dissipative branches) following the imposition of a rigid relative velocity step input $v_x = -3.2$ $\mathrm{m}\,\mathrm{s}^{-1}$, for constant values of $V\ped{r} = 16$ and $V_x = 12.8$ $\mathrm{m}\,\mathrm{s}^{-1}$, and starting from an initially unloaded tangential configuration. As expected, as the parameter $s$ decreases from 0.8 to 0.4, slightly higher longitudinal forces are produced over shorter time scales.

\section{Conclusions}\label{sect:conc}

This paper presented a general framework for the modelling of sliding and rolling friction between viscoelastic contact pairs. By combining linear viscoelastic rheologies with dynamic friction formulations based on bristle-like representations, a class of viscoelasto-kinematic equations was derived in the form of semilinear partial differential equations governing the evolution of frictional forces, bristle deformations, and internal state variables at the contact interface.

A central contribution of the work is the systematic extension of dynamic friction models, traditionally formulated as lumped systems of ordinary differential equations, to a distributed setting capable of accounting for the advection of material particles within the contact region. Starting from a general class of linear viscoelastic constitutive laws equivalent to Generalised Kelvin-Voigt rheologies, the corresponding viscoelasto-kinematic equations were derived and combined with established friction laws, leading to a unified PDE description of both sliding and rolling contact. This formulation naturally accommodates contact pairs in which the two bodies possess different elastic and viscoelastic properties, thereby extending previous analyses restricted to purely elastic systems.

From a mathematical standpoint, it was shown that the resulting governing equations retain a symmetric or symmetrisable hyperbolic structure under the assumption of small deformations and displacement gradients. Moreover, through a suitable transformation into Riemann variables, the system can be recast as an interconnection of transport equations propagating along characteristic directions determined by the elastic interaction between the contacting bodies. These results demonstrate that the introduction of linear viscoelasticity does not alter the fundamental hyperbolic character of frictional contact processes, but rather enriches their constitutive behaviour through additional internal dynamics.
Conversely, from a physical modelling perspective, the proposed framework removes several simplifying assumptions that are commonly adopted in existing rolling contact formulations. In particular, it allows both contacting bodies to possess arbitrary linear viscoelastic constitutive behaviour, without requiring one body to be rigid or both materials to be elastically similar. Consequently, the resulting equations remain applicable to a significantly broader class of engineering contacts, including tyre-road interaction, polymer contacts, rubber-covered rollers, wheel-rail systems with compliant layers, and many soft-contact interfaces.

The application of the theory to simple, albeit representative, sliding and rolling line contact problems in the absence of spin highlighted several interesting features. In the sliding case, mixed distributed formulations arise naturally whenever both contacting bodies are deformable, thereby challenging the traditional association between sliding friction and finite-dimensional models. In particular, the analysis conducted in the paper showed that the steady-state response is largely governed by the elastic partition of the deformation between the two bodies, whereas higher-order viscoelastic rheologies mainly influence transient behaviour and dissipation. In rolling contact, the presence of advection in both bodies preserves the fully distributed nature of the process and leads to force-generation mechanisms that differ fundamentally from those encountered in sliding contact, despite the strong mathematical analogies between the two regimes. Overall, the numerical examples evidenced that sliding and rolling friction, although often treated separately in the literature, may be interpreted within a common viscoelasto-kinematic framework. More generally, friction models should not be classified solely according to the constitutive laws adopted at the interface, but also according to the kinematic mechanisms governing the transport of material particles through the contact region. From this perspective, finite-dimensional and distributed formulations do not represent fundamentally different modelling paradigms, but different manifestations of the same underlying viscoelasto-kinematic structure. 

Finally, the present work also opens several directions for future research. In particular, the incorporation of more advanced constitutive descriptions, including nonlinear viscoelasticity and viscoplasticity and the development of efficient numerical schemes for large-scale simulations represent natural continuations of the proposed theory. Equally important is the experimental validation of the framework on practical systems involving highly viscoelastic materials, such as tyres, elastomeric components, and soft robotic interfaces, as well as more complex and realistic contact configurations accounting for spin. These developments may contribute to a deeper understanding of frictional contact phenomena and to the formulation of physically grounded friction models applicable across a broad range of engineering disciplines.

\section*{Funding declaration}
This research was financially supported by the project FASTEST (Reg. no. 2023-06511), funded by the Swedish Research Council. 


\section*{Compliance with Ethical Standards}

The author declares that there are no conflicts of interest directly or indirectly related to this work.

\section*{Data availability}
The datasets used and/or analysed during the current study are available from the corresponding author on reasonable request.

\section*{Author Contribution declaration}
L.R. is the sole author and contributor to the manuscript.

\appendix

\section{Technical results}\label{app:techn}
This appendix provides additional technical results. More specifically, Sect.~\ref{sect:appVarying} demonstrates that the hyperbolic nature of the equations is retained in the case of a variable contact area, whereas Sect.~\ref{sect:wellP} proves the well-posedness of the hyperbolic PDE system~\eqref{eq:PDEsystem}-\eqref{eq:funcH2} assuming time-independent contact areas and velocities $\bm{V}_1(\bm{x},t) = \bm{V}_1(\bm{x})$, $\bm{V}_2(\bm{x},t) = \bm{V}_2(\bm{x})$. Finally, Sect.~\ref{app:noncoaxial} discusses the extension of the viscoelasto-kinematic framework to the non-coaxial case.

\subsection{Hyperbolicity in the case of a time-varying contact area}\label{sect:appVarying}
The present appendix investigates the hyperbolic character of the PDE system~\eqref{eq:PDEsystem}-\eqref{eq:funcH2} in the situation of a time-varying contact region. The discussion is adapted from \cite{Tribology}. Accordingly, to cope with the fact that the domain $\mathring{\mathscr{C}}_\infty \triangleq \cup_{t \in \mathbb{R}} \mathring{\mathscr{C}} (t) \times \{t\}$ is noncylindrical, the following Assumption is introduced \cite{HeHsiao}.
\begin{assumption}\label{ass:1}
There exists a cylindrical domain $\mathring{\tilde{\mathscr{C}}}_\infty \triangleq \mathring{\tilde{\mathscr{C}}} \times \mathbb{R}$ and a level-preserving $C^\infty$-diffeomorphism $\bm{\Phi} : (\bm{\xi},\tau) = (\xi,\eta,\tau)= \bm{\Phi}(\bm{x},t) = [\bm{\Phi}_{\bm{x}}(\bm{x},t)^{\mathrm{T}} \; t]^{\mathrm{T}} =  [\Phi_x(\bm{x},t)\; \Phi_y(\bm{x},t)\; t]^{\mathrm{T}}$ from $\mathscr{C}_\infty \to \tilde{\mathscr{C}}_\infty$, with inverse denoted by $\bm{\Xi} : (\bm{x},t) = \bm{\Xi}(\bm{\xi},\tau) =[\bm{\Xi}_{\bm{\xi}}(\bm{\xi},\tau)^{\mathrm{T}}\; \tau]^{\mathrm{T}} = [\Xi_{\xi}(\bm{\xi},\tau)\; \Xi_{\eta}(\bm{\xi},\tau)\; \tau]^{\mathrm{T}}$.
\end{assumption}
Invoking Assumption~\ref{ass:1}, and defining $\bm{w}(\bm{\xi},\tau), \bm{w}_0(\bm{\xi},\tau)  \in \mathbb{R}^{2(n+1)}$ and $\mathbb{R}^2 \ni \bm{W}_1(\bm{\xi},\tau) = [W_{1\xi}(\bm{\xi},\tau)\; W_{1\eta}(\bm{\xi},\tau)]^{\mathrm{T}}$, $\mathbb{R}^2 \ni \bm{W}_2(\bm{\xi},\tau) = [W_{2\xi}(\bm{\xi},\tau)\; W_{2\eta}(\bm{\xi},\eta)]^{\mathrm{T}}$ as
\begin{subequations}
\begin{align}
\bm{w}(\bm{\xi},\tau) & = \eval{\bm{u}(\bm{x},t)}_{(\bm{x},t) = \bm{\Xi}(\bm{\xi},\tau)}, \\
\bm{w}_{0}(\bm{\xi}) & = \eval{\bm{u}_{0}(\bm{x})}_{(\bm{x},0) = \bm{\Xi}(\bm{\xi},0)}, \\
\bm{W}_1(\bm{\xi},\tau)&  = \eval{\biggl(\bigl(\bm{V}_1(\bm{x},t) \cdot \nabla_{\bm{x}}\bigr) \bm{\Phi}_{\bm{x}}(\bm{x},t)+ \dpd{\bm{\Phi}_{\bm{x}}(\bm{x},t)}{t}\biggr)}_{(\bm{x},t) = \bm{\Xi}(\bm{\xi},\tau)}, \label{eq:vTauFixed} \\
\bm{W}_2(\bm{\xi},\tau)&  = \eval{\biggl(\bigl(\bm{V}_2(\bm{x},t) \cdot \nabla_{\bm{x}}\bigr) \bm{\Phi}_{\bm{x}}(\bm{x},t)+ \dpd{\bm{\Phi}_{\bm{x}}(\bm{x},t)}{t}\biggr)}_{(\bm{x},t) = \bm{\Xi}(\bm{\xi},\tau)},
\end{align}
\end{subequations}
the original problem described by Eqs.~\eqref{eq:PDEsystem}-\eqref{eq:funcH2} becomes
\begin{align}\label{eq:inTermsOfq4}
\dpd{\bm{w}(\bm{\xi},\tau)}{\tau} + \mathbf{\Upsilon}_\xi(\bm{\xi},\tau)\dpd{\bm{w}(\bm{\xi},\tau)}{\xi} +\mathbf{\Upsilon}_\eta(\bm{\xi},\tau)\dpd{\bm{w}(\bm{\xi},\tau)}{\eta}  & = \tilde{\bm{g}}\bigl(\bm{w}(\bm{\xi},\tau),\bm{\xi},\tau), \quad
(\bm{\xi},\tau) \in \mathring{\tilde{\mathscr{C}}} \times [0,T],
\end{align}
with the matrices $\mathbf{\Upsilon}_\xi(\bm{\xi},\tau), \mathbf{\Upsilon}_\eta(\bm{\xi},\tau) \in \mathbf{Sym}_{2(n+1)}(\mathbb{R})$ given by
\begin{subequations}
\begin{align}
\mathbf{\Upsilon}_\xi(\bm{\xi},\tau)  &\triangleq \begin{bmatrix}\mathbf{S}_1W_{1\xi}(\bm{\xi},\tau)+\mathbf{S}_2W_{2\xi}(\bm{\xi},\tau) & \mathbf{0} & \mathbf{0} \\ \mathbf{0} & \mathbf{I}_{2n_1}W_{1\xi}(\bm{\xi},\tau) & \mathbf{0} \\ \mathbf{0} & \mathbf{0} & \mathbf{I}_{2n_2}W_{2\xi}(\bm{\xi},\tau) \end{bmatrix}, \\
\mathbf{\Upsilon}_\eta(\bm{\xi},\tau) & \triangleq \begin{bmatrix}\mathbf{S}_1W_{1\eta}(\bm{\xi},\tau)+\mathbf{S}_2W_{2\eta}(\bm{\xi},\tau) & \mathbf{0} & \mathbf{0} \\ \mathbf{0} & \mathbf{I}_{2n_1}W_{1\eta}(\bm{\xi},\tau) & \mathbf{0} \\ \mathbf{0} & \mathbf{0} & \mathbf{I}_{2n_2}W_{2\eta}(\bm{\xi},\tau) \end{bmatrix},
\end{align}
\end{subequations}
and $\tilde{\bm{g}} : \mathbb{R}^{2(n+2)}\times \mathbb{R}_{\geq 0} \to \mathbb{R}^{2(n+1)}$ defined as
\begin{align}\label{eq:funcH23}
\tilde{\bm{g}}(\bm{x},\bm{\xi},\tau) \triangleq \eval{\tilde{\bm{h}}(\bm{u},\bm{x},t)}_{(\bm{u},\bm{x},\tau) = (\bm{w},\bm{\Xi}(\bm{\xi},\tau))}.
\end{align}
By inspection of Eqs.~\eqref{eq:inTermsOfq4}-\eqref{eq:funcH23}, it may be easily concluded that the existence of the diffeomorphism $(\bm{\xi},\tau) = \bm{\Phi}(\bm{x},t)$ according to Assumption~\ref{ass:1} preserves the hyperbolic character of the PDE system. This observation is formalised below in Lemma~\ref{lemma:hyper2}.  
\begin{lemma}[Hyperbolicity]\label{lemma:hyper2}
Under Assumptions~\ref{ass:comm} and~\ref{ass:s,allGrad}, the PDE system~\eqref{eq:inTermsOfq4}-\eqref{eq:funcH23} is symmetric hyperbolic.
\begin{proof}
The proof is identical to that of Lemma~\ref{lemma:hyper}.
\end{proof}
\end{lemma}
Furthermore, exactly as done in Sect.~\ref{sect:Riemann}, the hyperbolic system~\eqref{eq:inTermsOfq4}-\eqref{eq:funcH23} may be reformulated in terms of Riemann variables. Following a similar rationale, it is possible to demonstrate the correctness of the BCs~\eqref{eq:BCs} also in the case of a time-varying contact area.

\subsection{Well-posedness}\label{sect:wellP}
This appendix is concerned with the well-posedness of the following semilinear hyperbolic PDE system, expressed in Riemann variables and postulated on a time-invariant domain:
\begin{subequations}\label{eq:quasiStationary0}
\begin{align}
& \dpd{\tilde{\bm{u}}(\bm{x},t)}{t} + \tilde{\mathbf{\Lambda}}_x(\bm{x})\dpd{\tilde{\bm{u}}(\bm{x},t)}{x} + \tilde{\mathbf{\Lambda}}_y(\bm{x})\dpd{\tilde{\bm{u}}(\bm{x},t)}{y} = \tilde{\bm{h}}\bigl(\tilde{\bm{u}}(\bm{x},t),\bm{x},t\bigr),\quad (\bm{x},t) \in \mathring{\mathscr{C}}\times (0,T), \\
& \tilde{f}_1(\bm{x},t) = 0, \quad (\bm{x},t) \in \tilde{\mathscr{L}}_1\times (0,T), \label{eq:BCquasistesd1} \\
& \tilde{f}_2(\bm{x},t) = 0, \quad (\bm{x},t) \in \tilde{\mathscr{L}}_2 \times (0,T), \\
& \bm{\zeta}_1(\bm{x},t) = \bm{0}, \quad (\bm{x},t) \in \mathscr{L}_1\times (0,T), \\
& \bm{\zeta}_2(\bm{x},t) = \bm{0}, \quad (\bm{x},t) \in \mathscr{L}_2\times (0,T),  \label{eq:BCquasistesd4}\\
& \tilde{\bm{u}}(\bm{x},0) = \tilde{\bm{u}}_0(\bm{x}), \quad \bm{x} \in \mathring{\mathscr{C}},
\end{align}
\end{subequations}
with the matrices $\tilde{\mathbf{\Lambda}}_x(\bm{x})$ and $\tilde{\mathbf{\Lambda}}_y(\bm{x})$ as in Eq.~\eqref{eq:Lambdatilde}, but constant in time.
To prove existence and uniqueness results, the PDE~\eqref{eq:quasiStationary0} is recast in abstract form as
\begin{align}
\dod{\tilde{\bm{u}}(t)}{t}& = \mathscr{A}\tilde{\bm{u}}(t) + \tilde{\bm{H}}\bigl(\tilde{\bm{u}}(t),t\bigr), \quad t \in (0,T),\\
\tilde{\bm{u}}(0) & = \tilde{\bm{u}}_0,
\end{align}
with $\tilde{\bm{H}} : L^2(\mathring{\mathscr{C}};\mathbb{R}^{2(n+1)})\times[0,T] \to L^2(\mathring{\mathscr{C}};\mathbb{R}^{2(n+1)})$ defined as $(\tilde{\bm{H}}(\tilde{\bm{u}},t))(\bm{x}) \triangleq \tilde{\bm{h}}(\tilde{\bm{u}}(\bm{x}),\bm{x},t)$, the unbounded operator $(\mathscr{A},\mathscr{D}(\mathscr{A}))$ given by
\begin{align}
(\mathscr{A}\bm{v})(\bm{x}) & \triangleq \bigl(\diag\{\tilde{\mathscr{A}}_1,\tilde{\mathscr{A}}_2,\mathscr{A}_1,\mathscr{A}_2\}\bm{v}\bigr)(\bm{x}), \\
\mathscr{D}(\mathscr{A}) & \triangleq \mathscr{D}(\tilde{\mathscr{A}}_1) \times \mathscr{D}(\tilde{\mathscr{A}}_2) \times \mathscr{D}(\mathscr{A}_1)\times\mathscr{D}(\mathscr{A}_2),
\end{align}
and the operators $(\tilde{\mathscr{A}}_i,\mathscr{D}(\tilde{\mathscr{A}}_i))$ and $(\mathscr{A}_i,\mathscr{D}(\mathscr{A}_i))$, $i \in \{1,2\}$, reading respectively
\begin{subequations}
\begin{align}
(\tilde{\mathscr{A}}_iv)(\bm{x}) & \triangleq -\bigl(\tilde{\bm{V}}_i(\bm{x})\cdot \nabla_{\bm{x}}\bigr)v(\bm{x}), \\
\mathscr{D}(\tilde{\mathscr{A}}_i) & \triangleq \Bigl\{v\in L^2(\mathring{\mathscr{C}};\mathbb{R}) \mathrel{\Big|} (\tilde{\bm{V}}_i\cdot\nabla_{\bm{x}})v \in  L^2(\mathring{\mathscr{C}};\mathbb{R}), \; \eval[0]{v}_{\tilde{\mathscr{L}}_i} = 0 \Bigr\}, \quad i \in \{1,2\},
\end{align}
\end{subequations}
and
\begin{subequations}
\begin{align}
(\mathscr{A}_i\bm{v})(\bm{x}) & \triangleq -\bigl(\bm{V}_i(\bm{x})\cdot \nabla_{\bm{x}}\bigr)\bm{v}(\bm{x}), \\
\mathscr{D}(\mathscr{A}_i) & \triangleq \Bigl\{\bm{v}\in L^2(\mathring{\mathscr{C}};\mathbb{R}^{2n_i}) \mathrel{\Big|} (\bm{V}_i\cdot\nabla_{\bm{x}})\bm{v} \in  L^2(\mathring{\mathscr{C}};\mathbb{R}^{2n_i}), \; \eval[0]{\bm{v}}_{\mathscr{L}_i} = \bm{0} \Bigr\}, \quad i \in \{1,2\}.
\end{align}
\end{subequations}
Existence and uniqueness results for the solution of Eq.~\eqref{eq:quasiStationary0} are asserted by Theorem~\ref{thm:wellP}.
\begin{theorem}[Existence and uniqueness of solutions]\label{thm:wellP}
Suppose that $\bm{V}_i \in C^1(\mathscr{C};\mathbb{R}^2)$, $i \in \{1,2\}$, $\mathring{\mathscr{C}}\subset \mathbb{R}^2$ is bounded with boundary $\partial \mathscr{C}$ piecewise $C^1$.
Then, if $\tilde{\bm{H}} : L^2(\mathring{\mathscr{C}};\mathbb{R}^{2(n+1)})\times [0,T] \to L^2(\mathring{\mathscr{C}};\mathbb{R}^{2(n+1)})$ is continuous in $t$ on $[0,T]$, and uniformly Lipschitz continuous on $L^2(\mathring{\mathscr{C}};\mathbb{R}^{2(n+1)})$, the PDE~\eqref{eq:quasiStationary0} admits a unique \emph{mild solution} $\tilde{\bm{u}} \in C^0([0,T];L^2(\mathring{\mathscr{C}};\mathbb{R}^{2(n+1)}))$ for all ICs $\tilde{\bm{u}}_0 \in L^2(\mathring{\mathscr{C}};\mathbb{R}^{2(n+1)})$. Additionally, if $\tilde{\bm{H}} : L^2(\mathring{\mathscr{C}};\mathbb{R}^{2(n+1)})\times [0,T] \to L^2(\mathring{\mathscr{C}};\mathbb{R}^{2(n+1)})$ is continuously differentiable from $L^2(\mathring{\mathscr{C}};\mathbb{R}^{2(n+1)})\times [0,T] $ into $L^2(\mathring{\mathscr{C}};\mathbb{R}^{2(n+1)})$, and the IC $\tilde{\bm{u}}_0 \in \mathscr{D}(\mathscr{A})$, the solution is \emph{classical}, that is, $\tilde{\bm{u}} \in C^1([0,T];L^2(\mathring{\mathscr{C}};\mathbb{R}^{2(n+1)})) \cap C^0([0,T];\mathscr{D}(\mathscr{A}))$.
\begin{proof}
First, it may be realised that Eqs.~\eqref{eq:princVel} and~\eqref{eq:eigens}, in conjunction with $\bm{V}_i \in C^1(\mathscr{C};\mathbb{R}^2)$, $i \in \{1,2\}$, imply $\tilde{\bm{V}}_i \in C^1(\mathscr{C};\mathbb{R}^2)$, $i \in \{1,2\}$. 
Therefore, under the stated assumptions on $\bm{V}_i \in C^1(\mathscr{C};\mathbb{R}^2)$, $i \in \{1,2\}$, and $\partial \mathscr{C}$, $(\tilde{\mathscr{A}}_i,\mathscr{D}(\tilde{\mathscr{A}}_i))$, $i \in \{1,2\}$, generate a $C_0$-semigroup on $L^2(\mathring{\mathscr{C}};\mathbb{R})$, and $(\mathscr{A}_1,\mathscr{D}(\mathscr{A}_1))$, $(\mathscr{A}_2,\mathscr{D}(\mathscr{A}_2))$ generate $C_0$-semigroups on $L^2(\mathring{\mathscr{C}};\mathbb{R}^{2n_1})$ and $L^2(\mathring{\mathscr{C}};\mathbb{R}^{2n_2})$, respectively \cite{Bardos}. Hence, the operator $(\mathscr{A},\mathscr{D}(\mathscr{A}))$ generates a $C_0$-semigroup on $L^2(\mathring{\mathscr{C}};\mathbb{R}) \times L^2(\mathring{\mathscr{C}};\mathbb{R}) \times L^2(\mathring{\mathscr{C}};\mathbb{R}^{2n_1}) \times L^2(\mathring{\mathscr{C}};\mathbb{R}^{2n_2}) \cong L^2(\mathring{\mathscr{C}};\mathbb{R}^{2(n+1)})$. The claims follow then from an application of Theorems 6.1.2 and 6.1.5 from \cite{Pazy}, which ensure the existence and uniqueness of mild and classical solutions $\tilde{\bm{u}} \in C^0([0,T];L^2(\mathring{\mathscr{C}};\mathbb{R}^{2(n+1)}))$ and $\tilde{\bm{u}} \in C^1([0,T];L^2(\mathring{\mathscr{C}};\mathbb{R}^{2(n+1)})) \cap C^0([0,T];\mathscr{D}(\mathscr{A}))$.
\end{proof}
\end{theorem}
The assumptions of Theorem~\ref{thm:wellP} obviously hold for the case of frictionless contact, as described by Eq.~\eqref{eq:h0frictionless}. Global mild and classical solutions may instead be obtained for the LuGre and FrBD models with non-regularised and regularised friction laws, respectively, in the absence of spin or when the rigid relative velocity $\bm{v}\ped{r}(\bm{z},\bm{\zeta},\bm{x},t)$ is approximated as $\bm{v}\ped{r}(\bm{0},\bm{0},\bm{x},t)$ in $\mathbf{M}^{-2}(\bm{v}\ped{r})\norm{\mathbf{M}(\bm{v}\ped{r})\bm{v}\ped{r}}_{2,\varepsilon}$ (see, e.g., \cite{FrBDroll,FrBDvisc}), or, as more common in the literature, cross-terms involving spin and internal variables are disregarded. The latter is indeed the standard approach in rolling contact mechanics, especially when dealing with wheel-rail interactions \cite{KalkerBook,2000RCP}. Alternatively, by adopting sufficiently smooth friction laws, more regular solutions can be recovered under non-uniform Lipschitz conditions by specifying additional compatibility conditions and exploiting the dissipativity of the nonlinear terms, applying similar techniques as those presented in \cite{Krstic2,Coron}.

\subsection{Extension to non-coaxial elastic stiffness matrices}
\label{app:noncoaxial}

Under Assumption~\ref{ass:comm}, the analysis conducted in the main part of the paper requires that the elastic stiffness matrices $\bar{\mathbf K}_{0,1}$ and $\bar{\mathbf K}_{0,2}$ commute.
This expedient allows the same symmetric matrices $\mathbf{S}_1$ and $\mathbf{S}_2$ to describe both the partition of the relative deformation and the principal transport operator of the force equation, enabling an orthogonal transformation into principal force components.

This appendix shows that, if coaxiality/commutativity is not imposed, the correctly reformulated system remains symmetrisable hyperbolic. However, the matrices appearing in the deformation-partition relations and in the force-transport operator no longer coincide. Consequently, the non-coaxial case requires a modification of Eqs.~\eqref{eq:PDEsystem}-\eqref{eq:funcH2} and does not generally admit the orthogonal characteristic representation developed in Sect.~\ref{ects:hyper}. In this context, it is crucial to observe that the matrices $\mathbf{S}_1, \mathbf{S}_2 \in \mathbf{M}_2(\mathbb{R})$ introduced in Eq.~\eqref{eq:matricesS} may be rewritten as
\begin{subequations}\label{eq:matricesSnonComm}
\begin{align}
\mathbf{S}_1 & \triangleq \bigl(\bar{\mathbf{K}}_{0,1} + \bar{\mathbf{K}}_{0,2}\bigr)^{-1}\bar{\mathbf{K}}_{0,2} = \bar{\mathbf{K}}_{0,1}^{-1}\bar{\mathbf{K}}_0, \\
\mathbf{S}_2 & \triangleq \bigl(\bar{\mathbf{K}}_{0,1} + \bar{\mathbf{K}}_{0,2}\bigr)^{-1}\bar{\mathbf{K}}_{0,1} = \bar{\mathbf{K}}_{0,2}^{-1}\bar{\mathbf{K}}_0,
\end{align}
\end{subequations}
and still satisfy $\mathbf{S}_1\mathbf{S}_2 = \mathbf{S}_2\mathbf{S}_1$ and $\mathbf{S}_1 + \mathbf{S}_2 = \mathbf{I}_2$. Consequently, the relationships~\eqref{eq:z1z2_z} remain valid without requiring coaxiality/commutativity. By contrast, Eq.~\eqref{eq:PDEf} becomes
\begin{align}\label{eq:PDEfnonComm}
\begin{split}
\dpd{\bm{f}(\bm{x},t)}{t} + \Bigr[\bigl(\bm{V}_1(\bm{x},t)\cdot \nabla_{\bm{x}}\bigr)\mathbf{S}_1^{\mathrm{T}} + \bigl(\bm{V}_2(\bm{x},t)\cdot \nabla_{\bm{x}}\bigr)\mathbf{S}_2^{\mathrm{T}}\Bigl]\bm{f}(\bm{x},t) & = \tilde{\bm{h}}_0\bigl(\bm{f}(\bm{x},t),\bm{\zeta}(\bm{x},t),\bm{x},t\bigr), \\
& \qquad \bm{x}\in \mathring{\mathscr{C}}(t), \; t \in (0,T),
\end{split}
\end{align}
with $\tilde{\bm{h}}_0 : \mathbb{R}^{2(n+2)}\times \mathbb{R}_{\geq 0} \to \mathbb{R}^2$ now defined as
\begin{align}\label{eq:tildeHnonComm}
\begin{split}
\tilde{\bm{h}}_0(\bm{f},\bm{\zeta},\bm{x},t) &\triangleq \bar{\mathbf{K}}_0\bm{h}_0\Bigl(\bar{\mathbf{K}}_0^{-1}(\bm{f}-\bar{\mathbf{C}}\bm{\zeta}), \bm{f}, \bm{v}\ped{r}\bigl(\bar{\mathbf{K}}_0^{-1}(\bm{f}-\bar{\mathbf{C}}\bm{\zeta}),\bm{\zeta},\bm{x},t\bigr)\Bigr) \\
& \quad + \mathbf{S}_1^{\mathrm{T}}\mathbf{C}_1\bm{h}_1(\bm{f},\bm{\zeta})-\mathbf{S}_2^{\mathrm{T}}\mathbf{C}_2\bm{h}_2(\bm{f},\bm{\zeta}),
\end{split}
\end{align}
and clearly $\mathbf{S}_1^{\mathrm{T}}\mathbf{S}_2^{\mathrm{T}} = \mathbf{S}_2^{\mathrm{T}}\mathbf{S}_1^{\mathrm{T}}$ and $\mathbf{S}_1^{\mathrm{T}} + \mathbf{S}_2^{\mathrm{T}} = \mathbf{I}_2$. Since the PDEs~\eqref{eq:PDEzeta1zeta2} for the internal state variables remain unchanged, introducing the vector $\mathbb{R}^{2(n+1)} \ni \bm{u}(\bm{x},t) \triangleq [\bm{f}^{\mathrm{T}}(\bm{x},t)\; \bm{\zeta}^{\mathrm{T}}(\bm{x},t)]^{\mathrm{T}}$ allows writing the resulting system formally as in Eq.~\eqref{eq:PDEsystem}, but with the matrices $\mathbf{\Lambda}_x(\bm{x},t), \mathbf{\Lambda}_y(\bm{x},t) \in \mathbf{M}_{2(n+1)}(\mathbb{R})$ given by
\begin{subequations}\label{eq:LambndasNonComm}
\begin{align}
\mathbf{\Lambda}_x(\bm{x},t) \triangleq \begin{bmatrix}\mathbf{S}_1^{\mathrm{T}}V_{1x}(\bm{x},t)+\mathbf{S}_2^{\mathrm{T}}V_{2x}(\bm{x},t) & \mathbf{0} & \mathbf{0} \\ \mathbf{0} & \mathbf{I}_{2n_1}V_{1x}(\bm{x},t) & \mathbf{0} \\ \mathbf{0} & \mathbf{0} & \mathbf{I}_{2n_2}V_{2x}(\bm{x},t) \end{bmatrix}, \\
\mathbf{\Lambda}_y(\bm{x},t) \triangleq \begin{bmatrix}\mathbf{S}_1^{\mathrm{T}}V_{1y}(\bm{x},t)+\mathbf{S}_2^{\mathrm{T}}V_{2y}(\bm{x},t) & \mathbf{0} & \mathbf{0} \\ \mathbf{0} & \mathbf{I}_{2n_1}V_{1y}(\bm{x},t) & \mathbf{0} \\ \mathbf{0} & \mathbf{0} & \mathbf{I}_{2n_2}V_{2y}(\bm{x},t) \end{bmatrix}.
\end{align}
\end{subequations}
Although not generally symmetric, the matrices $\mathbf{\Lambda}_x(\bm{x},t), \mathbf{\Lambda}_y(\bm{x},t) \in \mathbf{M}_{2(n+1)}(\mathbb{R})$ in Eq.~\eqref{eq:LambndasNonComm} admit a common symmetric positive-definite symmetriser.

\begin{lemma}[Symmetrisable hyperbolicity]
\label{lem:noncoaxial_hyperbolicity}
Suppose that $\mathbf{Sym}_2(\mathbb{R}) \ni \bar{\mathbf{K}}_{0,1}, \bar{\mathbf{K}}_{0,2} \succ \mathbf{0}$.
Then, under Assumption~\ref{ass:s,allGrad}, the PDE system~\eqref{eq:PDEsystem} with~\eqref{eq:tildeHnonComm} and~\eqref{eq:LambndasNonComm} is symmetrisable hyperbolic.
\begin{proof}
First, the matrix $\mathbf{Sym}_{2(n+1)}(\mathbb{R}) \ni \mathbf{G} \succ \mathbf{0}$ is defined as
\begin{align}\label{eq:noncoaxial_symmetriser}
\mathbf{G} & \triangleq \begin{bmatrix} \bar{\mathbf{K}}_0^{-1} & \mathbf{0} & \mathbf{0} \\ \mathbf{0} & \mathbf{I}_{2n_1} & \mathbf{0} \\ \mathbf{0} & \mathbf{0} & \mathbf{I}_{2n_2}\end{bmatrix}.
\end{align}
Equation~\eqref{eq:matricesSnonComm} also yields
\begin{subequations}
\begin{align}
\mathbf{S}_1^{\mathrm{T}} & = \bar{\mathbf{K}}_0\bar{\mathbf{K}}_{0,1}^{-1}, \label{eq:S1T}\\
\mathbf{S}_2^{\mathrm{T}} & = \bar{\mathbf{K}}_0\bar{\mathbf{K}}_{0,2}^{-1}.
\end{align}
\end{subequations}
Consequently,
\begin{subequations}
\begin{align}
\bar{\mathbf{K}}_0^{-1}\mathbf{S}_1^{\mathrm{T}} & = \bar{\mathbf{K}}_{0,1}^{-1}, \\
\bar{\mathbf{K}}_0^{-1}\mathbf{S}_2^{\mathrm{T}} & = \bar{\mathbf{K}}_{0,2}^{-1},
\end{align}
\end{subequations}
with $\bar{\mathbf{K}}_0^{-1}\mathbf{S}_1^{\mathrm{T}}, \bar{\mathbf{K}}_0^{-1}\mathbf{S}_2^{\mathrm{T}} \in \mathbf{Sym}_2(\mathbb{R})$. Therefore, using Eq.~\eqref{eq:LambndasNonComm}, it may be immediately inferred that
\begin{subequations}
\begin{align}
\mathbf{G}\mathbf{\Lambda}_x(\bm{x},t) & = \mathbf{\Lambda}_x^{\mathrm{T}}(\bm{x},t)\mathbf{G}, \\
\mathbf{G}\mathbf{\Lambda}_y(\bm{x},t) & = \mathbf{\Lambda}_y^{\mathrm{T}}(\bm{x},t)\mathbf{G}, \quad (\bm{x},t) \in \mathscr{C}\times[0,T].
\end{align}
\end{subequations}
Hence, for all $\mathbb{R}^2 \ni \bm{\alpha}=(\alpha_x,\alpha_y)$ and $(\bm{x},t) \in \mathscr{C}\times[0,T]$, the matrix $\mathbf{G}[\alpha_x\mathbf{\Lambda}_x(\bm{x},t) + \alpha_y\mathbf{\Lambda}_y(\bm{x},t)] \in \mathbf{Sym}_{2(n+1)}(\mathbb{R})$, proving the claim.
\end{proof}
\end{lemma}
According to Lemma~\ref{lem:noncoaxial_hyperbolicity} above, the removal of the coaxiality Assumption~\ref{ass:comm} does not alter the essentially hyperbolic character of the viscoelasto-kinematic equations, which was established in Sect.~\ref{ects:hyper}. The main difference with the coaxial case resides in the role of the matrices $\mathbf{S}_1$, $\mathbf{S}_2$, $\mathbf{S}_1^{\mathrm{T}}$, and $\mathbf{S}_2^{\mathrm{T}}$: whilst $\mathbf{S}_1$ and $\mathbf{S}_2$ continue to describe the elastic partition of the deformation between the two bodies in frictional contact, the characteristic transport structure of the resulting PDEs is determined by their transposes, $\mathbf{S}_1^{\mathrm{T}}$ and $\mathbf{S}_2^{\mathrm{T}}$.
In this context, it is essential to observe that $\mathbf{S}_1^{\mathrm{T}}$ and $\mathbf{S}_2^{\mathrm{T}}$ commute even when $\bar{\mathbf{K}}_{0,1}$ and $\bar{\mathbf{K}}_{0,2}$ do not. Moreover, Eqs.~\eqref{eq:noncoaxial_symmetriser} imply
\begin{subequations}
\begin{align}
\bar{\mathbf{K}}_0^{-1}\mathbf{S}_1^{\mathrm{T}} & = \bigl(\mathbf{S}_1^{\mathrm{T}}\bigr)^{\mathrm{T}}\bar{\mathbf{K}}_0^{-1} = \mathbf{S}_1\bar{\mathbf{K}}_0^{-1}, \\
\bar{\mathbf{K}}_0^{-1}\mathbf{S}_2^{\mathrm{T}} & = \bigl(\mathbf{S}_2^{\mathrm{T}}\bigr)^{\mathrm{T}}\bar{\mathbf{K}}_0^{-1}=\mathbf{S}_2\bar{\mathbf{K}}_0^{-1}.
\end{align}
\end{subequations}
Therefore, $\mathbf{S}_1^{\mathrm{T}}$ and $\mathbf{S}_2^{\mathrm{T}}$ are self-adjoint with respect to the inner product induced by $\bar{\mathbf{K}}_0^{-1}$. Since $\mathbf{S}_2^{\mathrm{T}}=\mathbf{I}_2-\mathbf{S}_1^{\mathrm{T}}$, the spectral theorem in the $\bar{\mathbf{K}}_0^{-1}$-weighted inner product implies that the two force-transport matrices may be simultaneously diagonalised by a common matrix $\mathbf{P}\in \mathbf{GL}_2(\mathbb{R})$, whose columns may be chosen $\bar{\mathbf{K}}_0^{-1}$-orthonormal. In particular,
\begin{subequations}
\begin{align}
\mathbf{P}^{-1}\mathbf{S}_1^{\mathrm{T}}\mathbf{P} & =\begin{bmatrix}\lambda_1(\mathbf{S}_1) & 0 \\ 0 & \lambda_2(\mathbf{S}_1) \end{bmatrix}, \\
\mathbf{P}^{-1}\mathbf{S}_2^{\mathrm{T}}\mathbf{P} &=\begin{bmatrix}\lambda_1(\mathbf{S}_2) & 0 \\ 0 & \lambda_2(\mathbf{S}_2) \end{bmatrix} = \mathbf{I}_2-\begin{bmatrix}\lambda_1(\mathbf{S}_1) & 0 \\ 0 & \lambda_2(\mathbf{S}_1) \end{bmatrix}, \\
\mathbf{P}^{\mathrm{T}}\bar{\mathbf{K}}_0^{-1}\mathbf{P} & = \mathbf{I}_2. \label{eq:PMP}
\end{align}
\end{subequations}
Consequently, even in the non-commuting or non-coaxial case, it is possible to restate the resulting PDE system in Riemann variables using a transformation of the form $\mathbb{R}^2 \ni \tilde{\bm{f}}(\bm{x},t) \triangleq \mathbf{P}^{-1}\bm{f}(\bm{x},t)$. Indeed, defining $ \mathbf{Sym}_2(\mathbb{R}) \ni \mathbf{D} \triangleq \bar{\mathbf{K}}_0^{\frac{1}{2}} \bar{\mathbf{K}}_{0,1}^{-1} \bar{\mathbf{K}}_0^{\frac{1}{2}} \succ \mathbf{0}$, it follows from Eq.~\eqref{eq:S1T} that 
\begin{align} 
\mathbf{S}_1^{\mathrm{T}} = \bar{\mathbf{K}}_0^{\frac{1}{2}} \mathbf{D} \bar{\mathbf{K}}_0^{-\frac{1}{2}}. \label{eq:noncoaxial_similarity_M} 
\end{align} 
Hence, $\mathbf{S}_1^{\mathrm{T}}$ is similar to the symmetric positive-definite matrix $\mathbf{D}$. Furthermore, since $\mathbf{0} \prec \bar{\mathbf{K}}_{0,1}^{-1} \prec \bar{\mathbf{K}}_0^{-1}$, congruence with $\bar{\mathbf{K}}_0^{\frac{1}{2}}$ gives 
\begin{align} 
\mathbf{0} \prec \mathbf{D} = \bar{\mathbf{K}}_0^{\frac{1}{2}} \bar{\mathbf{K}}_{0,1}^{-1} \bar{\mathbf{K}}_0^{\frac{1}{2}} \prec \mathbf{I}_2. \label{eq:noncoaxial_D1_bounds} 
\end{align} 
Since similar matrices have the same eigenvalues, it follows that $\lambda_1(\mathbf{S}_1^{\mathrm{T}}), \lambda_2(\mathbf{S}_1^{\mathrm{T}}) \in(0,1)$. Consequently, the principal velocities remain convex combinations of the two material transport velocities, as given in Eq.~\eqref{eq:princVel}. Finally, selecting $\mathbf{P}$ to satisfy Eq.~\eqref{eq:PMP}, introducing the matrix $\mathbf{Q} \in \mathbf{GL}_{2(n+1)}(\mathbb{R})$ as
\begin{align}
\mathbf{Q} \triangleq \begin{bmatrix} \mathbf{P} & \mathbf{0} & \mathbf{0} \\ \mathbf{0} & \mathbf{I}_{2n_1} & \mathbf{0} \\ \mathbf{0} & \mathbf{0} & \mathbf{I}_{2n_2}\end{bmatrix},
\end{align}
and applying the transformation $\mathbb{R}^{2(n+1)} \ni \tilde{\bm{u}}(\bm{x},t) \triangleq \mathbf{Q}^{-1}\bm{u}(\bm{x},t) = [\tilde{\bm{f}}^{\mathrm{T}}(\bm{x},t)\; \bm{\zeta}^{\mathrm{T}}(\bm{x},t)]^{\mathrm{T}}$ yields a system in Riemann variables whose principal part is formally identical to that in Eq.~\eqref{eq:PDEsystemR}, and possessing a real diagonal principal part in the generalised Riemann variables. Well-posedness results may therefore be established by invoking again Theorem~\ref{thm:wellP}.

\section{FrBD$_{n+1}$-GKV model for viscoelastic contact pairs}\label{app:FrBD}
This appendix provides additional details on the two-dimensional FrBD$_{n+1}$-GKV models for viscoelastic contact pairs, obtained by specifying the function $\bm{h}_0 : \mathbb{R}^6 \to \mathbb{R}^2$ as in Eq.~\eqref{eq:h0FrBD}. \emph{Mutatis mutandis}, their derivation may be conducted exactly as in \cite{FrBDvisc}, and is not repeated here. Instead, Sect.~\ref{sect:appRq} directly provides the full set of equations in a form that is amenable to the study of passivity, which is subsequently proved in Sect.~\ref{sect:passivity} under standard assumptions.

\subsection{Model equations}\label{sect:appRq}
The equations for the FrBD$_{n+1}$-GKV are stated in the sequel considering time-independent contact areas and velocities $\bm{V}_1(\bm{x},t) = \bm{V}_1(\bm{x})$, $\bm{V}_2(\bm{x},t) = \bm{V}_2(\bm{x})$; the extension to the time-varying case is however immediate. 
Accordingly, assuming that the rheologies of the bristles attached to the upper and lower bodies are governed by GKV elements with $n_1$ and $n_2$ branches, the equations for the internal variables $\mathbb{R}^{2n_1} \ni \bm{\zeta}_1(\bm{x},t) \triangleq [\bm{z}_{1,1}^{\mathrm{T}}(\bm{x},t)\; \dots \; \bm{z}_{n_1,1}^{\mathrm{T}}(\bm{x},t)]^{\mathrm{T}}$ and $\mathbb{R}^{2n_2} \ni \bm{\zeta}_2(\bm{x},t) \triangleq [\bm{z}_{1,2}^{\mathrm{T}}(\bm{x},t)\; \dots \; \bm{z}_{n_2,2}^{\mathrm{T}}(\bm{x},t)]^{\mathrm{T}}$ may be inferred in components as
\begin{subequations}\label{eq:PDEsinternalstates}
\begin{align}
\begin{split}
\dfrac{\D_1 \bm{z}_{i,1}(\bm{x},t)}{\D t} & = \dpd{\bm{z}_{i,1}(\bm{x},t)}{t} + \bigl(\bm{V}_1(\bm{x})\cdot\nabla_{\bm{x}}\bigr)\bm{z}_{i,1}(\bm{x},t) \\
& = -\bm{\tau}_{i,1}^{-1}\bm{z}_{i,1}(\bm{x},t) + \bar{\mathbf{C}}_{i,1}^{-1}\bm{f}(\bm{x},t), \quad i \in \{1,\dots,n_1\}, 
\end{split}\\
\begin{split}
\dfrac{\D_2 \bm{z}_{i,2}(\bm{x},t)}{\D t} & = \dpd{\bm{z}_{i,2}(\bm{x},t)}{t} + \bigl(\bm{V}_2(\bm{x})\cdot\nabla_{\bm{x}}\bigr)\bm{z}_{i,2}(\bm{x},t) \\
& = -\bm{\tau}_{i,2}^{-1}\bm{z}_{i,2}(\bm{x},t) - \bar{\mathbf{C}}_{i,2}^{-1}\bm{f}(\bm{x},t), \quad i \in \{1,\dots,n_2\}, \; (\bm{x},t) \in \mathring{\mathscr{C}} \times (0,T), 
\end{split}
\end{align}
\end{subequations}
where $\bm{\tau}_{i,1} \in \mathbf{M}_2(\mathbb{R})$, $i \in \{1,\dots,n_1\}$, and $\bm{\tau}_{i,2} \in \mathbf{M}_2(\mathbb{R})$, $i \in \{1,\dots,n_2\}$, are the matrices of relaxation times for the two bodies, respectively:
\begin{subequations}
\begin{align}
\bm{\tau}_{i,1} & \triangleq \bar{\mathbf{K}}_{i,1}^{-1}\bar{\mathbf{C}}_{i,1}, \quad i \in \{1,\dots,n_1\}, \\
\bm{\tau}_{i,2} & \triangleq \bar{\mathbf{K}}_{i,2}^{-1}\bar{\mathbf{C}}_{i,2}, \quad i \in \{1,\dots,n_2\}, 
\end{align}
\end{subequations}
$\mathbf{Sym}_2(\mathbb{R}) \ni \bar{\mathbf{K}}_{i,1} \succ \mathbf{0}$, $i \in \{0,\dots,n_1\}$, $\mathbf{Sym}_2(\mathbb{R}) \ni \bar{\mathbf{K}}_{i,2} \succ \mathbf{0}$, $i \in \{0,\dots,n_2\}$, are the corresponding normalised stiffness matrices, and $\mathbf{Sym}_2(\mathbb{R}) \ni \bar{\mathbf{C}}_{i,1} \succ \mathbf{0}$, $i \in \{1,\dots,n_1\}$, $\mathbf{Sym}_2(\mathbb{R}) \ni \bar{\mathbf{C}}_{i,2} \succ \mathbf{0}$, $i \in \{1,\dots,n_2\}$, are matrices of normalised damping coefficients.
Equations~\eqref{eq:PDEsinternalstates} may be easily reformulated in terms of Riemann variables using the transformation $\bm{f}(\bm{x},t) = \mathbf{R}\tilde{\bm{f}}(\bm{x},t)$.

Consequently, inserting Eqs.~\eqref{eq:PDEsinternalstates} into
\begin{align}\label{eq:PDEf_FrBD}
\begin{split}
 \dpd{\bm{f}(\bm{x},t)}{t}   + \Bigr[\bigl(\bm{V}_1(\bm{x})\cdot \nabla_{\bm{x}}\bigr)\mathbf{S}_1 & + \bigl(\bm{V}_2(\bm{x})\cdot \nabla_{\bm{x}}\bigr)\mathbf{S}_2\Bigl]\bm{f}(\bm{x},t)  =  \bar{\mathbf{K}}_0\mathbf{\Psi}\bigl(\bm{v}\ped{r}(\bm{x},t)\bigr)\bm{f}(\bm{x},t)-\bar{\mathbf{K}}_0\bm{v}\ped{r}(\bm{x},t)  \\
& \quad  - \bar{\mathbf{K}}_0\sum_{i=1}^{n_1}\dfrac{\D_1 \bm{z}_{i,1}(\bm{x},t)}{\D t} + \bar{\mathbf{K}}_0\sum_{i=1}^{n_2}\dfrac{\D_2 \bm{z}_{i,2}(\bm{x},t)}{\D t}, \quad (\bm{x},t) \in \mathring{\mathscr{C}} \times (0,T),
\end{split}
\end{align}
with $\mathbf{\Psi} : \mathbb{R}^2\to \mathbf{M}_2(\mathbb{R})$ reading
\begin{align}
\mathbf{\Psi}(\bm{v}\ped{r}) \triangleq -\mathbf{M}^{-2}(\bm{v}\ped{r})\norm{\mathbf{M}(\bm{v}\ped{r})\bm{v}\ped{r}}_{2,\varepsilon}.
\end{align}
Alternatively, in Riemann coordinates, Eq.~\eqref{eq:PDEf_FrBD} becomes
\begin{align}\label{eq:PDEf_FrBD_riemm}
\begin{split}
 \dpd{\tilde{\bm{f}}(\bm{x},t)}{t}  & + \begin{bmatrix} \tilde{\bm{V}}_1(\bm{x})\cdot\nabla_{\bm{x}} & 0 \\ 0 & \tilde{\bm{V}}_2(\bm{x})\cdot\nabla_{\bm{x}} \end{bmatrix}\tilde{\bm{f}}(\bm{x},t)  =  \mathbf{R}^{\mathrm{T}}\bar{\mathbf{K}}_0\mathbf{\Psi}\bigl(\bm{v}\ped{r}(\bm{x},t)\bigr)\mathbf{R}\tilde{\bm{f}}(\bm{x},t)-\mathbf{R}^{\mathrm{T}}\bar{\mathbf{K}}_0\bm{v}\ped{r}(\bm{x},t)  \\
& \quad  - \mathbf{R}^{\mathrm{T}}\bar{\mathbf{K}}_0\sum_{i=1}^{n_1}\dfrac{\D_1 \bm{z}_{i,1}(\bm{x},t)}{\D t} + \mathbf{R}^{\mathrm{T}}\bar{\mathbf{K}}_0\sum_{i=1}^{n_2}\dfrac{\D_2 \bm{z}_{i,2}(\bm{x},t)}{\D t}, \quad (\bm{x},t) \in \mathring{\mathscr{C}}\times (0,T).
\end{split}
\end{align}
It is straightforward to realise that the PDEs~\eqref{eq:PDEf_FrBD} and~\eqref{eq:PDEsinternalstates} may be recast in the form~\eqref{eq:quasiStationary0}. Well-posedness results may then be established by invoking Theorem~\ref{thm:wellP} under appropriate assumptions on the system's coefficients and input velocities. More specifically, for $\mathbf{M}\in C^0(\mathbb{R}^2;\mathbf{Sym}_2(\mathbb{R}))$, it is sufficient to require $\bm{V}_1,\bm{V}_2 \in C^1(\mathscr{C};\mathbb{R}^2)$, $\bm{v}\ped{r} \in C^0(\mathscr{C}\times[0,T];\mathbb{R}^2)$ and $\mathbf{Q}^{\mathrm{T}}\bm{u}_0 \in L^2(\mathring{\mathscr{C}};\mathbb{R}^{2(n+1)})$ for mild solutions $\mathbf{Q}^{\mathrm{T}}\bm{u} \in C^0([0,T];L^2(\mathring{\mathscr{C}};\mathbb{R}^{2(n+1)}))$. For classical solutions $\mathbf{Q}^{\mathrm{T}}\bm{u} \in C^1([0,T];L^2(\mathring{\mathscr{C}};\mathbb{R}^{2(n+1)})) \cap C^0([0,T];\mathscr{D}(\mathscr{A}))$, it may additionally be required that $\mathbf{M}\in C^1(\mathbb{R}^2;\mathbf{Sym}_2(\mathbb{R}))$, $\bm{v}\ped{r} \in C^1(\mathscr{C}\times[0,T];\mathbb{R}^2)$, and $\mathbf{Q}^{\mathrm{T}}\bm{u}_0\in \mathscr{D}(\mathscr{A})$.

\subsection{Passivity}\label{sect:passivity}
Frictional processes are inherently dissipative, converting mechanical energy into heat at the interface of contacting bodies. In mathematical and systems-theoretic terms, this behaviour is captured by the notion of passivity. Concerning the FrBD$_{n+1}$-GKV models for viscoelastic contact pairs, passivity is formalised in Definition~\ref{def:passivity} below.
\begin{definition}[Passivity]\label{def:passivity}
The system described by the PDEs~\eqref{eq:PDEf_FrBD} and~\eqref{eq:PDEsinternalstates} with BCs~\eqref{eq:BC1_tilfed0},~\eqref{eq:BC2_tilfed0},~\eqref{eq:BC1_zeta}, and~\eqref{eq:BC2_zeta}, and  output~\eqref{eq:Force} and~\eqref{eq:Mz} (alternatively~\eqref{eq:Force} and~\eqref{eq:Mzapprox}) is called \emph{passive} if, for all inputs $\bm{v}\ped{r} \in C^1(\mathscr{C}\times\mathbb{R}_{\geq 0};\mathbb{R}^2) \cap L^\infty(\mathscr{C}\times\mathbb{R}_{\geq 0};\mathbb{R}^2)$, velocities $\bm{V}_1, \bm{V}_2 \in C^1(\mathscr{C};\mathbb{R}^2)$, and ICs $\mathbf{Q}^{\mathrm{T}}\bm{u}_0 \in \mathscr{D}(\mathscr{A})$, there exists a storage function $W : L^2(\mathring{\mathscr{C}};\mathbb{R}^{2(n+1)}) \to \mathbb{R}_{\geq 0}$ such that 
\begin{align}\label{eq:Fvres}
\begin{split}
&-\int_0^t \bigl\langle p(\cdot)\bm{f}(\cdot,t), \bm{v}\ped{r}(\cdot,t)\bigr\rangle_{L^2(\mathring{\mathscr{C}};\mathbb{R}^2)} \geq W\bigl(\bm{u}(\cdot,t)\bigr)-W\bigl(\bm{u}_0(\cdot)\bigr), \quad t \in [0,T].
\end{split}
\end{align}
\end{definition}
To prove that passivity holds for the FrBD$_{n+1}$-GKV models, it may be first noted that, under Assumption~\ref{ass:comm}, the matrix $\mathbf{R}$ introduced in Eq.~\eqref{eq:MR}  also diagonalises $\bar{\mathbf{K}}_0$, and automatically $\bar{\mathbf{K}}_0^{-1}$. In particular,
\begin{align}
\mathbf{R}^{\mathrm{T}}\bar{\mathbf{K}}_0^{-1}\mathbf{R} = \tilde{\mathbf{K}}_0^{-1},
\end{align}
with $\mathbf{Sym}_2(\mathbb{R}) \ni \tilde{\mathbf{K}}_0 \succ \mathbf{0}$ defined as
\begin{align}
\tilde{\mathbf{K}}_0 \triangleq \mathbf{R}^{\mathrm{T}}\bar{\mathbf{K}}_0\mathbf{R} = \begin{bmatrix} \tilde{k}_1 & 0 \\ 0 & \tilde{k}_2 \end{bmatrix}.
\end{align}
Accordingly, the energy stored in the elastic branches of the GKV elements reads
\begin{align}\label{eq:energy1}
\begin{split}
\mathcal{E}\ped{e}\bigl(\bm{f}(\cdot,t)\bigr) & =\dfrac{1}{2}  \iint_{\mathscr{C}} p(\bm{x})\bm{f}^{\mathrm{T}}(\bm{x},t)\bar{\mathbf{K}}_{0,1}^{-1}\bm{f}(\bm{x},t) \dif \bm{x} +\dfrac{1}{2}  \iint_{\mathscr{C}}p(\bm{x}) \bm{f}^{\mathrm{T}}(\bm{x},t)\bar{\mathbf{K}}_{0,2}^{-1}\bm{f}(\bm{x},t) \dif \bm{x} \\
& = \dfrac{1}{2} \iint_{\mathscr{C}}p(\bm{x})\bm{f}^{\mathrm{T}}(\bm{x},t)\bar{\mathbf{K}}_0^{-1}\bm{f}(\bm{x},t) \dif \bm{x} = \dfrac{1}{2} \iint_{\mathscr{C}}p(\bm{x})\tilde{\bm{f}}^{\mathrm{T}}(\bm{x},t)\tilde{\mathbf{K}}_0^{-1}\tilde{\bm{f}}(\bm{x},t) \dif \bm{x}.
\end{split}
\end{align}
On the other hand, the energy stored in the dissipative branches may be calculated as
\begin{align}\label{eq:energy2}
\mathcal{E}_{\textnormal{d},j}\bigl(\bm{\zeta}_j(\cdot,t)\bigr) =\dfrac{1}{2} \sum_{i=1}^{n_j}\iint_{\mathscr{C}} p(\bm{x})\bm{z}_{i,j}^{\mathrm{T}}(\bm{x},t)\bar{\mathbf{K}}_{i,j}\bm{z}_{i,j}(\bm{x},t) \dif \bm{x}, \quad j \in \{1,2\}.
\end{align}
The storage function in Definition~\ref{def:passivity} may be assembled by summing the energy contributions computed according to Eqs.~\eqref{eq:energy1} and~\eqref{eq:energy2}. Under Assumption~\ref{ass:comm}, the result is formalised, in Lemma~\ref{lemma:DissF202} below.

\begin{lemma}[Passivity of the FrBD$_{n+1}$-GKV models for viscoelastic contact pairs]\label{lemma:DissF202}
Under Assumption~\ref{ass:comm}, suppose that $p \in C^1(\mathscr{C};\mathbb{R}_{\geq 0})$ satisfies
\begin{align}\label{eq:condP}
\nabla_{\bm{x}}\cdot p(\bm{x})\bm{V}_i(\bm{x}) \leq 0, \quad \bm{x} \in \mathscr{C}, \; i \in \{1,2\}.
\end{align}
Then, the system described by the PDEs~\eqref{eq:PDEf_FrBD} and~\eqref{eq:PDEsinternalstates} with BCs~\eqref{eq:BC1_tilfed0},~\eqref{eq:BC2_tilfed0},~\eqref{eq:BC1_zeta}, and~\eqref{eq:BC2_zeta}, and output~\eqref{eq:Force} and~\eqref{eq:Mz} (alternatively~\eqref{eq:Force} and~\eqref{eq:Mzapprox}) is passive with storage function
\begin{align}\label{eq:VdissF23}
\begin{split}
W\bigl(\bm{u}(\cdot,t)\bigr) & \triangleq \mathcal{E}\ped{e}\bigl(\bm{f}(\cdot,t)\bigr) +\mathcal{E}_{\textnormal{d},1}\bigl(\bm{\zeta}_1(\cdot,t)\bigr)+\mathcal{E}_{\textnormal{d},2}\bigl(\bm{\zeta}_2(\cdot,t)\bigr)= \dfrac{1}{2}\iint_{\mathscr{C}} p(\bm{x}) \bm{f}^{\mathrm{T}}(\bm{x},t)\bar{\mathbf{K}}_0^{-1}\bm{f}(\bm{x},t) \dif \bm{x} \\
& \quad + \dfrac{1}{2}\sum_{i=1}^{n_1}\iint_{\mathscr{C}} p(\bm{x}) \bm{z}^{\mathrm{T}}_{i,1}(\bm{x},t)\bar{\mathbf{K}}_{i,1}\bm{z}_{i,1}(\bm{x},t)\dif \bm{x}+ \dfrac{1}{2}\sum_{i=1}^{n_2}\iint_{\mathscr{C}} p(\bm{x}) \bm{z}^{\mathrm{T}}_{i,2}(\bm{x},t)\bar{\mathbf{K}}_{i,2}\bm{z}_{i,2}(\bm{x},t)\dif \bm{x}. 
\end{split}
\end{align}
\begin{proof}
First, it may be observed that the condition stated in Eq.~\eqref{eq:condP} also implies
\begin{align}\label{eq:condP2}
\nabla_{\bm{x}}\cdot p(\bm{x})\tilde{\bm{V}}_i(\bm{x}) \leq 0, \quad \bm{x} \in \mathscr{C}, \; i \in \{1,2\}.
\end{align}
Moreover, differentiating Eq.~\eqref{eq:VdissF23} along the dynamics~\eqref{eq:PDEf_FrBD} and~\eqref{eq:PDEsinternalstates}, and recalling Eq.~\eqref{eq:energy1}, yields
\begin{align}\label{eq:WKVS}
\begin{split}
\dod{W\bigl(\bm{u}(\cdot,t)\bigr)}{t} &= -\bigl\langle p(\cdot)\bm{f}(\cdot,t),\bm{v}(\cdot,t)\bigr\rangle_{L^2(\mathring{\mathscr{C}};\mathbb{R}^2)} +\iint_{\mathscr{C}}p(\bm{x})\bm{f}^{\mathrm{T}}(\bm{x},t)\mathbf{\Psi}\bigl(\bm{v}(\bm{x},t)\bigr)\bm{f}(\bm{x},t) \dif \bm{x} \\
& \quad - \sum_{i=1}^{n_1} \iint_{\mathscr{C}}p(\bm{x})\bm{f}^{\mathrm{T}}(\bm{x},t)\dfrac{\D_1 \bm{z}_{i,1}(\bm{x},t)}{\D t} \dif \bm{x} +\sum_{i=1}^{n_2} \iint_{\mathscr{C}}p(\bm{x})\bm{f}^{\mathrm{T}}(\bm{x},t)\dfrac{\D_2 \bm{z}_{i,2}(\bm{x},t)}{\D t} \dif \bm{x} \\
& \quad + \sum_{i=1}^{n_1} \iint_{\mathscr{C}} p(\bm{x}) \biggl( \bm{f}(\bm{x},t)- \bar{\mathbf{C}}_{i,1}\dfrac{\D_1 \bm{z}_{i,1}(\bm{x},t)}{\D t}\biggr)^{\mathrm{T}}\dfrac{\D_1 \bm{z}_{i,1}(\bm{x},t)}{\D t}\dif \bm{x} \\
& \quad  -\sum_{i=1}^{n_2} \iint_{\mathscr{C}} p(\bm{x}) \biggl( \bm{f}(\bm{x},t)+ \bar{\mathbf{C}}_{i,2}\dfrac{\D_2 \bm{z}_{i,2}(\bm{x},t)}{\D t}\biggr)^{\mathrm{T}}\dfrac{\D_2 \bm{z}_{i,2}(\bm{x},t)}{\D t}\dif \bm{x} \\
& \quad - \dfrac{1}{2}\iint_{\mathscr{C}}p(\bm{x})\bigl(\tilde{\bm{V}}_1(\bm{x})\cdot\nabla_{\bm{x}}\bigr)\dfrac{\tilde{f}_1^2(\bm{x},t)}{\tilde{k}_1} \dif \bm{x}-\dfrac{1}{2}\iint_{\mathscr{C}}p(\bm{x})\bigl(\tilde{\bm{V}}_2(\bm{x})\cdot\nabla_{\bm{x}}\bigr)\dfrac{\tilde{f}_2^2(\bm{x},t)}{\tilde{k}_2} \dif \bm{x} \\
& \quad  - \dfrac{1}{2}\sum_{i=1}^{n_1} \iint_{\mathscr{C}}p(\bm{x})\bm{V}_1(\bm{x})\cdot\nabla_{\bm{x}}\Bigl(\bm{z}_{i,1}^{\mathrm{T}}(\bm{x},t)\bar{\mathbf{K}}_{i,1}\bm{z}_{i,1}(\bm{x},t)\Bigr) \dif \bm{x} \\
 & \quad - \dfrac{1}{2}\sum_{i=1}^{n_2} \iint_{\mathscr{C}}p(\bm{x})\bm{V}_2(\bm{x})\cdot\nabla_{\bm{x}}\Bigl(\bm{z}_{i,2}^{\mathrm{T}}(\bm{x},t)\bar{\mathbf{K}}_{i,2}\bm{z}_{i,2}(\bm{x},t)\Bigr) \dif \bm{x}, \quad t\in (0,T).
\end{split}
\end{align}
Integrating by parts the last four terms in Eq.~\eqref{eq:WKVS} and enforcing the BCs~\eqref{eq:BC1_tilfed0},~\eqref{eq:BC2_tilfed0},~\eqref{eq:BC1_zeta}, and~\eqref{eq:BC2_zeta} produces 
\begin{subequations}\label{eq:byParts}
\begin{align}
\begin{split}
& - \dfrac{1}{2}\iint_{\mathscr{C}}p(\bm{x})\bigl(\tilde{\bm{V}}_i(\bm{x})\cdot\nabla_{\bm{x}}\bigr)\dfrac{\tilde{f}_i^2(\bm{x},t)}{\tilde{k}_i} \dif \bm{x} = -\dfrac{1}{2}\int_{\tilde{\mathscr{T}}_i}p(\bm{x})\dfrac{\tilde{f}_i^2(\bm{x},t)}{\tilde{k}_i} \tilde{\bm{V}}_i(\bm{x}) \cdot \hat{\bm{n}}_{\partial \mathscr{C}}(\bm{x}) \dif L \\
& \quad  + \dfrac{1}{2} \iint_{\mathscr{C}} \bigl(\nabla_{\bm{x}}\cdot p(\bm{x})\tilde{\bm{V}}_i(\bm{x})\bigr)\dfrac{\tilde{f}_i^2(\bm{x},t)}{\tilde{k}_i} \dif \bm{x} \leq \dfrac{1}{2} \iint_{\mathscr{C}} \bigl(\nabla_{\bm{x}}\cdot p(\bm{x})\tilde{\bm{V}}_i(\bm{x})\bigr)\dfrac{\tilde{f}_i^2(\bm{x},t)}{\tilde{k}_i} \dif \bm{x}, \quad i \in \{1,2\},
\end{split} \\
\begin{split}
& - \dfrac{1}{2}\sum_{i=1}^{n_j} \iint_{\mathscr{C}}p(\bm{x})\bm{V}_j(\bm{x})\cdot\nabla_{\bm{x}}\Bigl(\bm{z}_{i,j}^{\mathrm{T}}(\bm{x},t)\bar{\mathbf{K}}_{i,j}\bm{z}_{i,j}(\bm{x},t)\Bigr) \dif \bm{x}  \\
& \quad = - \dfrac{1}{2}\sum_{i=1}^{n_j} \int_{\mathscr{T}_j}p(\bm{x}) \bm{z}_{i,j}^{\mathrm{T}}(\bm{x},t)\bar{\mathbf{K}}_{i,j}\bm{z}_{i,j}(\bm{x},t) \bm{V}_j(\bm{x}) \cdot \hat{\bm{n}}_{\partial \mathscr{C}}(\bm{x}) \dif L \\
& \qquad + \dfrac{1}{2}\iint_{\mathscr{C}}\bigl(\nabla_{\bm{x}} \cdot p(\bm{x})\bm{V}_j(\bm{x})\bigr)\bm{z}_{i,j}^{\mathrm{T}}(\bm{x},t)\bar{\mathbf{K}}_{i,j}\bm{z}_{i,j}(\bm{x},t) \dif \bm{x} \\
& \qquad \leq \dfrac{1}{2}\iint_{\mathscr{C}}\bigl(\nabla_{\bm{x}} \cdot p(\bm{x})\bm{V}_j(\bm{x})\bigr)\bm{z}_{i,j}^{\mathrm{T}}(\bm{x},t)\bar{\mathbf{K}}_{i,j}\bm{z}_{i,j}(\bm{x},t) \dif \bm{x}, \quad j \in \{1,2\}.
\end{split}
\end{align}
\end{subequations}
Consequently, the following bound may be deduced from Eqs.~\eqref{eq:WKVS} and~\eqref{eq:byParts}
\begin{align}\label{eq:WKVS2}
\begin{split}
\dod{W\bigl(\bm{u}(\cdot,t)\bigr)}{t} &\leq -\bigl\langle p(\cdot)\bm{f}(\cdot,t),\bm{v}(\cdot,t)\bigr\rangle_{L^2(\mathring{\mathscr{C}};\mathbb{R}^2)} +\iint_{\mathscr{C}}p(\bm{x})\bm{f}^{\mathrm{T}}(\bm{x},t)\mathbf{\Psi}\bigl(\bm{v}(\bm{x},t)\bigr)\bm{f}(\bm{x},t) \dif \bm{x} \\
& \quad - \sum_{i=1}^{n_1} \iint_{\mathscr{C}} p(\bm{x})\dfrac{\D_1 \bm{z}_{i,1}^{\mathrm{T}}(\bm{x},t)}{\D t}\bar{\mathbf{C}}_{i,1}\dfrac{\D_1 \bm{z}_{i,1}(\bm{x},t)}{\D t}\dif \bm{x} \\
& \quad - \sum_{i=1}^{n_2} \iint_{\mathscr{C}} p(\bm{x})\dfrac{\D_2 \bm{z}_{i,2}^{\mathrm{T}}(\bm{x},t)}{\D t}\bar{\mathbf{C}}_{i,2}\dfrac{\D_2 \bm{z}_{i,2}(\bm{x},t)}{\D t}\dif \bm{x} \\
& \quad +\dfrac{1}{2} \iint_{\mathscr{C}} \bigl(\nabla_{\bm{x}}\cdot p(\bm{x})\tilde{\bm{V}}_1(\bm{x})\bigr)\dfrac{\tilde{f}_1^2(\bm{x},t)}{\tilde{k}_1} \dif \bm{x} + \dfrac{1}{2} \iint_{\mathscr{C}} \bigl(\nabla_{\bm{x}}\cdot p(\bm{x})\tilde{\bm{V}}_2(\bm{x})\bigr)\dfrac{\tilde{f}_2^2(\bm{x},t)}{\tilde{k}_2} \dif \bm{x} \\
& \quad  + \dfrac{1}{2} \sum_{i=1}^{n_1}\iint_{\mathscr{C}}\bigl(\nabla_{\bm{x}} \cdot p(\bm{x})\bm{V}_1(\bm{x})\bigr) \bm{z}_{i,1}^{\mathrm{T}}(\bm{x},t)\bar{\mathbf{K}}_{i,1}\bm{z}_{i,1}(\bm{x},t) \dif \bm{x} \\
& \quad  + \dfrac{1}{2} \sum_{i=1}^{n_2}\iint_{\mathscr{C}}\bigl(\nabla_{\bm{x}} \cdot p(\bm{x})\bm{V}_2(\bm{x})\bigr) \bm{z}_{i,2}^{\mathrm{T}}(\bm{x},t)\bar{\mathbf{K}}_{i,2}\bm{z}_{i,2}(\bm{x},t) \dif \bm{x}, \quad t \in (0,T).
\end{split}
\end{align}
The second, third, and fourth terms appearing on the right-hand side of Eq.~\eqref{eq:WKVS2} are always negative semidefinite. Therefore, if the inequality~\eqref{eq:condP} is satisfied, it may be concluded that 
\begin{align}\label{eq:lastdd}
-\bigl\langle p(\cdot)\bm{f}(\cdot,t),\bm{v}(\cdot,t)\bigr\rangle_{L^2(\mathring{\mathscr{C}};\mathbb{R}^2)} \geq \dod{W\bigl(\bm{u}(\cdot,t)\bigr)}{t}, \quad t\in (0,T).
\end{align}
Finally, integrating the above Eq.~\eqref{eq:lastdd} proves the claim.
\end{proof}
\end{lemma}

Some conclusive observations are collected below.
\begin{remark}[Inequality~\eqref{eq:condP}]
The inequalities in Eq.~\eqref{eq:condP} are similar to those of Lemmata 4.1 and 4.2 of \cite{FrBDvisc}, and, when $\bm{V}_1(\bm{x})$ and $\bm{V}_2(\bm{x})$ have the same direction, may be verified simultaneously. When the transport velocities are constant in space, as it happens in the absence of spin, they are obviously satisfied by a constant pressure distribution. In viscoelastic rolling contact, the contact pressures may even decrease along the rolling direction, rendering the storage function $W(\bm{u}(\cdot,t))$ in Eq.~\eqref{eq:VdissF23} a Lyapunov function. 
\end{remark}

\begin{remark}[Extension to the non-coaxial case]
\label{rem:noncoaxial_passivity}
The passivity result established in Lemma~\ref{lemma:DissF202} extended to the non-coaxial case considered in Appendix~\ref{app:noncoaxial}. Indeed, introducing the generalised Riemann variables $\mathbb{R}^2 \ni \tilde{\bm{f}}(\bm{x},t) \triangleq \mathbf{P}^{-1}\bm{f}(\bm{x},t)$, recalling Eq.~\eqref{eq:PMP} and exploiting the fact that $\mathbf P^{-1}= \mathbf{P}^{\mathrm{T}}\bar{\mathbf{K}}_0^{-1}$, the energy stored in the elastic branches of the GKV elements may be expressed as
\begin{align}\label{eq:energy1P}
\begin{split}
\mathcal{E}\ped{e}\bigl(\bm{f}(\cdot,t)\bigr) & = \dfrac{1}{2} \iint_{\mathscr{C}}p(\bm{x})\bm{f}^{\mathrm{T}}(\bm{x},t)\bar{\mathbf{K}}_0^{-1}\bm{f}(\bm{x},t) \dif \bm{x} = \dfrac{1}{2} \iint_{\mathscr{C}}p(\bm{x})\tilde{\bm{f}}^{\mathrm{T}}(\bm{x},t)\tilde{\bm{f}}(\bm{x},t) \dif \bm{x},
\end{split}
\end{align}
from which an estimate analogous to~\eqref{eq:lastdd} may be deduced by specifying the appropriate BCs and under analogous assumptions as Lemma~\ref{lemma:DissF202}. 
\end{remark}

\section{Numerical implementation and convergence analysis}\label{app:numerical}
\label{app:numerical_convergence}

This appendix documents the numerical method used to solve the one-dimensional sliding and rolling contact problems introduced in Sect.~\ref{sect:appl}, and describes the verification and grid-convergence analyses. Specifically, Sect.~\ref{app:semi_discrete} provides exhaustive details about the numerical implementation, whereas convergence studies are presented in Sect.~\ref{app:numCond}. 

\subsection{Semi-discrete formulation}
\label{app:semi_discrete}

The spatial contact interval $[0,L]$ was partitioned into $N-1$ uniform cells, with nodal coordinates
\begin{equation}
    \xi_j=(j-1)\Delta\xi,
    \quad \textnormal{where}\quad
    \Delta\xi=\frac{L}{N-1},
    \label{eq:uniform_grid}
\end{equation}
and $j\in \{1,\dots,N\}$, $N\in\{51,101,201,401\}$. 
After spatial discretisation, the PDEs were written as a finite-dimensional ODE system and integrated in MATLAB\textsuperscript{\textregistered} using the variable step, variable-order solver \texttt{ode15s}, with relative and absolute tolerances $\mathtt{RelTol}=10^{-9}$ and $\mathtt{AbsTol}=10^{-11}$.
Although \texttt{ode15s} is an implicit adaptive solver and is not subject to
the stability restriction of an explicit upwind method, its maximum internal step is limited according to $\Delta t\ped{max} =0.8\frac{\Delta\xi}{c\ped{max}}$, where $c\ped{max}\in \mathbb{R}_{>0}$ denotes the largest magnitude amongst the transport velocities of the PDE variables. This restriction ensures adequate temporal resolution of material transport over one spatial cell and makes the spatial refinement comparison less sensitive to the adaptive time-step selection.

In particular, for the sliding problem in Eq.~\eqref{eq:PDEsinternalstatesS}, the nondimensional bristle force is transported with velocity $sV_x$, the internal variables associated with body~1 obey local ODEs, and the internal variables associated with body~2 are transported with velocity $V_x$. Consequently,
\begin{align}
    c\ped{max}=\max\{sV_x,V_x\}=V_x.
    \label{eq:cmax_sliding}
\end{align}
The BCs~\eqref{eq:BCfxSlid} and~\eqref{eq:BCz_2Slid} were imposed strongly at the first (inflow) grid point, corresponding to $\xi=0$.

On the other hand, for the rolling problem in Eq.~\eqref{eq:fxRoll}, the nondimensional bristle force is transported with velocity $V\ped{r}+s v_x$, whilst the internal variables of bodies~1 and~2 are transported with velocities $V\ped{r}$ and $V_x$, respectively. Accordingly,
\begin{equation}
    c\ped{max}=\max\Bigl\{\abs{(1-s)V\ped{r}+s V_x},\abs{V\ped{r}},\abs{V_x}\Bigr\}.
    \label{eq:cmax_rolling}
\end{equation}
The BCs~\eqref{eq:BCfxRoll},~\eqref{eq:BCz_1Roll}, and~\eqref{eq:BCz_2Roll} were again are imposed strongly at the first (inflow) grid point.

\subsection{Numerical convergence study}\label{app:numCond}
In the following, numerical convergence studies are conducted concerning both the steady-state problem (Appendix~\ref{app:elastic_verification}) and the transient one (Appendix~\ref{app:transient_convergence}). The analyses adopt the same model parameters as in Tables~\ref{tab:parameters1} and~\ref{tab:parameters2}.

\subsubsection{Verification in the purely elastic limit for the stationary solutions}
\label{app:elastic_verification}

The accuracy of the implementation was first verified in the purely elastic limit, for which the closed-form steady-state solutions~\eqref{eq:fxSlidSolEl} and~\eqref{eq:fxRollSolEl} are available.

Specifically, for a numerical stationary force $f_{x}^h(\xi) \in \mathbb{R}$, the relative spatial $L^2$- and $L^\infty$-errors were defined as
\begin{subequations}\label{eq:relative_Linf_error}
\begin{align}
 e\ap{rel}_{L^2}(h) & \triangleq   \norm{e\ap{rel}(\cdot,h)}_{L^2((0,L);\mathbb{R})} = \dfrac{\norm{f_x^h(\cdot)-f_x(\cdot)}_{L^2((0,L);\mathbb{R})}}{\norm{f_x(\cdot)}_{L^2((0,L);\mathbb{R})}}, \\
e\ap{rel}_{L^\infty}(h) & \triangleq  \norm{e\ap{rel}(\cdot,h)}_{L^\infty([0,L];\mathbb{R})} = \dfrac{\norm{f_x^h(\cdot)-f_x(\cdot)}_{L^\infty([0,L];\mathbb{R})}}{\norm{f_x(\cdot)}_{L^\infty([0,L];\mathbb{R})}}. 
\end{align}
\end{subequations}
All the integrals in Eqs.~\eqref{eq:relative_Linf_error} were evaluated using the composite trapezoidal rule.
The observed order between two successive grids is
\begin{align}
    q_k =\dfrac{\log\bigl(e_k\ap{rel}/e_{k+1}\ap{rel}\bigr)}{\log\bigl(\Delta\xi_k/\Delta\xi_{k+1}\bigr)},\label{eq:observed_order}
\end{align}
where $\mathbb{R}_{\geq 0} \ni e_k\ap{rel}\triangleq e\ap{rel}_{L^2}(\Delta \xi_k)$. Since the spatial derivatives are approximated using a first-order upwind formula, $q_k\rightarrow1$ is expected in the asymptotic refinement regime, provided that the temporal and stationary iteration errors are sufficiently smaller than the spatial error.

The numerical values of the quantities defined in Eqs.~\eqref{eq:relative_Linf_error} and~\eqref{eq:observed_order} are reported in Table~\ref{tab:elastic_convergence} for $s=0.4$. Note that the values for $q_k$ in the last column of Table~\ref{tab:elastic_convergence} confirm the observations reported above, indicating convergence towards the expected first-order asymptotic regime. 

\begin{table}[]
    \centering
    \caption{Elastic-limit verification against the analytical stationary
    solutions. The errors are evaluated using the relative spatial
    $L^2$- and $L^\infty$-spatial norms. The quantity $q$ denotes the
    observed convergence order in the relative $L^2$ norm}
    \label{tab:elastic_convergence}
    \begin{tabular}{llcccc}
        \toprule
        \textbf{Case} & $N$ & $\Delta\xi$ &
        $e\ap{rel}_{L^2}$ &
        $e\ap{rel}_{L^\infty}$ & $q$ \\
        \midrule
        \textbf{Sliding}
        & 51  & $4.0\times10^{-3}$ & $3.0745\times10^{-2}$ & $1.7291\times10^{-1}$ & -- \\
        & 101 & $2.0\times10^{-3}$ & $1.8002\times10^{-2}$ & $1.0109\times10^{-1}$ & $0.772$ \\
        & 201 & $1.0\times10^{-3}$ & $9.7582\times10^{-3}$ & $5.7947\times10^{-2}$ & $0.883$ \\
        & 401 & $5.0\times10^{-4}$ & $5.1011\times10^{-3}$ & $3.0836\times10^{-2}$ & $0.936$ \\
        \midrule
        \textbf{Rolling}
        & 51  & $2.0\times10^{-3}$ & $1.1760\times10^{-2}$ & $1.2909\times10^{-2}$ & -- \\
        & 101 & $1.0\times10^{-3}$ & $5.9426\times10^{-3}$ & $6.5491\times10^{-3}$ & $0.985$ \\
        & 201 & $5.0\times10^{-4}$ & $2.9873\times10^{-3}$ & $3.2981\times10^{-3}$ & $0.992$ \\
        & 401 & $2.5\times10^{-4}$ & $1.4977\times10^{-3}$ & $1.6550\times10^{-3}$ & $0.996$ \\
        \bottomrule
    \end{tabular}
\end{table}

\begin{figure}
\centering
\includegraphics[width=1\linewidth]{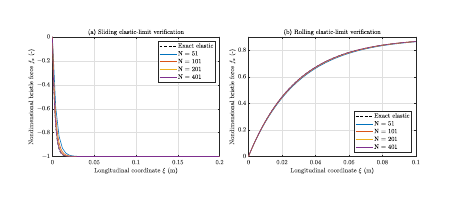} 
\caption{Trends of the nondimensional bristle forces computed numerically, $f_x^h(\xi)$, compared to the exact elastic solutions, for $N\in\{51,101,201,401\}$ and $s=0.4$: (a) sliding; (b) rolling.}
\label{fig:conv1}
\end{figure}

The trends of the nondimensional bristle forces computed numerically, $f_x^h(\xi)$, are compared to the exact elastic solutions in Fig.~\ref{fig:conv1}. In particular, inspection of Fig.~\ref{fig:conv1} reveals a faster convergence for the rolling contact case (Fig.~\ref{fig:conv1}(b)), although the results obtained in the sliding scenario (Fig.~\ref{fig:conv1}(a)) are already satisfactory for $N = 101$.

\begin{figure}
\centering
\includegraphics[width=1\linewidth]{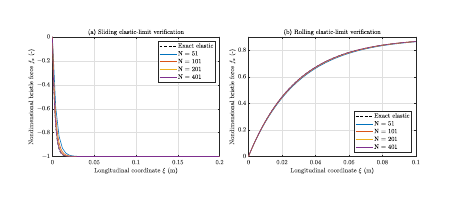} 
\caption{Trends of $L^2$-spatial and time errors for $N\in\{51,101,201,401\}$ and $s=0.4$: (a) elastic steady-state error $e\ap{rel}_{L^2}(h)$; (b) transient convergence error $e^F_{L^2}(h)$.}
\label{fig:conv2}
\end{figure}

The convergence of the error $e\ap{rel}_{L^2}(h)$ in logarithmic scale is visualised in Fig.~\ref{fig:conv2}(a). In particular, for rolling, the observed order $q_k$ increases from $0.985$ to $0.996$; for sliding, it increases from $0.772$ to $0.936$, indicating convergence towards the expected first-order asymptotic regime. These results are consistent with the first-order upwind spatial discretisation.

\subsubsection{Transient convergence analysis}
\label{app:transient_convergence}

Since an exact expression is not available for the full viscoelastic transient problems, the convergence of the numerical solutions was assessed by a grid-refinement study. Because the maximum internal integration step was reduced proportionally to $\Delta\xi$, the transient analysis presented in the following constitutes a coupled space-time refinement test. In this context, all force histories were sampled at the same output times, and the solution $F_{x,\textnormal{ref}}(t)\in \mathbb{R}$ obtained with the finest grid, corresponding to $N\ped{ref}=401$, was used as the reference curve. Accordingly, the relative $L^2$- and $L^\infty$-time errors between the numerical approximations $F_x^h(t) \in \mathbb{R}$ and the reference value were computed as
\begin{subequations}\label{eq:transient_L2_error}
\begin{align}
   e^F_{L^2}(h) & \triangleq \norm{e^F(\cdot,h)}_{L^2((0,T);\mathbb{R})} =  \dfrac{\norm{F_{x}^h(\cdot)-F_{x,\textnormal{ref}}(\cdot)}_{L^2((0,T);\mathbb{R})}}{\norm{F_{x,\textnormal{ref}}(\cdot)}_{L^2((0,T);\mathbb{R})}}, \\
  e^F_{L^\infty}(h) & \triangleq \norm{e^F(\cdot,h)}_{L^\infty([0,T];\mathbb{R})} =\dfrac{\norm{F_{x}^h(\cdot)-F_{x,\textnormal{ref}}(\cdot)}_{L^\infty([0,T];\mathbb{R})}}{\norm{F_{x,\textnormal{ref}}(\cdot)}_{L^\infty([0,T];\mathbb{R})}}.
\end{align}
\end{subequations}
The relative differences in the peak and terminal forces were also evaluated:
\begin{subequations}\label{eq:quasiErrors}
\begin{align}
    e^F\ped{peak}(h) & \triangleq \dfrac{\abs{\max_{t\in[0,T]}\abs{F_{x}^h(t)}-\max_{t\in[0,T]}\abs{F_{x,\textnormal{ref}}(t)}}}{\max_{t\in[0,T]}\abs{F_{x,\textnormal{ref}}(t)}}, \label{eq:peak_error} \\
e^F_T(h) & \triangleq \dfrac{ \abs{F_{x}^h(T)}-\abs{F_{x,\textnormal{ref}}(T)}}{\abs{F_{x,\textnormal{ref}}(T)}}.
    \label{eq:final_error}
\end{align}
\end{subequations}
The values of the errors defined in Eqs.~\eqref{eq:transient_L2_error} and~\eqref{eq:quasiErrors} for different spatial grid resolutions are listed in Table~\ref{tab:transient_convergence} for $s = 0.4$.

\begin{table}[]
    \centering
    \caption{Transient grid-refinement analysis of the resultant force $F_x$ for $s=0.4$. The
    solution obtained with $N_{\mathrm{ref}}=401$ is used as the reference.
    The reported quantities are the relative $L^2$- and
    $L^\infty$-time norms and the relative differences in the peak and
    terminal resultant forces. The zero entries for $N=401$ follow from the
    definition of the self-convergence measure}
    \label{tab:transient_convergence}
    \begin{tabular}{llccccc}
        \toprule
        \textbf{Case} & $N$ & $\Delta\xi$ &
        $e^F_{L^2}$ &
        $e^F_{L^\infty}$ &
        $e^F\ped{peak}$ &
        $e^F_T$ \\
        \midrule
        \textbf{Sliding}
        & 51  & $4.0\times10^{-3}$ & $9.0823\times10^{-3}$ & $9.1059\times10^{-3}$ & $9.0941\times10^{-3}$ & $9.0941\times10^{-3}$ \\
        & 101 & $2.0\times10^{-3}$ & $3.8934\times10^{-3}$ & $3.9046\times10^{-3}$ & $3.8985\times10^{-3}$ & $3.8985\times10^{-3}$ \\
        & 201 & $1.0\times10^{-3}$ & $1.2979\times10^{-3}$ & $1.3020\times10^{-3}$ & $1.2997\times10^{-3}$ & $1.2997\times10^{-3}$ \\
        & 401 & $5.0\times10^{-4}$ & $0$ & $0$ & $0$ & $0$ \\
        \midrule
        \textbf{Rolling}
        & 51  & $2.0\times10^{-3}$ & $1.0577\times10^{-2}$ & $1.1281\times10^{-2}$ & $1.0427\times10^{-2}$ & $1.0427\times10^{-2}$ \\
        & 101 & $1.0\times10^{-3}$ & $4.5455\times10^{-3}$ & $4.8781\times10^{-3}$ & $4.4818\times10^{-3}$ & $4.4818\times10^{-3}$ \\
        & 201 & $5.0\times10^{-4}$ & $1.5173\times10^{-3}$ & $1.6354\times10^{-3}$ & $1.4961\times10^{-3}$ & $1.4961\times10^{-3}$ \\
        & 401 & $2.5\times10^{-4}$ & $0$ & $0$ & $0$ & $0$ \\
        \bottomrule
    \end{tabular}
\end{table}

\begin{figure}
\centering
\includegraphics[width=1\linewidth]{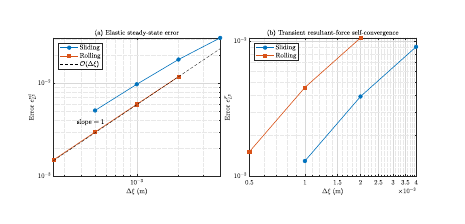} 
\caption{Trends of the longitudinal forces computed numerically, $F_x^h(t)$, for $N\in\{51,101,201,401\}$ and $s=0.4$: (a) sliding; (b) rolling.}
\label{fig:conv3}
\end{figure}

The trends of the longitudinal forces computed numerically, $F_x^h(t)$, are visualised in Fig.~\ref{fig:conv3}, where a rapid convergence may be observed for both the sliding and rolling contact cases. 
Finally, Fig.~\ref{fig:conv2}(b) illustrates a systematic decrease of the relative $L^2$-time differences under refinement. Since the solution $F_{x,\textnormal{ref}}(t)$ is used as the numerical reference, the finest-grid difference is zero by construction and is omitted from the logarithmic plot.

\end{document}